\documentclass[12pt]{article}
\usepackage{amsmath, amsthm, amssymb, enumerate, amsfonts, mathrsfs, graphicx, wrapfig,float,subfig,tikz}
\usepackage{amsrefs}
\usepackage{pgf}
\usepackage{stmaryrd} 
\usetikzlibrary{arrows}
\usetikzlibrary{snakes, shapes,arrows,patterns,decorations.pathreplacing,matrix}

\nocite{*}

\newtheorem{definition}{Definition}
\newtheorem{theorem}{Theorem}

\newcommand{\tr}{\operatorname{tr}}

\newcommand{\op}{\operatorname}

\newcommand{\comment}[1]{}
\title{Cosmology: macro and micro}
\author{Jonathan Holland and George Sparling\\Laboratory of Axiomatics, University of Pittsburgh}
\begin{document}
\maketitle
\pagestyle{empty}
\subsection*{Abstract}
A new approach to cosmology and space-time is developed, which emphasizes the description of the matter degrees of freedom of Einstein's theory of gravity by a family of K\"{a}hler-Einstein Fano manifolds.
\section*{Introduction}
Galileo Galilei formulated the principle of relativity in his account of physics in a moving ship, the thought experiment of the advocate, Salviati [32].
\\\\
\textit{"Shut yourself up with some friend in the main cabin below decks on some large ship, and have with you there some flies, butterflies, and other small flying animals. Have a large bowl of water with some fish in it; hang up a bottle that empties drop by drop into a wide vessel beneath it. With the ship standing still, observe carefully how the little animals fly with equal speed to all sides of the cabin. The fish swim indifferently in all directions; the drops fall into the vessel beneath; and, in throwing something to your friend, you need throw it no more strongly in one direction than another, the distances being equal; jumping with your feet together, you pass equal spaces in every direction. When you have observed all these things carefully (though doubtless when the ship is standing still everything must happen in this way), have the ship proceed with any speed you like, so long as the motion is uniform and not fluctuating this way and that.   You will discover not the least change in all the effects named, nor could you tell from any of them whether the ship was moving or standing still. " }
\eject\noindent
Johannes Kepler  found universal structure in the orbits of the planets [47]:\\\\
\textit{"The square of the periodic times are to each other as the cubes of the mean distances."}\\\\
He sought to bring order to the universe, using certain regular solids to try to parametrize the orbits.\\\\
\textit{"On the regular figures, the harmonic proportions they create, their source, their classes, their order, and their distinction into knowability and representability."}
\\\\
Sir Isaac Newton accounted for the motion of the planets and the trajectories of falling apples with his universal laws of gravity [54].   Newton's theory left two key issues unresolved:  first it lacked a detailed theory of matter, required in principle, since the matter influences the  gravitational field; second he was unable to address the issue of initial conditions: once the planets were in place he could predict their orbits, but he could not  explain how they got there.    \\\\
\textit{"Gravity explains the motions of the planets, but it cannot explain who sets the planets in motion." }\\\\
\textit{"This most beautiful system of the sun, planets and comets, could only proceed from the counsel and dominion of an intelligent and powerful Being."}\\\\
\textit{"To myself I am only a child playing on the beach, while vast oceans of truth lie undiscovered before me."}\\
\\Also Newton's theory suffered from the problem of instantaneous action at arbitrarily large distances, which might seem unphysical.  \\\\
\textit{"It is inconceivable that inanimate Matter should, without the Mediation of something else, which is not material, operate upon, and affect other matter without mutual Contact".} 
\eject\noindent
\textit{ "That Gravity should be innate, inherent and essential to Matter, so that one body may act upon another at a distance thro' a Vacuum, without the Mediation of any thing else, by and through which their Action and Force may be conveyed from one to another, is to me so great an Absurdity that I believe no Man who has in philosophical Matters a competent Faculty of thinking can ever fall into it. Gravity must be caused by an Agent acting constantly according to certain laws; but whether this Agent be material or immaterial, I have left to the Consideration of my readers."}\\\\
This latter problem became acute when Albert Einstein formulated his version of the relativity principle and introduced the concept of the maximal speed of propagation for causal influences [24].   Einstein realized that his ideas applied also to gravity:  indeed from the modern viewpoint the maximal propagation speed is that of the gravitational disturbances [25].  
\\\\
Hermann Minkowski supplied the mathematical framework for Einstein's theory, developing the idea of space-time, unifying the three dimensions of space with the one dimension of time in a four-dimensional arena [50].  \\\\
\textit{"The views of space and time which I wish to lay before you have sprung from the soil of experimental physics, and therein lies their strength. They are radical. Henceforth space by itself, and time by itself, are doomed to fade away into mere shadows, and only a kind of union of the two will preserve an independent reality."}\\\\
Then over a period of about seven years, Einstein first understood that the force of gravity could be incorporated in the metrical geometry of space-time and then eventually hit on his theory of general relativity [26].  This theory has stood to this day.   His field equations are conceptually extremely simple:  they equate the Einstein tensor of the gravitational field (a purely geometrical concept, based on the space-time metric and calculable directly, given the metric) with the energy-momentum tensor of the matter.    The key mathematical fact bolstering his theory was that both sides of this relation were divergence free.   The key physical fact was that in the weak field limit his theory incorporated Newton's theory, with corrections which in particular explained the precession of the perihelion of Mercury.  Einstein's  theory was supported immediately by the Lagrangian formulation of the theory by David Hilbert [40]. 
\\\\However, like Newton's theory, Einstein's theory is incomplete:  a priori one can take any metric,  work out its Einstein tensor and declare that tensor (perhaps constrained somewhat by positivity conditions) to be also the matter tensor, reducing the field  equations to  a tautology.   Also the same question about initial conditions lingers.   Einstein tried to address this question by applying his theory to the large scale structure of space-time, thus inaugurating the mathematical study of cosmology [27].   Einstein's universe was static: the product of a spatial three-sphere of constant curvature with a trivial time direction:  thus he could sidestep the question of origins: perhaps the universe was always the way it is.  Unfortunately his theory was incorrect, disproved experimentally, when it was discovered by Edwin Hubble that the universe is expanding, so the problem of initial conditions persists [42].   Now the expansion is believed to be accelerating [60, 61].  In the present work, we embark on a new, comprehensive approach to analyzing the structure of Einstein's relativity.   This approach has three main axes.    First we generalize the idea of cosmological model, going significantly beyond the conventional framework.   Remarkably, in the first steps of this process, we encounter a plethora of beautiful mathematics, involving the deepest ideas of modern mathematics.   Second we uncover a duality inherent in general relativity: roughly speaking,  this duality exchanges the canonical one-form and the Schouten one-form.   Third, rather than bringing the study of local general relativity to bear on cosmology, we turn the procedure around: we use our cosmologies, to analyze the local structure of space-time.  In the words of William Blake [8]:\\\\
\textit{"To see a world in a grain of sand\\
And a heaven in a wild flower,\\
Hold infinity in the palm of your hand,\\
And eternity in an hour".}
\\\\
We first discuss the last point.    When studying the structure of curves in low dimensions, it is frequently useful to approximate a given curve near a given point of the curve by another simple curve, for example an osculating circle.  This simpler curve encapsulates properties of the the given curve and the variation of these properties along the curve gives significant information.  This idea was perfected by \'{E}lie Cartan,  in his  conformal connection:  he  imagines the local $n$-dimensional geometry possessing an osculating compact space of constant curvature, such as a sphere and the "rolling" of the sphere along curves in the space encodes the conformal geometry of space-time [12]. 
\eject\noindent  Our strategy departs from that of Cartan in just the same way as we depart from standard cosmology:  the standard cosmologies are of constant curvature, ours are not.  Instead they depend on a certain finite number of parameters: we adapt these parameters to encode at  the point of osculation the Ricci curvature of the space-time and its first two divergences at that point.   Thus we construct a new way of viewing a space-time: on the one hand we have the intricate space-time of Einstein.  On the other we have for each of Einstein's space-times a family of osculating space-times, each actually a cosmology, each depending on a modicum of parameters.     So we have achieved a \emph{geometrization of matter}.   For this we rely on a feature of the Cartan conformal connection that in the connection itself the Ricci tensor appears explicitly, but the Weyl tensor appears only in the curvature.  So we  turn Einstein's equations around:  we aim to dictate the required matter theory by  developing the theory of the osculating cosmologies. 
\\\\
This may be construed as giving a formulation of the "multi-verse"  much liked by inflation theorists: one could perhaps imagine a process where one of these osculating cosmologies liberates itself from the surrounding space-time and creates a new one.  Here the fact that the "coupling constants"  of the osculating cosmologies vary, is natural, since these constants are determined by the Ricci curvature of their envelope, which will usually vary with space-time point.\\\\
At this point the reader may be puzzled by the question of which cosmologies will be used.   To answer this we need to describe our cosmologies and how we project that they describe our present cosmological arena.  To put it most succinctly, our cosmologies will describe the period immediately before and perhaps a little after, the big-bang.  \\\\The current standard models of the universe are all conformally flat.  This appears to accurately describe the large scale structure of the universe, even though, on small scales, the matter structure guarantees that the universe is not exactly conformally flat.   However, even if we do not live now in a conformally flat regime, Sir Roger Penrose has proposed that in the distant past the universe must have had nearly vanishing Weyl curvature [58].   Briefly, his argument is that initially the matter was in a very high entropy state, so to allow room for the universe to evolve, the gravitational degrees of freedom had to be suppressed. 
\eject\noindent  Then the evolution of the universe proceeds by transferring entropy from the matter into the gravitational degrees of freedom, allowing lower entropy structures such as galaxies and humans to form.   There is then a connection between the arrow of time and his "Weyl curvature hypothesis": that the initial state of the universe  had to have nearly vanishing Weyl curvature.   \\\\Paul Tod, building on ideas of Helmut Friedrich and others, showed that it is then consistent to imagine the big bang as purely a singularity of the metric, the conformal structure being smooth [31, 67].   In this case the space-time manifold may be consistently continued to a region prior to the big bang, at least as a conformal manifold.  Making this precise, Tod hypothesizes that at the big bang hypersurface, $\mathcal{B}$, say,   the intrinsic conformal metric and the trace-free part of the second fundamental form of $\mathcal{B}$ are regular on $\mathcal{B}$ (these form the conformal initial data):  equivalently the hypersurface twistor structure of $\mathcal{B}$ is smooth (in this latter formulation, this allows for the possibility that parts of $\mathcal{B}$ might not be space-like).  He then shows that an ordinary space-time can evolve consistently from such an initial configuration.
\\\\
Accordingly we take as our mathematical departure point that a cosmology is, by definition, a conformally flat space-time.  However we will always bear in mind that we would like to allow for small deviations from conformal flatness, so \emph{we will aim at structures that are stable under small deformations}. \\\\ The simplest conformally flat geometry is easy to construct, in any dimension $n$:   we take a real vector space $\mathbb{V}$ of dimension $n + 2$, equipped with a non-degenerate symmetric bilinear form $g$ of signature $(2, n)$.   The symmetry group of $\mathbb{V}$ is the  orthogonal group of $g$, $\mathcal{O}(g)$.  Then our space-time $\mathbb{M}$ is the space of rays of the null cone $g(x, x) = 0$ of the origin of $\mathbb{V}$.  The induced conformal structure of $\mathbb{M}$ is represented by the tensor $g(dx, dx)$ and is non-singular, of Lorentzian signature $(1, n-1)$ and conformally flat.  The group $\mathcal{O}(g)$ then acts naturally on $\mathbb{M}$, as its group of conformal transformations.   Then the general metric compatible with this conformal structure may be written $(f(x))^{-1}g(dx, dx)$; here $f(x)$ is a function on the null-cone (with the vertex removed) that is homogeneous of degree two: $f(tx) = t^2 f(x)$, for any real $t \ne 0$.   We usually assume, for convenience, that the function $f$ is the restriction to the null-cone of a homogeneous function defined on $\mathbb{V}$ itself.  These general metrics define our cosmologies, on the manifold $\mathbb{M}_f$, obtained by deleting the hyper-surface $f = 0$ from $\mathbb{M}$.  
\eject\noindent
A specific choice of the metric breaks the conformal invariance.   The simplest choice is to take $f(x) = g(a, x)^2$, where $a$ is a fixed non-zero element of $\mathbb{V}$.   Then the metric is globally an affine Minkowski space-time in the case that $a$ is null, $g(a, a) = 0$;   then the symmetry group is reduced to the Poincar\'{e} group.  In the case that $a$ is timelike, $g(a, a) >  0$ the metric is of  constant curvature, with positive cosmological constant  and is globally a de-Sitter space [63].  In the case that $a$ is space-like, $g(a, a) <  0$ the metric is of  constant curvature, with negative cosmological constant  and is globally an anti-de-Sitter space [9].   In the non-flat cases, the symmetry group is reduced to the orthogonal group of the space of vectors of $\mathbb{V}$ orthogonal to the vector $a$.  These three cases form the backdrop of most modern cosmologies.  
\\\\
In our work, published earlier this year, we considered the case (in dimension four) that $f(x)$ has the form $g(a, x)^2 + g(b, x)^2$, where the vectors $a$ and $b$  in $\mathbb{V}$ are timelike and orthogonal; this reduces the symmetry group to the product of orthogonal groups, one in two dimensions, the other in $n$ dimensions [41].   In the case that $g(a, a) = g(b, b)$ this reduces to the Einstein universe.   We analyzed the case $g(a, a) \ne g(b, b)$.   A remarkable feature of our metric is the automatic emergence of a sine-Gordon matter scalar field, which might be identified with a Higgs field, properly coupled to the metric.  Also there is a natural timelike conformal Killing vector.  These properties make the model potentially ideal for a description of the putative inflationary regime at the beginning of the universe [1, 19, 20, 30, 34, 35, 38].
\\\\
Since the publication of our work, we found that a similar model has been constructed previously by cosmologists: called "natural inflation, " it is due particularly to Kathleen Freese and her co-authors  [30].  We are unsure whether or not our model is exactly equivalent to the natural inflation model, or perhaps slightly more general: our model depends on two parameters, one a "cosmological constant" and the other the mass of the sine-Gordon soliton;  these are correlated by an energy inequality.  We are not quite clear on the number of parameters in the natural inflation model.   However, if we assume that ours is at least as general, it is (apparently) currently not ruled out by experiment and indeed, we believe, predicts that gravitational waves  of the kind reported by BICEP2 will be confirmed to be real [6].
\eject\noindent
From the mathematical viewpoint, the most striking aspect of our model, which contrasts starkly with the (relatively featureless) standard models, is the natural appearance of elliptic functions: indeed, perhaps the most famous elliptic function, the Gauss mean value function, arises as a measure of the lifetime of our universe [33].  This feature prompted us to investigate further.  In so doing it turned out to be natural to analyze the general quadratic $f(x)$: so $f(x) = h(x, x)$, where $h$ is a non-zero symmetric bilinear form on $\mathbb{V}$, linearly independent of the metric $g$.  Also, although, ultimately, we are concerned with real space-times, it emerges that the various structures are best understood when we work over the complex field.   This gives us our basic family of cosmologies,  on the projective space of $\mathbb{V}$,  a complex vector space:  
\[ G = \frac{g(dx, dx)}{h(x, x)}, \hspace{10pt} x \in \mathbb{V}, \hspace{10pt} g(x, x) = 0, \hspace{10pt} h(x, x) \ne 0.\]
Then the hypersurface $h$ plays the role of conformal infinity.    It is generically non-null, but goes null on the triple quadric intersection  $g(x, x) = h(x, x) =  k(x, x) = 0$, where $k = hg^{-1} h$.  This family of examples includes, of course all the previous examples as special cases.     The non-trivial mathematical structure present in these cosmologies begins to reveal itself first  that the triple quadric intersection where the conformal infinity goes null is precisely a $K3$ manifold, which supports a unique Calabi-Yau metric.    Of all its many remarkable features, perhaps the most dramatic is  that its second integral cohomology group has dimension $22$ and is the lattice [22 , 45]:
\[ E_8(-1)\oplus E_8(-1)\oplus U\oplus U\oplus U.\]
Here $E_8(-1)$ is the standard $E_8$ self-dual lattice with the quadratic form changed to its negative, and $U$ is the two dimensional hyperbolic plane.   Physically, following Dirac,  the second cohomology group is linked to the idea of magnetic charge [20].  \emph{So we see in the basic structure of matter, as viewed in the prism of Einstein's theory, a deep connection with the largest exceptional Lie group, even at a single point of space-time}

The overall structure is amazingly rich:  the conformal infinity of the cosmology the space $g(x, x) = h(x, x) =0$ is a Fano three-manifold, which generically, at least, supports a unique K\"{a}hler-Einstein metric.    The set of projective lines on the Fano manifold is the Jacobian of a hyperelliptic curve of genus two.  This hyperelliptic curve is that of the quadric pencil associated to $g$ and $h$.   
\eject\noindent An important special case is when the quadric $h$ is of the Battaglini type, meaning that the associated quadratic form factors as the Kulkarni--Nomizu product of the two quadrics in the twistor space itself, $h =q_1\owedge q_2$ [3, 4, 5].   The Battaglini quadrics are of co-dimension one in the space of all quadrics.  Geometrically, under the identification of the space of lines in the projective twistor space with the Klein quadric, the points of associated quadric line complex are lines whose intersections with the two quadrics $q_1$ and $q_2$ form a harmonic range.  
\\\\This we take as important, firstly because the Battaglini quadrics possess an extra degree of symmetry not present in the general quadric $h$ (the hyperellipic curve has an extra involution) and following particularly, Kepler, philosophically, we would like our cosmology to have as much structure as nature allows!  Second because we are using our cosmologies to describe the microscopic structure of matter and quadrics in twistor space can be used, in particular to describe the world-lines of massive particles, since the lines on quadrics describe the conformal geodesics of Minkowksi space-time and in particular cases ordinary timelike geodesics [43].  Also sources, such as magnetic charge are nicely described by quadrics.
\\\\
The following text is divided into four chapters with an Appendix. 
\\\\
In the first chapter, the general  idea of duality is exposed for space-times: it arises quite naturally for any space-time possessing a Weyl structure, when appropriately formulated.   Then, in four dimensions, the structure of metrics conformal to the conformally flat space-time of Klein are analyzed in detail.    
\\\\
In the second chapter the osculation theory is developed, using both the tractor calculus and the Fefferman-Graham theory and the key theorem is proved that  at every point of space-time there is a canonical osculating cosmology [28, 29, 36, 66].   It emerges that the osculating cosmologies have exactly the right structure to encode the $Q$-curvature and the end of this chapter is devoted to a discussion of the apparently parallel role of conformally invariant gauge variance in both the  theory of gravity and of electromagnetism, pointing the way to a possible unified theory completing an approach first put forward by Hermann Weyl [68, 69, 70]. 
\eject\noindent
In the third chapter,  the quadric cosmologies used in the osculation theory are studied in their own right.   First it is shown (in any dimension) that their geodesics form a remarkable completely integrable system, closely related to a famous discovery of Carl Jacob Jacobi, developed further particularly by J\"{u}rgen Moser [46, 51].  Then, specializing to dimension four, we  discuss the basic algebraic geometry and differential geometry of these cosmologies.
\\\\
Finally, in the fourth chapter,   we focus briefly on the special features of the Battaglini cosmologies.  A simple key fact that we prove is the quadric intersection studied by Alan Nadel  is of Battaglini type [52].   A last section discusses some issues that may be addressed in future work.
\\\\
In the appendix, we write out the details of the condition that an hyper-elliptic curve have an extra involution and compare our approach to a different approach using the invariants of Jun-Ichi Igusa [44].  Also we explicitly determine the Kummer surface in twistor space associated to a Battaglini cosmology.  Finally there is a short discussion of an invariant approach to the cross-ratio and the associated $j$ invariant.

\eject\noindent
\section{Cosmologies: the overall approach}
\subsection{Weyl structures, duality and the Schouten tensor}
Let $\mathbb{X}$ be a smooth $n$-manifold.  A Cartan-Penrose structure for $\mathbb{X}$ is a smooth vector bundle $\mathcal{V}$ over $\mathbb{X}$, of fibre dimension $n + 2$,  equipped with a metric $G$ of signature $(p + 1, q + 1)$ and with a metric preserving connection $d$.   We associate to this structure its local twistor bundle, $\mathcal{T}$, whose fiber at any point  is an irreducible (non-trivial) real representation $\mathbb{T}$ of the Clifford algebra of $G$ at that point.  Globally there may be an obstruction to  the existence of this bundle: this can be circumvented by passing to an appropriate cover of the original manifold, if needed.    A  conformal structure  for $(\mathbb{X}, \mathcal{V}, d)$ is a smooth section $x$ of the projective bundle of  $\mathbb{V}$, that is null: $G(x,x) = 0$.  Regarding $x$ as an operator on $\mathcal{T}$, it is defined up to scale and obeys the equation $x^2 = 0$.    The conformal metric determined by $x$ is $(dx)^2$; the canonical one-form associated to $x$ is $\theta = x dx = -  (dx) x$.  Note that under the transformation $x \rightarrow tx$, where $t$ is a non-zero smooth function, we have $\theta \rightarrow t^2 \theta$ and $(dx)^2\rightarrow t^2 (dx)^2$, since the terms involving $dt$ cancel.   We say that $x$ is a regular section of the projective bundle of $\mathcal{V}$,  if $\theta$ has maximal rank (equivalently if $g$ is invertible), except perhaps on an hypersurface in the space-time.    Given the conformal structure, we can develop the usual Cartan connection theory.   Here, however, we wish to directly fix the energy-momentum content of the space-time, in a natural way, so in the language of the Cartan theory, we want to directly specify the Schouten tensor [62].   This leads us to define a Weyl structure as a smooth two-dimensional sub-bundle $\mathcal{W}$  of  the bundle $\mathcal{V}$ of induced signature $(1,1 )$.   In terms of the Clifford algebra, we then have two sections $x$ and $y$, each with square zero, each defined up to scale,  determined by the null directions of $\mathcal{W}$, which do not anti-commute:
\[ x^2 = y^2 = 0, \hspace{10pt} xy + yx \ne 0.\]
Then, without loss of generality, we can reduce the scaling freedom to $x \rightarrow tx$ and $y \rightarrow t^{-1} y$, by the requirement:
\[ xy + yx = 1.\]
We now have two canonical one-forms, $\theta = xdx $ and $\theta' = y dy$ and two "metrics"   $g =  (dx)^2$ and $g' =  (dy)^2$.
At worst after deleting an hypersurface in the space-time,   we require that at least one of the metric tensors $g$ and $g'$ be of maximum rank.   
\eject\noindent
Put $\gamma = xy$ and $\gamma' = yx$.   Then we have $\gamma^2 = \gamma$, $(\gamma')^2 = \gamma'$ and $\gamma + \gamma' = 1$.  Then the Schouten tensor $\mathcal{S}$ is by definition: 
\[  \mathcal{S} = \textrm{tr} (d\gamma)^2 = \textrm{tr} (d\gamma')^2.\]   
(This rather natural definition may differ by a factor of two and possibly a minus sign, relative to other definitions; see below).
So the two conformal structures have the same Schouten tensor.   The space-time is said to be Einstein-Weyl if and only if $\mathcal{S}$ vanishes.  Note that the Schouten tensor is scale invariant, even though we have not committed ourselves to a specific metric $g$ or $g'$ in either conformal class.  In terms of the metric $G$ and the vectors $x$ and $y$,  we have the formulas:
\[ G(x, x) = G(y, y) = 0, \hspace{10pt}G(x, y) = 1, \]
\[ g = G(dx, dx), \hspace{10pt} g' = G(dy, dy), \]
\[ \mathcal{S} = 2 G(dx, dy) + G(x, dy)^2 + G(y, dx)^2.\]
We have constructed a natural duality underlying the Einstein theory:  for every metric, there is a dual metric   with the same Schouten tensor.   Finally, in this language we have a clear definition of a cosmology:
\begin{itemize} \item A cosmology is a Weyl structure with a flat connection $d$.
\end{itemize}
Here for the conformal  metric to have Lorentzian signature, we would require that $p = 1$ or $q = 1$.
We say that the resulting Weyl cosmology is Einstein-Weyl if and only if  $\mathcal{S}$  vanishes.
\subsection{Complex cosmologies in four dimensions}
We adapt the discussion of the previous section to the complex holomorphic case in four dimensions,  (the source of our key examples), bearing in mind that ultimately we will need to impose a reality condition.   As such our cosmologies will be automatically real analytic.   This is just for convenience of exposition, however: it would be possible to restructure the discussion to focus, for example, on the real smooth case.   We also use only  the semi-spinor bundle, which entails using just the products $xy$ and $yx$ of the previous section, not just $x$ and $y$ individually.  The semi-spinor bundle carries an irreducible representation of the even part of the Clifford algebra.  
\eject\noindent
Accordingly,  a complex Weyl cosmology  $(\mathbb{X}, \mathcal{T}, d, \gamma^\pm, S)$ consists of the following holomorphic data:
\begin{itemize} \item $\mathbb{X}$, the space-time, a connected four-dimensional complex manifold.
\item $\mathcal{T}$, a complex vector bundle of fibre dimension four over the space-time $\mathbb{X}$.
\item $d$, a flat connection for the bundle $\mathcal{T}$: its curvature endomorphism vanishes:
\[ d^2 = 0.\]
\item $\gamma^\pm$,  complementary endomorphisms of $\mathcal{T}$, each a projection operator of trace two:
\[ \gamma^+\gamma^+ = \gamma^+,  \hspace{10pt} \gamma^-\gamma^- = \gamma^-,  \hspace{10pt}  \gamma^- \gamma^+  =  \gamma^+ \gamma^- = 0,   \hspace{10pt} \textrm{tr}(\gamma^\pm) = 2, \]
\[ \gamma^+ + \gamma^- = I.\]
\end{itemize}
Here and in the following, $\textrm{tr}$ denotes the trace of an endomorphism.  Also $I$ denotes the identity automorphism of $\mathcal{T}$.
\begin{itemize}
\item $\mathcal{S} = \textrm{tr}((d\gamma^+)^2) = \textrm{tr}((d\gamma^-)^2) $. \\So $\mathcal{S}$ is a covariant symmetric tensor field on $\mathbb{X}$ of two vector arguments.  $\mathcal{S}$ is called the Schouten tensor of the Weyl cosmology.
\end{itemize}
Given a Weyl cosmology $(\mathbb{X}, \mathcal{T}, d, \gamma^\pm, \mathcal{S})$, the dual cosmology is:  $(\mathbb{X}, \mathcal{T}, d, \gamma^\mp, \mathcal{S})$.  Then the dual of the dual cosmology is the original cosmology.  

 Associated to a Weyl cosmology $(\mathbb{X}, \mathcal{T}, d, \gamma^\pm, S)$ is a natural decomposition of the bundle $\mathcal{T}$ and of its connection $d$:
\[ \mathcal{T} = \mathcal{T}^+ +  \mathcal{T}_-, \]
\[  \mathcal{T}^+ = \ker(\gamma^-) = \textrm{im}(\gamma^+), \hspace{10pt}   \mathcal{T}_- = \ker(\gamma^+) = \textrm{im}(\gamma^-), \]
\[ d = d_+  + d_-  + \theta + \theta', \]
\[  d_+ = \gamma^+\circ  d\circ \gamma^+, \hspace{10pt} d_- = \gamma^-\circ  d\circ \gamma^-, \]
\[  \theta = \gamma^+\circ d \circ \gamma^-, \hspace{10pt} \theta' = \gamma^- \circ d \circ \gamma^+, \]
\[  \gamma^+ \circ d = d_+ + \theta, \hspace{10pt}  \gamma^- \circ d = d_- + \theta'.\] 
Then the quantities $\theta$ and $\theta'$ are one-forms on $\mathcal{X}$  with values in the endomorphisms of $\mathcal{T}$.  Also $d_+$ provides a connection for the bundle $\mathcal{T}^+$, whereas $d_-$ gives a connection for $\mathcal{T}_-$.    The two-dimensional bundle $\mathcal{T}^+$ is called the unprimed spin bundle, with dual the unprimed co-spin bundle, denoted $\mathcal{T}_+$.   The two-dimensional bundle $\mathcal{T}_-$ is called the primed co-spin bundle, with dual the primed spin bundle, denoted $\mathcal{T}^-$.  The spinor algebra is the tensor algebra generated by the basic bundles $\mathcal{T}^\pm$ and $\mathcal{T}_\pm$.    Then the one-form $\theta$ gives a global section of $\mathcal{T}^+\otimes \mathcal{T}^-$, whereas the one-form $\theta'$ gives a global section of $\mathcal{T}_+\otimes \mathcal{T}_-$.  Denote by $\rho$ and $\rho'$ the integer-valued functions on $\mathbb{X}$, which assign to each point $x$ of $\mathbb{X}$, the ranks $\rho_x$ and $\rho'_x$ of the one-forms $\theta$ and $\theta'$, respectively.  Then the functions $\rho$ and $\rho'$ are lower semi-continuous, taking values $0, 1, 2, 3$ or $4$.   Define:
\[ \mathbb{U} = \rho^{-1}(\{4\}), \hspace{10pt} \mathbb{U}' = (\rho')^{-1}(\{4\}).\]
Then $\mathbb{U}$ and $\mathbb{U'}$ are open subsets of $\mathbb{X}$.  A priori,  either or both of these sets could be empty.
To avoid trivialities, we impose a rank condition on our Weyl cosmologies:
\begin{itemize}
\item We require that the union of the open subsets $\mathbb{U}$ and $\mathbb{U'}$ be dense  in $\mathbb{X}$. 
\end{itemize}
On $\mathbb{U}$ there is then an associated conformal structure defined by $\theta$, denoted $g$, such that a non-zero vector $v$ at a point of $\mathbb{U}$ is null if and only if the rank of the endomorphism $\theta(v)$ is one.  Dually, on $\mathbb{U}'$ there is then an associated conformal structure defined by $\theta'$, denoted $g'$, such that a non-zero vector $v$ at a point of $\mathbb{U}'$ is null if and only if the rank of the endomorphism $\theta'(v)$ is one.   In each case the associated spin connection is torsion-free and preserves the conformal structure.   We say that the resulting Weyl cosmology  $(\mathbb{X}, \mathcal{T}, d, \gamma^\pm, \mathcal{S})$ is Einstein-Weyl if and only if the Schouten tensor $\mathcal{S}$ vanishes.

We can now write the decomposition of the local twistor connection, with respect to $\gamma$,  as follows:
\[ dZ = d(Z^+,\hspace{5pt}  Z_{-}) = \left(dZ^+ + \theta^{+-}Z_-,\hspace{5pt}  dZ_- + \rho_{-+}Z^+\right).\]
Here $Z \in\mathcal{T}$, $Z^+ \in  \mathcal{T}^+ $ and $Z_-\in   \mathcal{T}_-$.
The various entities on the right-hand side are to be considered as spinors, so it is legitimate to write the spin connection uniformly as $d$. Computing the twistor curvature $d^2$, which is known to be zero and using the curvature relations proved above, we get:
\[ \hspace{-50pt} 0 = d^2 Z =  \left(d(d Z^+ + \theta^{+-}Z_-) + \theta^{+-}(dZ_- + \rho_{-+}Z^+), \hspace{5pt} d( d Z_- + \rho_{-+}Z^+ )+ \rho_{-+}(d Z^+ + \theta^{+-}Z_-)\right).\]
\[ = \left((d\theta^{+-}) Z_-,  \hspace{5pt}  (d\rho_{-+}) Z^+\right).\]
So we infer the equations of vanishing torsion and the Bianchi identity:
\[ d\theta^{+-} = 0, \hspace{10pt} d\rho_{-+} = 0.\]
To clarify, we write the spinor indices explicitly: 
\[ Z = (Z^A, \hspace{5pt} Z_{A'}), \]
\[ dZ = d\left(Z^A,\hspace{5pt}  Z_{A'}\right) = \left(dZ^A + \theta^{AB'}Z_{B}',\hspace{5pt}  dZ_{A'} + \rho_{A'B}Z^{B}\right).\]
\[ d^2 Z_{A'} = -  \mathcal{R}_{A'}^{B'} Z_{B'}, \hspace{10pt} d^2 Z^{A} =   \mathcal{R}_{B}^{A} Z^{B},\]
\[ \mathcal{R}^{A}_B = - \theta^{AC'} \rho_{BC'}, \hspace{10pt}  \mathcal{R}^{A'}_{B'} = - \theta^{A'C} \rho_{B'C}, \] 
\[ d\theta^a = 0, \hspace{10pt} d\rho_a =0.\]
In this language, the projection operator acts as the endomorphism:
\[ \gamma^+ = \hspace{3pt} \begin{array}{|cc|} \delta_B^A&0\\0&0\end{array}\hspace{3pt}.\]
In this language, we have for the covariant derivative of $\gamma^+$:
\[ d\gamma^+ = \hspace{3pt} \begin{array}{|cc|} d\delta_B^A&0\\0&0\end{array}\hspace{3pt} + \hspace{3pt} \begin{array}{|cc|} 0&\theta^{AC'}\\\rho_{A'C}&0\end{array}\hspace{3pt} \begin{array}{|cc|} \delta_B^C&0\\0&0\end{array}\hspace{3pt}  -  \hspace{3pt} \begin{array}{|cc|} \delta_C^A&0\\0&0\end{array}\hspace{3pt}  \begin{array}{|cc|} 0&\theta^{CB'}\\\rho_{C'B}&0\end{array}\hspace{3pt} =   \hspace{3pt} \begin{array}{|cc|} 0&- \theta^{AB'} \\\rho_{A'B}&0\end{array}\hspace{3pt},  \]
\[ (d\gamma^+)^2  =  \hspace{3pt} \begin{array}{|cc|} 0&- \theta^{AC'} \\\rho_{A'C}&0\end{array}\hspace{3pt}\begin{array}{|cc|} 0&- \theta^{CB'} \\\rho_{C'B}&0\end{array}\hspace{3pt} =   \hspace{3pt} \begin{array}{|cc|}- \theta^{AC'}\rho_{C'B}&0\\0 & - \theta^{CB'} \rho_{A'C}\end{array}\hspace{3pt}.\]
In particular we have that the trace of $(d\gamma^+)^2$ determines the Schouten tensor, the symmetric tensor $\rho_{(ab)}$, where $\rho_a = \theta^b \rho_{ab}$:
\[ \textrm{tr}((d\gamma^+)^2) = - 2\theta^a \rho_a. \]
\eject\noindent
\subsection{Twistor space and its Grassmanians}

The basic space is twistor space, a complex four-dimensional vector space $\mathbb T$ [53,  56, 57, 58, 59].  For real spacetimes (of Lorentzian signature), $\mathbb T$ is a spin representation of the conformal group $\op{SU}(2,2)$.  At least initially, it is convenient to confine attention to the complex, in which case the relevant spin group is $\op{SL}(4,\mathbb C)$.  Moreover, we will work projectively, being concerned with the three-dimensional complex projective space $\mathbb P\mathbb T$.  The most convenient group to work with is then $\op{GL}(\mathbb T)$, which does not involve specifying a volume form on $\mathbb T$ to reduce to the special linear group.

The Pl\"ucker embedding of the Grassmannian $\op{Gr}(2,\mathbb T)$ of two-dimensional linear subspaces of $\mathbb T$ into the projective space over $\Omega^2\mathbb T$ (the exterior square of the twistor space) associates to a two-dimensional linear subspace $\mathbb X\subset\mathbb T$ the one-dimensional linear subspace of $\Omega^2\mathbb T$ given by $\Omega^2\mathbb X$.  The image of the Pl\"ucker embedding is a four-dimensional hypersurface in $\mathbb P(\Omega^2\mathbb T)$.  This image coincides with the non-degenerate quadric hypersurface defined by the equation $X\wedge X=0$.  It is called the {\em Klein quadric}, and here is denoted by $\mathbb K$.

The Klein quadric serves as the conformally flat background for all of the cosmological metrics [48].  We describe here the conformal structure on $\mathbb K$.  Let $\mathbb L=\Omega^4\mathbb T$ be the one-dimensional space of densities associated with the twistor space.  The wedge product defines a symmetric bilinear $\mathbb L$-valued form on $\Omega^2\mathbb T$ of which the Klein quadric is the null cone:
$$\mathbb K = \{ X\in\mathbb P(\Omega^2\mathbb T)\mid X\wedge X=0\}.$$
The symmetric bilinear form $dX\wedge dX$ defines the conformal structure on $\mathbb K$.  This is homogeneous of degree two, and the conformally flat cosmologies that we consider are all obtained by dividing this conformal metric by a function that is homogeneous of degree two to obtain something that scales appropriately so that it restricts to an actual metric on $\mathbb K$.

The conformal structure on $\mathbb K$ can also be understood by means of the following purely synthetic construction of projective geometry.  The {\em Klein correspondence} is the isomorphisms of generalized flag varieties associated to the group of linear automorphisms of $\mathbb T$ and the group of conformal automorphisms of $\mathbb K$:
\begin{itemize}
\item A point $X$ of $\mathbb K$ by definition corresponds to a line $X_{\mathbb T}$ in $\mathbb{PT}$.  Conversely, a line $\ell$ in $\mathbb{PT}$ corresponds to a point $\ell_{\mathbb K}$ in $\mathbb K$.
\item A point $Z$ of $\mathbb{PT}$ gives, on $\mathbb K$, an $\alpha$-plane $Z_{\mathbb K}$: a $2$-plane consisting of all elements of the form $Z\wedge Z'$ for $Z'\in\mathbb T$.
\item A $2$-plane $W$ in $\mathbb{PT}$, which can be naturally identified with it annihilator in the dual projective space $\mathbb{PT}^*$, corresponds to a $\beta$-plane $W_{\mathbb K}$ in $\mathbb{K}$, consisting of all $X\in\mathbb{K}$ such that $W\lrcorner X=0$.
\item A partial flag $(Z,W)$ in $\mathbb{PT}$ consisting of a twistor $Z$ and dual twistor $W$ such that $Z$ is on the $2$-plane $W$ can be identified with a {\em line} $(Z,W)_{\mathbb K}$ on $\mathbb K$.  This is the set of points $X$ of $\mathbb K$ such that $X_{\mathbb T}$ contains the twistor $Z$ and is contained within the $2$-plane $W$.
\end{itemize}
It is the last of these that defines a null geodesic on $\mathbb K$.   We may summarize some of the descriptions of the Klein space-time  $\mathbb{K}$ as follows:
\begin{itemize}\item $\mathbb{K}$  is the Grassmannian of all two-dimensional complex subspaces of  $\mathbb{T}$.
\item Equivalently $\mathbb{K}$  is the space of lines in $\mathbb{PT}$.
\item Equivalently $\mathbb{K}$ is a copy of the Klein quadric.
\item Dually, $\mathbb{K}$  is the Grassmannian of all two-dimensional complex subspaces of  $\mathbb{T}^*$.
\item Equivalently $\mathbb{K}$ is the space of lines in $\mathbb{PT}^*$. 
\item Equivalently  $\mathbb{K}$ is a copy of the dual Klein quadric.\end{itemize}
If $x$ and $y$ are distinct two-dimensional subspaces of $\mathbb{X}$, then $\Omega^2(x)$ and $\Omega^2(y)$ of the  Klein space-time are  null related, provided  $x + y \ne \mathbb{T}$, so provided the space $x \cap y$ is one-dimensional, so provided the (distinct) projective lines determined by the subspaces $x$ and $y$ intersect in a unique projective twistor, and the span of $x$ and $y$ is then the annihilator of a unique projective dual twistor.  Then the (five-dimensional) space of (unparametrized) null geodesics is represented by the space of pairs $(Z, W)$ with $Z$ a one-dimensional subspace of $\mathbb{T}$ and $W$ a three-dimensional subspace, such that $Z \subset W$.  The space of complete flags,  $(Z, x, W)$,  with $(Z, W)$ a null geodesic and  with $Z \subset x \subset W$ where  $x$ is a two-dimensional subspace of $\mathbb{T}$,  is the six-dimensional space of all pairs $(x, s)$ with $s$ a null geodesic in the space-time and $x$ a point of the null geodesic $s$. 
\eject\noindent
\subsection{Reality}
To produce a real space-time from the complex Klein space-time, we choose a conjugation on the basic twistor space $\mathbb{T}$.  There are a variety of ways of doing this;  for example, if we regard $\mathbb{T}$ as the complexification of an underlying real vector space, with conjugation given by ordinary complex conjugation, then the resulting real space-times turn out to have a real conformal structure of the ultra-hyperbolic signature $(2, 2)$.  To achieve instead the desired Lorentzian structure, we use a different conjugation, specifically  one mapping $\mathbb{T}$ to $\mathbb{T}^*$, $Z  \in \mathbb{T} \rightarrow \overline{Z} \in \mathbb{T}^*$, $Z^\alpha \rightarrow \overline{Z}_\alpha$, such that the resulting  pseudo-hermitian form $\overline{Z}.Z = \overline{Z}_\alpha Z^\alpha $ has signature $(2, 2)$.   This conjugation then extends to the tensor algebra of $\mathbb{T}$ and $\mathbb{T}^*$, such that its square is the identity.   In particular, note  that the complex line $\mathbb{L}$ has a natural unitary structure, such that $\epsilon_{\alpha\beta\gamma\delta} = \overline{\epsilon}_{\alpha\beta\gamma\delta} $.    The symmetry group of $\mathbb{T}$ is reduced to a  group isomorphic to the pseudo-unitary group $\mathbb{U}(2, 2; \mathbb{C})$, a real reductive Lie group of sixteen real dimensions.  \\\\   Given the twistor conjugation, the twistor space is divided into three parts:  $\mathbb{T}^\pm = \{ Z \in \mathbb{T}:  \pm Z.\overline{Z} > 0\}$ (each of real dimension eight) and $\mathbb{N} = \{Z\in \mathbb{T}: Z.\overline{Z} = 0\}$ (of real dimension seven).   The space $\mathbb{N}$ is called the null twistor space, its elements null twistors. The (complex) one-dimensional subspaces of $\mathbb{N}$ form $\mathbb{PN}$, the null projective twistor space, its elements null projective twistors.  $\mathbb{PN}$ is a five real-dimensional hypersurface in $\mathbb{PT}$, called a hyper-quadric (with a natural $\mathcal{CR}$-structure of non-singular indefinite Levi form). \\\\
The reality structure of $\Omega^2(\mathbb{T})$ is somewhat more subtle and is easiest to state at first in terms of the reality structure of $\mathbb{P}\Omega^2(\mathbb{T})$, the projective space of $\Omega^2(\mathbb{T})$.  If $ 0 \ne X\in \Omega^2(\mathbb{T})$ represents a point of    $\mathbb{P}\Omega^2(\mathbb{T})$, then its conjugate $\overline{X}$ represents an hyper-plane in $\mathbb{P}\Omega^2(\mathbb{T})$; put $\overline{X}'$ the polar point of the hyperplane $\overline{X}$ with respect to the (complex) Klein quadric.   Then the map $X \rightarrow \overline{X}'$ gives a conjugation on $\mathbb{P}\Omega^2\mathbb{T}$ and $X$ is said to be real if and only if $\overline{X}' = X$.   In tensor terms, given a skew twistor $X^{\alpha\beta}$, put $\overline{X}^{\alpha\beta} = \frac{1}{2}\epsilon^{\alpha\beta\gamma\delta} \overline{X}_{\gamma\delta} $.  This  conjugation maps $\Omega^2(\mathbb{T})$ to $\Omega^2(\mathbb{T})\otimes \mathbb{L}^{-1}$, so projects to give an idempotent conjugation from  $\mathbb{P}\Omega^2(\mathbb{T})$ to itself.   Then the real points of  $\mathbb{P}\Omega^2(\mathbb{T})$ are those that are self-conjugate: equivalently $0 \ne X^{\alpha\beta} \in \Omega^2(\mathbb{T})$ represents a real point of  $\mathbb{P}\Omega^2(\mathbb{T})$  if and only if $X^{\alpha\beta}$ and $\overline{X}^{\alpha\beta}$ are linearly dependent, if and only if a (necessarily non-zero) skew tensor $x^{\alpha\beta\gamma\delta}$ exists with $X^{\alpha\beta}  = \frac{1}{2}x^{\alpha\beta\gamma\delta} \overline{X}_{\gamma\delta}$.  
\eject\noindent 
Note that if $0 \ne X^{\alpha\beta}$ is real and $X^{\alpha\beta} \overline{X}_{\alpha\beta} = 0$, then $X \wedge X = 0$, so the real points of $\mathbb{P} \Omega^2(\mathbb{T})$ divide into three classes according as the real invariant $\xi = X^{\alpha\beta} \overline{X}_{\alpha\beta}$ is positive, negative or zero, with the zero case occurring, if and only if $X$ represents a point of $\mathbb{K}$. 
\\\\ Also if $\xi = X^{\alpha\beta} \overline{X}_{\alpha\beta}$ is non-zero, then we have, when $X^{\alpha\beta}$ is real:
\[ X^{\alpha\gamma} \overline{X}_{\beta \gamma} = \frac{1}{4} \xi \delta^{\alpha}_{\beta}, \]
\[ x^{\alpha\beta\gamma\delta} =   12\xi^{-1} X^{[\alpha\beta}X^{\gamma\delta]}.\]
Conversely, if $0 \ne X^{\alpha\beta}$ obeys the relation that $X^{\alpha\gamma} \overline{X}_{\beta\gamma}$ is a (necessarily real) multiple of the identity, then $X$ represents a  real point of $\mathbb{P}\Omega^2(\mathbb{T})$.
\\\\Summarizing: an element  $X$ of  $\Omega^2(\mathbb{T})$ is real if and only if $X^{\alpha\gamma} \overline{X}_{\beta\gamma}$ is a real multiple of the identity.  The space of such real points is seven real-dimensional and is invariant under the multiplication of $X$ by a complex scalar, so gives a five real-dimensional subset of $\mathbb{P}\Omega^2(\mathbb{T})$, topologically a real projective five-space, which, when the real scalar $\xi = X^{\alpha\beta} \overline{X}_{\alpha\beta}$ is non-zero,  has two open subsets, one for each sign of $\xi$.  These open subsets are separated by a closed four-dimensional subset, which we call $\mathbb{M}$, for which the scalar $\xi$ vanishes.  $\mathbb{M}$ is the desired real part of the Klein quadric.   The points of $\mathbb{M}$ constitute the two-dimensional subspaces of $\mathbb{T}$ that are totally null with respect to the  pseudo-hermitian form of $\mathbb{T}$; equivalently they are the complex projective lines in $\mathbb{PT}$ that lie entirely in $\mathbb{PN}$.  Expressed in terms of the skew tensor $X^{\alpha\beta}$ representing a point of $\Omega^2(\mathbb{T})$,  the reality condition defining $\mathbb{M}$ is $X^{\alpha\gamma} \overline{X}_{\beta\gamma} = 0$, as just discussed above: the space of solutions of the equation $X^{\alpha\gamma} \overline{X}_{\beta\gamma} = 0$ is six real-dimensional and represents a real  four-manifold inside the complex Klein space-time.   \\\\The complex conformal structure of $\mathbb{K}$ restricts to $\mathbb{M}$ in such a way that it is proportional to a real Lorentzian conformal structure for $\mathbb{M}$: in turn that conformal structure is conformally flat.   In fact this conformal structure is just that  determined by the restriction of the manifestly real tensor $dX^{\alpha\beta}d\overline{X}_{\alpha\beta}$ to the  real points.  The real null geodesics of $\mathbb{M}$ are represented by the projective null twistors: the $(1, 3)$ flag represented by a projective null twistor $Z$ is the pair $(Z, \overline{Z}^*)$, where $\overline{Z}^*$ is the annihilator of the one-dimensional subspace $\overline{Z}$ of $\mathbb{T}^*$.  The topology of $\mathbb{M}$ is the product $\mathbb{S}^1 \times \mathbb{S}^3$.
\eject\noindent
\subsection*{The complex Klein Cosmologies} The fundamental examples of complex cosmologies, here called the (complex) Klein cosmologies, may be constructed as Zariski open sub-manifolds of the Klein space-time, $\mathbb{K}$, as follows.  Using tensors, a point $X$ of $\Omega^2(\mathbb{T})$ is represented by the skew tensor $X^{\alpha\beta}$.  Fix an element $J$ of the projective space of $\Omega^2(\mathbb{T})$, represented by a non-zero skew tensor $J^{\alpha\beta}$.  Denote by $\mathbb{J}$ the intersection with $\mathbb{K}$ of the polar plane of $J$ with respect to the Klein quadric, so $\mathbb{J}$ is represented as the set of all $X \in \Omega^2(\mathbb{T})$, such that $X \wedge X = X \wedge J = 0$.   Note that, if $J$ does not lie on $\mathbb{K}$, then the space $\mathbb{J}$ is a non-singular three-dimensional quadric,  whereas if $J$ does lie on the Klein quadric, then $\mathbb{J}$ is a three-dimensional quadric cone, with vertex $J$.  Define $\mathcal{K}_J = \mathbb{K} - \mathbb{J}$, so $\mathcal{K}_J$ is a (Zariski) open subset of $\mathbb{K}$, with closure $\mathbb{K}$ itself.  The space $\mathcal{K}_J$ will constitute our complex cosmological space-time. \\\\ Put $\mathcal{T}_J = \mathcal{K}_J \times \mathbb{T}$,   the trivial bundle over $\mathcal{K}_J$ with fiber $\mathbb{T}$, equipped with the flat connection $d$, such that the global covariantly constant sections of $\mathcal{T}_J$ take the form $(X, Z)$, for fixed $Z \in \mathbb{T}$, as $X \in \mathcal{K}_J$ varies.  Given  $X \in \mathcal{K}_J$,  define an endomorphism $\gamma_J$ of the fibre $\mathbb{T}$ of $\mathcal{T}_J$, at $X$, by the formula:
\[ 2(\gamma_J)^\alpha_\beta  X^{[\beta\rho} J^{\sigma\tau]} = X^{\alpha[\rho} J^{\sigma\tau]}.\]
As $X \in \mathcal{K}_J$ varies, $\gamma_J$ gives  well-defined endomorphism of the bundle $\mathcal{T}_J$ and is a projection operator of trace two, whose image at the point $X$ is all $Z \in \mathbb{T}$, such that $Z \wedge X = 0$.    If we put $X'= J + \kappa X$, where $\kappa$ is chosen so that $X' \wedge X' = 0$ (so $\kappa = -  (J \wedge J)(2X \wedge J)^{-1}$ and $X' \in \mathcal{K}_J$ is, by definition, the reflection of $X \in \mathcal{K}_J$ in the plane $\mathbb{J}$),  the kernel of the endomorphism $\gamma_J$ is the set of all $Z \in\mathbb{T}$, such that $Z \wedge X' = 0$.    Also put $\gamma'_J = I - \gamma_J$, where $I$ is the identity automorphism of $\mathbb{T}$.   Then we have  symmetrical formulas:
\[ 2(\gamma_J)^\alpha_\beta  X^{[\beta\rho} (X')^{\sigma\tau]} = X^{\alpha[\rho} (X')^{\sigma\tau]}.\]
\[ 2(\gamma_J')^\alpha_\beta  (X')^{[\beta\rho} X^{\sigma\tau]} = (X')^{\alpha[\rho} X^{\sigma\tau]}.\]
Note that $(X')' = X$, for any $X \in \mathcal{K}_J$ and $(\gamma_J')' = \gamma_J$.
\begin{itemize}\item The Klein cosmology of the point $0 \ne J \in \Omega^2(\mathbb{T})$ is the quintet $(\mathcal{K}_J,  \mathcal{T}_J, d, \gamma_J, g_J)$, where the metric $g_J$ of the Klein cosmology is the metric $g_J$, given by the formula valid for any $X\in \mathcal{K}_J$:
\[ g_J (X\wedge J)^2  = dX \wedge dX.\]
Here $g_J$ takes values in $\mathbb{L}^{-1}$.  
\end{itemize}
\eject\noindent
If $J$ is off the Klein quadric, so $J \wedge J \ne 0$, then we may choose the scale of the metric in a natural way, giving the metric,  now called $G_J$, again valid for any $X\in \mathcal{K}_J$:
\[ G_J(X\wedge J)^2 =  (J\wedge J)(dX \wedge dX), \hspace{10pt} G_J = g_J (J\wedge J).\]
Then $G_J$ is an ordinary holomorphic metric and depends only on the image of $J$ in the projective space of $\Omega^2(\mathbb{T})$.\\\\
For the curvature of the Klein cosmology, we have two cases:
\begin{itemize} \item When $J \wedge J = 0$, the metric $g_J$ is flat: in fact, given any non-zero $\epsilon \in  \mathbb{L}$ the intrinsic geometry of the intersection of the hyperplane $X \wedge J = \epsilon$ with the Klein quadric is naturally a flat affine space-time.
\item  When $J \wedge J \ne 0$, the metric $G_J$ is a (conformally flat) metric of non-zero constant curvature.  The Einstein tensor is $-3$ times the metric.  The Ricci scalar is $12$.
\end{itemize}
The symmetry group of the Klein cosmology $\mathcal{K}_J$  is the subgroup of  $\mathbb{P}(\mathbb{K})$ that preserves the point $J$, regarded as a point of the projective space of $\Omega^2(\mathbb{T})$.
\begin{itemize} \item When $J$ is not in $\mathbb{K}$, so $J \wedge J \ne 0$, this is isomorphic to the quotient of the group $\mathbb{B}_2(\mathbb{C})$ of complex linear transformations in five dimensions, preserving a non-singular quadratic form, modulo its center.  The group is simple and has complex dimension ten.
\item When $J$ is on $\mathbb{K}$, so $J \wedge J = 0$, it is the complex Poincar\'{e} group, the ten-dimensional Lie group formed from the semi-direct product of the Lorentz group and the four-dimensional abelian group of translations.
\end{itemize}
Finally we record a useful formula for the endomorphism $\gamma_J$.   Let  $F^\alpha_\beta$ denote the vector fields representing the generators of the action of $\mathbb{GL}(\mathbb{T})$ on $\Omega^2(\mathbb{T})$.   Explicitly we have at the point $X^{\alpha\beta}$:
\[ F^\alpha_\beta = 2X^{\alpha\gamma} \partial_{\beta\gamma}, \hspace{10pt} \partial_{\alpha\beta} X^{\gamma\delta} = \delta_{[\alpha}^\gamma\delta_{\beta]}^\delta.\]
Note that the vector field $F$ preserves the relation $X \wedge X = 0$, so is tangent to the Klein space-time.
Then we have:
\[ \gamma_J = F(\ln(X \wedge J)).\]
\eject\noindent
\subsection{The real Klein cosmologies}
Given a twistor conjugation with signature $(2, 2)$, we obtain the real compact conformally flat conformal space-time $\mathbb{M}$,  as discussed above.   We wish the metric of the space-time to be real.   This may be arranged for the Klein cosmologies $\mathcal{K}_J$  by using a real $J \in \Omega^2(\mathbb{T})$.    When $J$ is real, put $\mathcal{M}_J$ the real part of $\mathcal{K}_J$.   Then the metrics $g_j$ and $G_J$ are real on $\mathcal{M}_J$ and are of constant real curvature: if $J \wedge J = 0$, then $\mathcal{M}_J$ is real Minkowski space-time.  If instead $J \wedge J \ne 0$, the space-time is de Sitter if $J^{\alpha\beta}  \overline{J}_{\alpha\beta} > 0$ and anti-de-Sitter if  $J^{\alpha\beta} \overline{J}_{\alpha\beta} < 0$.   These cosmologies are the standard ones.   The metric $g_J$ may be summarized in manifestly real form as:
\[ g_J  = \frac{dX^{\alpha\beta}d\overline{X}_{\alpha\beta}}{|X^{\gamma\delta} \overline{J}_{\gamma\delta}|^2}, \hspace{10pt} X^{\alpha\gamma}\overline{X}_{\beta\gamma} = 0, \hspace{10pt}  J^{\alpha\gamma}\overline{J}_{\beta\gamma} = \frac{j}{4} \delta^\alpha_\beta, \]
\[ 0 \ne X^{\alpha\beta} = - X^{\beta\alpha}, \hspace{10pt}  0 \ne J^{\alpha\beta} = - J^{\beta\alpha}, \hspace{10pt} j \in \mathbb{R}.\]
For completeness, we show how to write out the metric in co-ordinates.  We use an orthonormal basis $\{e_1, e_2, e_3, e_4\}$ for $\mathbb{T}$, with dual basis $\{e^1, e^2, e^3, e^4\}$ for $\mathbb{T}^*$, such that $e^1 = \overline{e_1},  e^2 = \overline{e_2}, e^3 = -\overline{e_3}$ and $e^4 = -  \overline{e_4}$.     The corresponding basis for $\Omega^2(\mathbb{T})$ is $\{e_{12}, e_{13}, e_{14}, e_{34}, e_{42}, e_{23}\}$, where $e_{ij}= e_i \wedge e_j$.  The dual basis is $\{2e^{12}, 2e^{13}, 2e^{14}, 2e^{34}, 2e^{42}, 2e^{23}\}$, where $e^{ij}= e^i \wedge e^j$.  We have the conjugates: $\overline{e_{12}} = e^{12}$, $\overline{e_{13}} = - e^{13}$, $\overline{e_{14}} = - e^{14}$,  $\overline{e_{23}} = - e^{23}$, $\overline{e_{24}} = -e^{24}$ and $\overline{e_{34}} = e^{34}$.     Also the corresponding basis for $\Omega^4(\mathbb{T})$ is $e_{1234} = e_1 \wedge e_2 \wedge e_3 \wedge e_4$, with $\overline{e_{1234}} =  e^1 \wedge e^2 \wedge e^3 \wedge e^4$.  \\\\
We may parametrize the space of non-zero real $X^{\alpha\beta}$ on the Klein quadric as follows:
\[ X = s(u e_{12} + \overline{u} e_{34} + pe_{13} - \overline{p} e_{42} +   qe_{14} - \overline{q} e_{23}), \]
\[ |u|^2 = |p|^2 + |q|^2 = 1.\]
Here $s, u, p$ and $q$ are complex numbers, with $s \ne 0$.  Also the quadruple $(s, u, p, q)$ represents the same point as does the quadruple $(-s, -u, -p, -q)$.  Then we have:
\[ |s|^{-2} dX^{\alpha\beta} d\overline{X}_{\alpha\beta} = |du|^2 - |dp|^2 - |dq|^2.\]
Note in particular that the terms involving $ds$ or $d\overline{s}$ cancel.  Also the space is topologically $\mathbb{S}^1 \times \mathbb{S}^3$, where the $\mathbb{S}^1$ factor is represented by the variable $u$ and the $\mathbb{S}^3$ factor by the pair $(p, q)$.
\eject\noindent
 Now introduce the real co-ordinates $(t, x, y, z)$, related to the variables  $(u, p, q)$ by the formula: $(u, p, q)  = (e^{it}, \cos(x) e^{iy}, \sin(x) e^{iz})$.  Then we have:
\[  |s|^{-2} dX^{\alpha\beta} d\overline{X}_{\alpha\beta} = dt^2 - dx^2 - \cos^2(x) dy^2 - \sin^2(x)dz^2.\]
The right-hand side represents the standard Einstein cylinder metric  of signature $(1, 3)$, defined on $\mathbb{S}^1 \times \mathbb{S}^3$.  As such, this metric is well-known to be conformally flat.   It has for its Einstein tensor the metric plus $2(dt)^2$.   \\\\Next we analyze the metrics, $g_J$, case by case, using this co-ordinate system.
\begin{itemize}\item If $j > 0$, without loss of generality, may choose $J = a(e_{12}  + e_{34})$, with $0 \ne a \in \mathbb{C}$. 
\end{itemize}
Then  we have:
\[  J^{\alpha\gamma}\overline{J}_{\beta\gamma}  = - \frac{1}{4}|a|^2[(e_1 \otimes e_2 - e_2 \otimes e_1  +  e_3 \otimes e_4 - e_4 \otimes e_3).(e^1 \otimes e^2 - e^2 \otimes e^1  +  e^3 \otimes e^4 - e^4 \otimes e^3)]^\alpha_\beta\]
\[  = \frac{|a|^2}{4} (e_1\otimes e^1 +  e_2\otimes e^2 +   e_3\otimes e^3 +   e_4\otimes e^4)^\alpha_\beta =    \frac{|a|^2}{4} \delta^\alpha_\beta.\]
So  $j = |a|^2$.  We have also:
\[ s^{-1} X^{\alpha\beta}\overline{J}_{\alpha\beta} =  \overline{a}(u e_{12} + \overline{u} e_{34})^{\alpha\beta}(e^{12} + e^{34})_{\alpha\beta} = \overline{a}\Re(u) = \overline{a} \cos(t).\] 
So the metric (of signature $(1, 3)$) becomes:
\[ g_J = \frac{dt^2 -   dx^2 -  \cos^2(x) dy^2 -  \sin^2(x) dz^2}{j \cos^2(t)}.\]
The Einstein tensor for this metric is easily computed and is $3j$ times the metric.   This is the metric of de Sitter space [62].  It has a three-dimensional space-like conformal infinity, in $\mathbb{K}$, given by  the vanishing of $ X^{\alpha\beta}\overline{J}_{\alpha\beta}$.
\begin{itemize}
 \item If $j < 0$, without loss of generality, we may choose $J = b(e_{13}  + e_{42})$, where $0 \ne b \in \mathbb{C}$.   
\end{itemize}
Then  we have:
\[  J^{\alpha\gamma}\overline{J}_{\beta\gamma}  =  \frac{1}{4}|b|^2[(e_1 \otimes e_3 - e_3 \otimes e_1  +  e_4 \otimes e_2 - e_2 \otimes e_4).(e^1 \otimes e^3 - e^3 \otimes e^1  +  e^4 \otimes e^2 - e^2 \otimes e^4)]^\alpha_\beta\]
\[  = - \frac{|b|^2}{4} (e_1\otimes e^1 +  e_2\otimes e^2 +   e_3\otimes e^3 +   e_4\otimes e^4)^\alpha_\beta =    - \frac{|b|^2}{4} \delta^\alpha_\beta.\]
So  $j = - |b|^2$.  We have also:
\[ s^{-1} X^{\alpha\beta}\overline{J}_{\alpha\beta} =  - \overline{b}(p e_{13} - \overline{p} e_{42})^{\alpha\beta}(e^{13} + e^{42})_{\alpha\beta} = - i\overline{b}\Im(p) =  - i\overline{b}\cos(x)\sin(y).\] 
So the metric (of signature $(1, 3)$) becomes:
\[ g_J =  \frac{dt^2 -   dx^2 -  \cos^2(x) dy^2 -  \sin^2(x) dz^2}{- j \cos^2(x)\sin^2(y)}.\]
The Einstein tensor for this metric is easily computed and is $3j$ times the metric.   This is then the metric of anti-de-Sitter space.   It has a three-dimensional time-like conformal infinity, given by  the vanishing of $ X^{\alpha\beta}\overline{J}_{\alpha\beta}$.
\begin{itemize}
 \item If $j = 0$, without loss of generality, we may choose $J = c(e_{1} - e_3)\wedge (e_2 - e_4)$, where $0 \ne c \in \mathbb{C}$.   
\end{itemize}
Then  we have:
\[  \hspace{-30pt} J^{\alpha\gamma}\overline{J}_{\beta\gamma}  =  - \frac{1}{4}|c|^2[(e_1 - e_3) \otimes (e_2 - e_4) - (e_2 - e_4) \otimes (e_1- e_3)).((e^1 + e^3)\otimes (e^2 + e^4) - (e^2+ e^4) \otimes (e^1+ e^3))]^\alpha_\beta  = 0.\]
So  $j = 0$, as required. We have also:
\[ s^{-1} X^{\alpha\beta}\overline{J}_{\alpha\beta} =  \overline{c}(u e_{12} + \overline{u} e_{34} + pe_{13} - \overline{p} e_{42} +   qe_{14} - \overline{q} e_{23})^{\alpha\beta}(e^{12} + e^{34} + e^{14} + e^{23})_{\alpha\beta}\]
\[ = \overline{c}(\Re(u) + \Im(q)).\]
So the metric (of signature $(1, 3)$) becomes:
\[g_J =   \frac{dt^2 -   dx^2 -  \cos^2(x) dy^2 -  \sin^2(x) dz^2}{|c|^2 (\cos(t) + \sin(x)\sin(z))^2}.\]
The Einstein tensor for this metric is easily computed and is zero.  This is then the metric of Minkowski space-time [49].   It has a three-dimensional null conformal infinity, given by  the vanishing of $ X^{\alpha\beta}\overline{J}_{\alpha\beta}$.\\\\
Summarizing: for all real non-zero $J^{\alpha\beta}$, the metric $g_J$ is of constant curvature and  has Ricci scalar $-12j$, where $j = J^{\alpha\beta}\overline{J}_{\alpha\beta}$.  All three constant curvature cosmologies arise.
\eject\noindent

\subsection{The conformal geodesics of the Klein space-time}
The conformal geodesics of the Klein space-time are represented by non-singular quadrics in the projective twistor space, equivalently by giving $\mathbb{T}$ a metric, a non-degenerate symmetric complex bilinear form.  When $\mathbb{T}$ has such a metric $G$, say, there are two one-dimensional families of two-dimensional subspaces of $\mathbb{T}$, which are isotropic (totally null) with respect to $G$.   These give a pair of curves in the space-time, say $\alpha^\pm_G$, such that if $x$ and $y$ are distinct points of these curves, then they are null related if and only  if one belongs to  $\alpha^+_G$ and the other to $\alpha^-_G$.  So each family foliates the quadric $G$ by projective lines, where the projective lines of one family do not meet each other and yet each line from one family meets every line of the other family.  Each such family is naturally a conic in the Klein quadric and each such family is a conformal geodesic.   All proper conformal geodesics are obtained in this way.   Altogether the family of proper conformal geodesics is nine-dimensional.    Each conformal geodesic determines uniquely a three-dimensional non-degenerate subspace of $\Omega^2(\mathbb{T})$, so a projective plane in  $\mathbb{P}(\Omega^2(\mathbb{T}))$,  such that the conformal geodesic is just the intersection of the three-space with the Klein quadric and conversely, non-degenerate three-planes in $\Omega^2(\mathbb{T})$ give rise to a conformal geodesic, on intersection with the Klein quadric.   If one such three-plane gives rise to one family of lines of a quadric $G$ in twistor space, then its orthogonal complement gives rise to the other family of the same quadric. 

Given a pair of quadrics in twistor space, so a pair of metrics $G$ and $H$ for $\mathbb{T}$,   then, regarding $G$ and $H$ as endomorphisms from $\mathbb{T}$ to $\mathbb{T}^*$, the endomorphism $J = G^{-1}H$ has a quartic characteristic polynomial, with at most four roots.  In the generic case of pairwise distinct roots $(a, b, c, d) \in \mathbb{C}^4$, with $abcd \ne 0$, we may simultaneously diagonalize $G$ and $H$, so in suitable co-ordinates, $Z = (p, q, r,s) \in \mathbb{C}^4$ for a point $Z$ of twistor space, we have:
\[ G(Z, Z)  = p^2 + q^2 + r^2 + s^2, \hspace{10pt} H(Z, Z) = ap^2 + b q^2 + c r^2 + ds^2.\]
The intersection of the two quadrics is an elliptic curve.   The equation $z^2 = \det(x G + yH)$, with $(x, y) \in \mathbb{C}^2$ and $z \in \mathbb{L}$ gives an isomorphic  elliptic curve:  in particular the two curves share the same $j$-invariant.  Geometrically the equation  $z^2 = \det(x G + yH)$ represents the space of projective lines that lie on the elements of the quadric pencil defined by $G$ and $H$; it maps canonically to the Riemann sphere of the ratio $x:y$ and the branch points of the projection are correspond to the singular elements of the pencil.
\eject\noindent
\subsection{The general Klein cosmologies}   
We generalize the Klein cosmologies in the simplest way possible: we retain the conformally flat Klein quadric background and consider the general holomorphic metric with that background.  So the metric, labelled $g_Q$ is given by the formula:
\[ g_{Q} = \frac{dX \wedge dX}{Q(X)}, \hspace{10pt} X\wedge X = 0.\]
Here the $\mathbb{L}$-valued function, $Q(X)$, is holomorphic on its domain and homogeneous of degree $2$.  Although not absolutely necessary, it will be convenient for us to regard the function $Q$ as being defined and holomorphic, not only on the Klein quadric, but also in a neighbourhood in $\Omega^2(\mathbb{T})$ of the quadric.  Note that we recover the cosmological metric $g_J$ by taking $Q(X)$ proportional to  $(X\wedge J)^2$.    Denote the domain of the metric $g_Q$ by $\mathbb{K}_Q$, so $\mathbb{K}_Q$ is the complement of the hypersurface $Q(X) =  0$.
The bundle $\mathcal{T}$ is the trivial bundle $\mathbb{K}_Q \times \mathbb{T}$, with the flat connection, denoted $d$.    

The key point rendering the calculations tractable is the following.   Introduce the generators of the general linear group, $F^\alpha_\beta$, given in ambient co-ordinates by the formula:
\[ F^\alpha_\beta = 2 X^{\alpha\gamma} \partial_{\beta\gamma}, \hspace{10pt} \partial_{\alpha\beta} X^{\gamma\delta} = \delta^{\gamma}_{[\alpha} \delta_{\beta]}^{\delta}.\]
More intrinsically, we have on the Klein quadric the duality relation, which determines the vector field $F^\alpha_\beta$:
\[ F^{\alpha}_\beta. dX^{\gamma\delta} =  - 2 \delta_{\beta}^{[\gamma} X^{\delta]\alpha}.\]
We have the formula:
\[ F^{\alpha}_\alpha. dX^{\gamma\delta} = 2 X^{\gamma\delta}.\]
This just indicates that the trace of $F$ is the twice the homogeneity operator.  Next we have the relations:
\[ X^{[\rho\sigma} F^{\alpha]}_{\beta} = 0, \]
\[  2X^{\beta \rho} F^{\alpha}_\beta. dX^{\gamma\delta} =   4 X^{ \rho[\gamma} X^{\delta]\alpha}  = 2X^{\alpha\rho} X^{\gamma\delta}  = X^{\alpha\rho} F_\sigma^\sigma.dX^{\gamma\delta}.\]
So we have:
\[ X^{\beta \rho} F^{\alpha}_\beta =  - \frac{1}{2} X^{\rho\alpha} F^\sigma_\sigma.\]
Now consider the contracted tensor product: $F_{\rho}^\alpha \otimes F^\rho_\beta$.  We have, using the relations just proved:
\[ X^{\gamma \delta} F_{\rho}^\alpha \otimes F^\rho_\beta    =  - 2X^{\rho[\gamma} F_{\rho}^{|\alpha|} \otimes F^{\delta]}_\beta\]
\[  =   - X^{\alpha[\gamma} F_{\rho}^{|\rho|} \otimes F^{\delta]}_\beta  =   \frac{1}{2} X^{\gamma\delta} F_{\rho}^{\rho} \otimes F^{\alpha}_\beta.\]
Accordingly, we infer the relation:
\[ 2F_{\rho}^\alpha \otimes F^\rho_\beta    =  F_{\rho}^{\rho} \otimes F^{\alpha}_\beta.\]
If now $Q$ is a function on the line bundle $X$, homogeneous of degree $2$, we have   $\frac{1}{2} F^\tau_\tau Q = 2Q$, so we get the relation:
\[  \left(\frac{1}{2}F_{\rho}^\alpha Q\right)  \left(\frac{1}{2}F^\rho_\beta Q\right) = \frac{1}{2} \left(\frac{1}{2} F^\tau_\tau Q\right) \left(\frac{1}{2} F^\alpha_{\beta} Q\right) = \frac{Q}{2} F^\alpha_{\beta} Q. \]
Put $ \gamma^{\alpha}_\beta  = \frac{1}{2} Q^{-1} F^\alpha_\beta Q $.   Then we have:
\[ \gamma^{\alpha}_\rho  \gamma^{\rho}_\beta =  \gamma^{\alpha}_\beta, \hspace{10pt} \gamma^\alpha_\alpha = 2.\]
So $\gamma$ is a projection operator of trace two and is homogeneous of degree zero in $X$, so represents a section of the bundle $\textrm{End}(\mathcal{T})$, the bundle of endomorphisms of $\mathcal{T}$.  Also we have:
\[ X^{[\rho\sigma} \gamma^{\alpha]}_\beta = 0, \]
\[ X^{\beta\rho} \gamma_\beta^\alpha =  \frac{1}{2} Q^{-1}X^{\beta\rho} F^\alpha_\beta Q   =  - \frac{1}{4}  X^{\rho\alpha} Q^{-1} F^\beta_\beta Q =   \frac{1}{2} X^{\alpha\rho}. \]
These formulas show, in particular,  that the endomorphism $\gamma$ has two dimensional image, the space of all $Z$ such that $Z \wedge X = 0$.    So we have proved:
 \begin{theorem}
The endomorphism $\gamma$ is a rank two projection operator: so $\tr\gamma=2$ and $\gamma^2=\gamma$.  The two dimensional image of $\gamma$ is the set of all $Z$ such that $Z\wedge X=0$.
\end{theorem}
The image of $\gamma$ is called the primed co-spin bundle, denoted $\mathcal{T}_-$ and the kernel of $\gamma$ is the unprimed spin bundle, denoted $\mathcal{T}^+$; so relative to $\gamma$, we may write the twistor $Z^\alpha$ as a split pair:
\[ Z^{\alpha} = (Z^A,  \hspace{4pt} Z_{A'}).\]
Here $Z^A$ lies in $\mathcal{T}^+$, so is the unprimed spinor part of $Z^\alpha$, whereas $Z_{A'}$ lies in $\mathcal{T}_-$, so is the primed co-spinor part.  
\eject\noindent  The (flat) connection $d$ now reads, for some one-forms $\theta^{AA'}$ and $\rho_{AA'}$:
\[ d(Z^A, \hspace{4pt} Z_{A'}) = (dZ^A + \theta^{AB'} Z_{B'},  \hspace{4pt} dZ_{A'} + \rho_{A'B} Z^B).\]
This equation serves to define the spin connection $dZ^A$, the primed co-spin connection $dZ_{A'}$ and the canonical one-forms $\theta^{AA'} = \theta^a$ and $\rho_{A'A} = \rho_a$.     Here we abbreviate pairs of indices, one unprimed and one primed, by lower case Latin letters, so $a = AA'$, $b = BB'$, etc., as convenient. \\\\The skew twistor $X^{\alpha\beta}$ represents the exterior product of the primed co-spin bundle with itself, so we have the formula:
\[ X^{\alpha\beta} = (0, 0, 0, x_{A'B'}), \hspace{10pt} 0 \ne x_{A'B'} = - x_{B'A'}.\]
Similarly, we may write its dual as:
\[ X_{\alpha\beta} = \frac{1}{2} \epsilon_{\alpha\beta\gamma\delta} X^{\gamma\delta} = (x_{AB}, 0, 0,  0),  \hspace{10pt} 0 \ne x_{AB} = - x_{BA}.\]
Here $x_{AB}$ takes values in $\mathbb{L}$.  
We use $x_{A'B'}$ and $x_{AB}$ to split the skew-symmetrizers:
\[   2\delta_{[A'}^{C'} \delta_{B']}^{D'}  = x_{A'B'}x^{C'D'}, \hspace{10pt}2\delta_{[A}^{C} \delta_{B]}^{D}  = x_{AB}x^{CD}. \]
Here $x^{A'B'}$ and $x^{AB}$ are skew and inverse to $x_{A'B'}$ and $x_{AB}$, respectively.  We employ these spinors to raise and lower spinor indices according to the scheme:
\[ v^A x_{AB} = v_B, \hspace{10pt} w^{A'} x_{A'B'} = w_{B'}, \hspace{10pt} x^{AB} y_B = y^A, \hspace{10pt}  x^{A'B'} z_{B'} = z^{A'}.\]
The non-zero components of the skew tensor $\epsilon_{\alpha\beta\gamma\delta}$ are determined by the relation:
\[ \epsilon_{AB}^{\hspace{14pt} C'D'} = x_{AB} x^{C'D'}.\] 
The connection $d$ extends naturally to the trivial line bundle $ \mathcal{L} = \mathbb{Q} \times \mathbb{L}$, such that we have $d \epsilon_{\alpha\beta\gamma\delta} = 0$.   Then we have, for some one-form $\alpha$, the formula:
\[ d x_{A'B'} = \alpha x_{A'B'}, \hspace{10pt} d x_{AB} = \alpha x_{AB}.\]
\eject\noindent
Now we compute $dX^{\alpha\beta}$ and $dX_{\alpha\beta}$.  We have:
\[ dX^{\alpha\beta} = (0, \theta^{AC'} x_{C'B'}, - \theta^{BC'} x_{C'A'},  dx_{A'B'})  = (0, \theta^{A}_{B'}, - \theta^{B}_{A'},  \alpha x_{A'B'}), \]
\[ dX_{\alpha\beta} = (dx_{AB}, - \theta^{A'C} x_{CB},  \theta^{B'C} x_{CA},  0)  = (\alpha x_{AB}, - \theta^{A'}_{B},  \theta^{B'}_{A},  0). \]
So the conformal structure of the space-time is:
\[  dX^{\alpha\beta} dX_{\alpha\beta} =  (0, \theta^{A}_{B'}, - \theta^{B}_{A'},  \alpha x_{A'B'}).(\alpha x_{AB}, - \theta^{A'}_{B},  \theta^{B'}_{A},  dx_{A'B'})\]
\[ = \theta^{A}_{B'} \theta^{B'}_{A}  + \theta^{B}_{A'} \theta^{A}_{B'}  = - 2g_{ab} \theta^a \theta^b,\]
\[ g_{ab} = x_{AB} x_{A'B'} = g_{ba}, \]
\[ dg_{ab} = 2\alpha g_{ab}.\]
Here we abbreviate pairs of indices one unprimed and one primed by lower case Latin letters, so $a = AA'$, $b = BB'$, etc., as convenient.  Since the conformal structure is known to be non-degenerate, we see that $\theta^a$ gives an isomorphism of the tangent bundle of $\mathbb{K}$ with the tensor product of the primed and unprimed spin bundles.  Then $\theta^a$ is called the canonical one-form.  Also the same equation shows that $g_{ab}$ represents the metric tensor.   We use $g_{ab}$ and its inverse to raise and lower vector indices and put:
\[ g = \theta^a \theta_a =   g_{ab}\theta^a \theta^b = - \frac{1}{2} dX^{\alpha\beta} dX_{\alpha\beta}.\]
Introduce the one-form $\theta^\alpha_\beta$:
\[ \theta^\alpha_\beta = X_{\beta \gamma} dX^{\alpha\gamma} =  (0, \theta^{A}_{C'}, - \theta^{C}_{A'},  \alpha x_{A'C'})(x_{BC}, 0, 0, 0)\]
\[ = (0, 0, \theta_{A'B}, 0).\]
The one-form  $\theta^\alpha_\beta$ is invariant under the scaling of $X$ up to a factor, so represents a (weighted) one-form on the space $\mathbb{Q}$ itself.    Then we have:
\[   4\theta^{[\alpha}_{[\gamma}\otimes \theta^{\beta]}_{\delta]} = X_{\gamma\delta} X^{\alpha\beta}g.\]
 Note that we have:
\[  F^{\alpha}_\beta. \theta^{\gamma}_{\epsilon}  = F^{\alpha}_\beta. X_{\epsilon\delta}dX^{\gamma\delta}   = F^{\alpha}_\beta. \theta^{\gamma}_{\epsilon} =   - 2 \delta_{\beta}^{[\gamma} X^{\delta]\alpha} X_{\epsilon\delta} =  X^{\alpha\gamma} X_{\beta\epsilon}. \]  
Thus the trace-free part of $F^{\alpha}_\beta$ is dual to  $\theta^{\gamma}_{\epsilon}$, so represents the space of tangent vectors to $\mathbb{Q}$.  
\\\\
Next we determine the curvature of the spinor connection:
\[ 0 = d^2 Z^\alpha = d(dZ^A + \theta^{AB'} Z_{B'}, \hspace{5pt} dZ_{A'} + \rho_{A'B} Z^B)\]
\[ = (d^2  Z^A  + \theta^{AC'} \rho_{C'B}Z^B + (d\theta^{AB'}) Z_{B'}, \hspace{5pt} d^2Z_{A'} - \theta^{B'C}\rho_{CA'}Z_{B'}  + (d\rho_{A'B}) Z^B)\]
So we have the curvature relations:
\[ d^2 Z^A = \mathcal{R}^A_B Z^B, \hspace{10pt}  d^2 Z_{A'} = - \mathcal{R}^{B'}_{A'} Z_{B'}, \]
\[  \mathcal{R}^A_B = -  \theta^{AC'} \rho_{C'B}, \hspace{10pt} \mathcal{R}^{A'}_{B'} = -  \theta^{A'C} \rho_{CB'}, \]
\[ d\theta^a =0, \hspace{10pt} d\rho_a = 0.\]
The equation $d\theta^a = 0$ signifies that the connection is torsion-free.  Applying $d$ to the equations $d x_{AB} = \alpha x_{AB}$ and $d x_{A'B'} = \alpha x_{A'B'}$ gives the relations:
\[ d\alpha   = - \mathcal{R}_{A}^{A}  = - \mathcal{R}_{A'}^{A'} = \theta^a \rho_a.\] 
So the spinor curvatures have equal traces and are trace-free if and only if $\alpha$ is a closed one-form if and only if  $\alpha$ is locally exact, if and only if the connection is locally Levi-Civita, for a representative of the conformal class.  \\\\
The embedding of the primed co-spin bundle inside the bundle $\mathcal{T}$ is given by the map $\pi_{A'} \rightarrow \pi_{B'} X^{B'\alpha}$, where $X^{B'\alpha} = (0, \delta^{B'}_{A'})$.  Then we have the relation:
\[  X^{\gamma\delta} = x_{A'B'} X^{A' \gamma} X^{B'\delta} = (0, 0, 0, x_{A'B'}). \]
We have:
\[ \theta^{\alpha B'} =  dX^{B'\alpha} = d(0, \delta^{B'}_{A'}) = (\theta^{AB'}, 0), \]
\[ dX^{\gamma\delta} =  2x_{A'B'} X^{A' [\gamma} \theta^{|B'|\delta]}   + \alpha X^{\gamma\delta} \]
\[ =  (0, x_{C'B'})(\theta^{DB'}, 0) - (0, x_{D'B'})(\theta^{CB'}, 0) +   \alpha X^{\gamma\delta}   = (0, \theta^C_{D'}, - \theta^{D}_{C'},  \alpha x_{C'D'}).\]
Next we have, for some skew twistor  $Y^{\alpha\beta}$, taking values in  $\mathcal{L}^{-1}$:
\[ 2\theta^{[A' |\gamma|}\otimes \theta^{B'] \delta}  = g x^{A'B'} Y^{\gamma\delta}.\]
Explicitly, we have:
\[ Y^{\alpha\beta} = (x^{AB}, 0, 0, 0).\]
Dually we have:
\[ Y_{\alpha\beta} =  \frac{1}{2} \epsilon_{\alpha\beta\gamma\delta} Y^{\gamma\delta} = (0, 0, 0, x^{A'B'}).\]
Here $Y_{\alpha\beta} $ is an ordinary skew tensor.  This gives the formula:
\[ X^{\alpha\gamma} Y_{\beta\gamma} = (0, 0, 0, \delta^{A'}_{B'}) = \gamma_{\beta}^\alpha = \frac{1}{2Q} F^\alpha_\beta Q  = Q^{-1} X^{\alpha\gamma} \partial_{\beta\gamma} Q. \]
This gives the relation,  valid for some scalar function $\sigma$ taking values in $\mathcal{L}^{-1}$. 
\[  Y_{\alpha\beta} = Q^{-1} \partial_{\alpha\beta} Q + \sigma X_{\alpha\beta}.\]
Then we have, since $X_{\alpha\beta} dX^{\alpha\beta} = 0$:
\[ Y_{\alpha\beta} dX^{\alpha\beta} = Q^{-1}(dX^{\alpha\beta}) \partial_{\alpha\beta} Q  = Q^{-1} dQ.\]
But we have also:
\[ Y_{\alpha\beta} dX^{\alpha\beta}   = (0, 0, 0, x^{A'B'}). (0, \theta^A_{B'}, - \theta^{B}_{A'}, \alpha x_{A'B'}) = 2\alpha.\]
So we have the relations:
\[ 2\alpha  =  Q^{-1} dQ, \] 
\[ d (g_{ab} Q^{-1})  = 0.\]
Putting $ \epsilon_{AB} = Q^{-\frac{1}{2}}x_{AB}$ and $ \epsilon_{A'B'} = Q^{-\frac{1}{2}}x_{A'B'}$,   we have also:
\[ d\epsilon_{AB} = 0, \hspace{10pt} d\epsilon_{A'B'} = 0.\]
We have proved that the spin connection determined by the projection $\gamma^\alpha_\beta$ is the Levi-Civita connection for the metric $Q^{-1} g_{ab}$.

The curvature tensor of the Levi-Civita connection $R_{ab}$  decomposes as:
\[ R^{ab} =  C^{ab} + 2\theta^{[a} P^{b]}.\]
Here $C^{ab}$ is the Weyl two-form, which vanishes here, since the metric is conformally flat.  The tensor $P_a$ is called the Schouten tensor and obeys the relation $\theta^a P_a = 0$.  Then we have:
\[ R^{AB} =  R^{BA} =  \frac{1}{2} \epsilon_{A'B'} R^{ab} =  \epsilon_{A'B'} \theta^{[a} P^{b]}  =  - \theta^{AA'} P_{A'}^{B}, \]
\[ R_B^A = - \theta^{A'B} P_{b}, \]
\[  R^{A'B'}  = R^{B'A'} = \frac{1}{2} \epsilon_{AB} R^{ab} =  \epsilon_{AB} \theta^{[a} P^{b]}  =  \theta^{AA'} P_{A}^{B'}, \]
\[ R_{B'}^{A'} = - \theta^{AB'} P_{b}. \]
So we have the equations: $\theta^{AA'}P_{A'B} = \theta^{AA'} \rho_{A'B}$ and  $\theta^{AA'}P_{AB'} = \theta^{AA'} \rho_{AB'}$, whence we have the key relation:
\[ \rho_a = P_a.\] 
This gives us a simple way of directly calculating the Schouten tensor, given $Q$: namely it is the endomorphism $- \gamma(d\gamma)$.  For example in the case of the Klein cosmologies, we have:
\[ \gamma^\alpha_\beta = 2(X\wedge Y)^{-1} X^{\alpha\gamma} \partial_{\beta\gamma} (X \wedge Y) =  2(X^{\rho\sigma} Y_{\rho \sigma})^{-1} X^{\alpha\gamma} Y_{\beta\gamma}, \]
\[ - (\gamma  (d\gamma))^\alpha_\beta   = 2(X^{\tau\upsilon} Y_{\tau \upsilon})^{-2}X^{\alpha\gamma} dX^{\rho\sigma} \left(Y_{\rho \sigma}  Y_{\beta\gamma} - 2Y_{\rho\gamma} Y_{\beta\sigma}\right)\]
\[   =  6(X^{\gamma\delta} Y_{\gamma\delta})^{-2}\left(X^{\alpha\gamma} dX^{\rho\sigma} Y_{[\rho\sigma} Y_{\beta\gamma]}  \right)\]
\[ = -  Y^{\rho\sigma} Y_{\rho\sigma}(X^{\gamma\delta} Y_{\gamma\delta})^{-2} \theta^\alpha_\beta.\]
Here we have:
\[ Y_{\alpha\beta} = \frac{1}{2} \epsilon_{\alpha\beta\gamma\delta}Y^{\gamma\delta}, \hspace{10pt} X_{\alpha\beta} = \frac{1}{2} \epsilon_{\alpha\beta\gamma\delta}X^{\gamma\delta}, \hspace{10pt}  \theta^\alpha_\beta = X_{\beta\gamma} dX^{\alpha\gamma}, \]
\[ 12 Y_{[\rho\sigma} Y_{\beta\gamma]}  = \epsilon_{\rho\sigma\beta\gamma} Y^{\alpha\beta}Y_{\alpha\beta}.\]
Taking the co-vector component of $-\gamma d\gamma$, we get:
\[ P_b = - Y^{\rho\sigma} Y_{\rho\sigma}(X^{\gamma\delta} Y_{\gamma\delta})^{-2} \theta_b = \theta^a P_{ab}, \]
\[ P_{ab} = -  g_{ab}Y^{\rho\sigma} Y_{\rho\sigma}.\]
This gives the Ricci scalar as $- 3P_e^e = 12 Y^{\rho\sigma} Y_{\rho\sigma}$, in agreement with our previous calculations.
\eject\noindent
\section{Osculation}
\subsection{The tractor calculus}
Let $M$ be a conformal manifold of dimension $n>2$, of signature $(p, q)$,  where  $p + q = n$.  So $M$ is equipped with a non-degenerate metric $g$ that is given only up to scale, $g\sim \lambda g$ for all positive scalars $\lambda$.  At each point $x\in M$, the specification of such a conformal class is simply a point $[g_x]$ in the projective space over $\op{Sym}^2T_x^*M$, the symmetric square of the cotangent bundle to $M$ at $x$.  As $x$ varies, the conformal class is identified with a distinguished line subbundle $\mathcal G$ of $\op{Sym}^2T_x^*M$.

The total space of the bundle $\mathcal G$ carries a taulogical degenerate metric $g_0$ defined on tangent vectors $X,Y\in T_{(x,g)}\mathcal G$ by
$$g_0(X,Y) = g(\pi_*X,\pi_*Y) $$
where $\pi:\mathcal G\to M$ is the bundle projection.   Dilations of the group of non-zero complex numbers act on $T^*M$ in the usual way, and this also yields an action $\mathcal G$ via $\rho_s g=s^2 g$.  The infinitesimal generator of the group of dilations is denoted by $X$.  So $g_0(X,\hspace{2pt} -)=0$ and $\mathscr{L}_X g_0=2g_0$.

A function $f$ defined on an open subset $\pi^{-1}U\subset\mathcal G$, where $U$ is an open subset of $M$, is called a conformal density with weight $w$ if $\delta_s^*f=s^w f$; equivalently, if $\mathscr{L}_Xf=wf$. We shall usually regard $w$ as being an integer, so it does not matter if the objects involved are real or complex.  The densities of conformal weight $w$ form a bundle over $M$, denoted $E[w]$.  We use conformally weighted bundles to twist other tensor bundles on $M$, so for instance $TM[w]=TM\otimes E[w]$ is the bundle of vectors with conformal weight $w$.

If $\tau$ is a section of $E[1]$, then the degenerate metric $\tau^{-2}g_0$ is homogeneous of degree zero and annihilated by $X$.  So $\tau^{-2}g_0$ descends to a proper metric $g_\tau$ on $M$ belonging to the conformal class.  Choosing a metric in the conformal class is (up to an overall sign) the same as specifying a section of $E[1]$.  We therefore call $E[1]$ the bundle of conformal scales.  Once such a background metric represented by $\tau$ has been chosen, all of the density bundles acquire a natural trivialization by the powers of $\tau$.  The Levi-Civita connection $\nabla_\tau$ associated with a conformal scale $\tau$ can then be extended to the bundles of density and their tensor products with other tensor bundles by imposing that $\tau$ is a parallel section of $E[1]$, so $\nabla_\tau\tau=0$.  There is a canonical section of $E[2]\otimes\op{Sym}^2T^*M$ that represents the conformal class of the metric; the inverse metric is a section of $E[-2]\otimes\op{Sym}^2TM$.
\eject
It is not just fixed conformal scales that concern us, but also what happens to all of the various objects when one conformal scale is replaced by another, $\hat{\tau}=\Omega^{-1}\tau$, where $\Omega$ is some regular function on $M$.  Then on sections of weight $w$, we have
\begin{align*}
\nabla_{\hat{\tau}}\sigma &= \nabla_{\tau}\sigma + w \Upsilon \sigma\\
\nabla_{\hat{\tau}}V &= \nabla_{\tau}V + (w+1)\Upsilon\otimes V - g_\tau(V)\otimes g_\tau^{-1}(\Upsilon) + (\Upsilon\cdot V)\op{Id}\\
\nabla_{\hat{\tau}}\alpha &= \nabla_{\tau}\alpha + (w-1)\Upsilon\otimes \alpha - \alpha\otimes\Upsilon + g_\tau^{-1}(\alpha,\Upsilon)g_\tau
\end{align*}
for $\sigma\in \mathscr E[w],V\in \mathscr TM[w], \alpha\in \mathscr T^*M[w]$, where $\Upsilon=\Omega^{-1}d\Omega$.  The Schouten tensor $P_\tau$ is a trace-adjusted multiple of the Ricci tensor with a simplified conformal transformation law.  It is defined by
$$P_\tau = \frac{1}{n-2}\left(\op{Ric}_\tau - \frac{\op{Sc}_\tau\,g_\tau}{2(n-1)}\right) $$
where $\op{Ric}_\tau$ is the Ricci curvature and $\op{Sc}_\tau$ is the scalar curvature.  Notice that $\op{tr}_{g_\tau}P_\tau=2^{-1}(n-1)^{-1}\op{Sc}_{\tau} $.  The Schouten tensor transforms according to
$$P_{\hat{\tau}} = P_\tau -\nabla_\tau\Upsilon+\Upsilon\otimes\Upsilon-\frac12 g_\tau^{-1}(\Upsilon,\Upsilon)g_\tau.$$

The manifold $M$ also carries a natural vector bundle of dimension $n+2$ called the {\em tractor bundle}, denoted by $\mathcal T$, that is equipped with a non-degenerate quadratic form $k$ of signature $(p+1, q + 1)$ and one-dimensional null subspace $\mathcal X$ [35, 65].  This space is associated to the standard flat model of a conformal sphere (the space of rays of the null cone of $k$) together with a marked point of contact with the manifold at the point of the conformal sphere represented by the ray $\mathcal X$.  The idea of a ``marked point of contact'' is specified by a linear isomorphism $Z:TM\to E[1]\otimes\mathcal X^\perp/\mathcal X$ of vector bundles on $M$.  The metric of the tractor bundle is related to the conformal structure of $M$ by requiring that the restriction of $k$ to $\mathcal X^\perp\times\mathcal X^\perp$, when it is composed with $Z$, should agree modulo $\mathcal X$ with the conformal structure of $M$, a density in $\op{Sym}^2T^*M[2]$.  This identification makes it natural to assign the line bundle $\mathcal X$ a conformal weight of $-1$ and to identify it with $E[-1]$.  This identification is encoded in a distinguished trivialization of the line bundle $\mathcal X[1]$: that is, a preferred section $X$ of $\mathcal X[1]$.
\eject
An explicit construction of the tractor bundle can be found  in Bailey, Eastwood, and Gover, as the space of non-holonomic two-jets of conformal scales that solve (to second order) the Einstein vacuum equation with cosmological constant [2]. The holonomic constraint on two-jets yields a natural conformally invariant linear connection in the tractor bundle.  This connection coincides with the normal conformal Cartan connection, which can here be taken to mean that it satisfies the following requirements:
\begin{itemize}
\item $\nabla$ preserves the metric $k$
\item $Z$ is the mapping $TM\to E[1]\otimes\mathcal X^\perp/\mathcal X$ given by $Z(v)=\nabla_vX\pmod{\mathcal X}$
\item The connection is {\em normal}: The curvature form $\Omega$, a two-form on $M$ with values in $\op{End}(\mathcal T)$, annihilates $\mathcal X$, has range contained in $\mathcal X^\perp$, and $\Omega\circ Z^{-1}$ is totally traceless.
\end{itemize}

In terms of a fixed conformal scale $\tau$, let
\begin{align*}
Z_\tau &= \nabla_\tau X \in T^*M\otimes \mathcal T[1]\\
Y_\tau &= -\frac{1}{n}\Delta_\tau X\in \mathcal T[-1]
\end{align*}
where $\nabla_\tau$ is the connection on weighted tractors obtained by coupling the Levi-Civita connection associated to the conformal scale $\tau$ to the (conformally invariant) tractor connection, and $\Delta_\tau=g_\tau^{-1}(\nabla_\tau, \nabla_\tau)$ is the Laplacian associated to the conformal scale $\tau$.  These define a frame of $\mathcal T$, in the sense that element of the tractor bundle has the form
$$\sigma X + \mu\cdot Z_\tau + \rho Y_\tau $$
where $\sigma\in E[-1]$, $\mu\in TM[-1]$, and $\rho\in E[1]$.  Moreover, the inverse metric of $k$ is given in this frame by
$$k^{-1} = 2XY_\tau + g^{-1}_\tau(Z_\tau,Z_\tau).$$
In particular, $Y_\tau$ is null, $k(X,Y_\tau)=1$, and $Z_\tau$ is in the orthgonal complement of the span of $Y_\tau$ and $X$.
Acting on these sections, the coupled tractor Levi-Civita connection is given by
\begin{equation*}
\nabla_\tau X = Z_\tau,\quad \nabla_\tau Z = -P_\tau X - Y_\tau g_\tau,\quad \nabla_\tau Y_\tau = g_\tau^{-1}(Z_\tau)\cdot P_\tau 
\end{equation*}
where in the last equation the contraction between $g_\tau^{-1}(Z_\tau)$ and $P_\tau$ is on one index.
\eject\noindent
The tractor $D$-operator is a conformally-invariant differential operator $D:\mathcal V[w]\to\mathcal T^*\otimes\mathcal V[w-1]$ acting on sections of any weighted tractor bundle $\mathcal V[w]$ via the formula
$$DV = (n+2w-2)wY_\tau\otimes V + (n+2w-2)g_\tau^{-1}(Z_\tau,\nabla_\tau V) - X\Box_\tau V$$
where $\Box_\tau = (\Delta_\tau + w\op{tr}_{g_\tau}P_\tau).$
\subsection{Osculation of conformal densities}
\begin{definition}
Two conformal densities $\sigma_1$ and $\sigma_2$ are said to {\em osculate} at a point of $M$ if the following three conditions hold at that point
\begin{align*}
\sigma_1&=\sigma_2\\
D\sigma_1 &= D\sigma_2\\
D^2\sigma_1&=D^2\sigma_2.
\end{align*}
\end{definition}

These equations are not entirely independent.  For any density, the tractor $D^2\tau = (D\otimes D)\tau$ is symmetric and trace-free.  Moreover, if the weight is regarded as fixed, $w$ say, then the contraction of any of these tensor with $X$ depends only on the weight and earlier derivatives, which gives two point constraints: $D_X\sigma=w(n-2+2w)\sigma$ and $D_XD\sigma=(w-1)(n-4+2w)D\sigma$.  In particular, for most weights $D_X\sigma$ is determined by $D^2\sigma$, although notably not in the case of conformal scales ($w=1$).

\begin{theorem}\label{curvatureconditions}
If two conformal scales $\tau_1$ and $\tau_2$ osculate at a point of $M$, then the following conditions hold at the point:
\begin{itemize}
\item $\tau_1=\tau_2$
\item $P_{\tau_1}=P_{\tau_2}$
\item $\op{div}_{\tau_1}P_{\tau_1}=\op{div}_{\tau_2}P_{\tau_2}$
\item $\op{div}^2_{\tau_1}P_{\tau_1}=\op{div}^2_{\tau_2}P_{\tau_2}$.
\end{itemize}
These conditions are necessary but not sufficient.  The equation $D\tau_1=D\tau_2$ contains the additional information of a $Y$ tractor, which is an $n$-dimensional affine degree of freedom.
\end{theorem}

In the theorem statement, $\op{div}_\tau$ denotes the divergence with respect to the Levi-Civita connection of the conformal scale $\tau$.  In particular, the third equation is an equation of $n$-dimensional covectors and the last equation is a scalar equation.

\begin{proof}
We expand out the osculation condition:
\begin{align*}
\tau^{-1}D\tau &= -\op{tr}_{g_\tau}P_\tau X +  nY_\tau\\
\tau^{-1}D^2\tau &= \left[\op{div}^2_\tau P_\tau + (n-1)\op{tr}\left(g_\tau^{-1}(P_\tau)\cdot g_\tau^{-1}(P_\tau)\right)\right]X^2 - \left[2(n-2)\op{div}_\tau P_\tau\right] XZ_\tau + \\
&\qquad+\left[n(n-2)P_\tau - (n-2)(\op{tr}_{g_\tau}P_\tau)g_\tau\right]Z_\tau^2.
\end{align*}
Observe that $\tau^{-2}k(D\tau,D\tau)=-2n\op{tr}_{g_\tau}P_\tau$.  So
$$n\tau Y_\tau = D\tau - (2n\tau)^{-1}k(D\tau,D\tau)X.$$
Then, because $Z_\tau$ is uniquely determined from its image in $\mathcal T$, $Y_\tau$ determines $Z_\tau$, and the coefficients of $D\tau$ and $D^2\tau$ can all be determined from invariant combinations with $D\tau$ and $X$.
\end{proof}
\subsection{Osculation and the Fefferman-Graham theory}
The apparatus of tractor calculus can also be completely built out of the ambient construction of Charles Fefferman and Robin Graham [28, 29].  This consists of an asymptotically vacuum metric $\tilde{g}$ defined on the manifold $\tilde{M}=\mathscr{G}\times (-1,1)$ that extends the degenerate metric $g_0$ from $\mathscr G$ and is homogeneous of degree two under the natural action of the group of non-zero complex numbers.  In particular, $\mathscr{G}\subset\tilde{M}$ is a null hypersurface.  All of the basic objects (densities and so forth) can be regarded as being extensible to some asymptotic order off this null hypersurface.  The metric $\tilde{g}$ can be made asymptotically vacuum to order at least $n/2$ near the null hypersurface, where $n$ is the dimension of $M$, in a parameter $Q=\tilde{g}(X,X)$ that is a degree $2$ defining function of $\mathscr{G}$ in $\tilde{M}$.

Conformally invariant powers of the Laplace operator, as described in Graham, Jenne, Mason, and Sparling, are differential operators that can be applied to conformal densitites of appropriate weights [37].  They arise as obstructions to the following Dirichlet problem:
\begin{itemize}
\item Given a section $\sigma$ of $E[w]$, construct a formal power series $\tilde{\sigma} = \sigma + Q\sigma_1 + \cdots + Q^{k-1}\sigma_k + O(Q^k)$ that solves the ambient Laplace equation $\tilde{\Delta}\tilde{\sigma} = 0$ modulo $O(Q^{k-1})$.
\end{itemize}
In dimension $n$, a harmonic extension is possible to order at least $k=w+n/2$.  At this value of $k$, when $n$ is even, there is an obstruction to continuing the procedure in the form of a linear differential operator applied to the original density $\sigma$.

\begin{theorem}
Suppose that $n+2w\ge 5$ where $n$ is the dimension of the space-time $M$ and $w$ is an integer weight.  Two densities $\sigma_1$ and $\sigma_2$ of conformal weight $w$ osculate at a point of $M$ if and only if the $2$-jets of their harmonic extensions agree at that point.
\end{theorem}

The inequality $n+2w\ge 5$ is not a sharp condition on the weight $w$ of the conformal densities or on the dimension $n$, but conveniently includes all of the dimensions and weights that are of interest to us (in particular, $w=1$ and $n\ge 3$).  The purpose of the inequality is to ensure that there are enough valid terms in the asymptotic expansion of the harmonic extensions of the densities so that two derivatives can unambigously be performed.

\begin{proof}
This theorem follows almost immediately from the relation of the tractor calculus to the ambient construction as described in detail in \S 3 of Gover and Peterson [36].  We recall briefly the relation they establish between the operator $D$ and the ambient connection $\tilde\nabla$.  A tensor field $V$ on $\tilde{M}$ is homogeneous of weight $w$ if $\tilde{\nabla}_XV=wV$. The restriction of $V$ to the null surface $\mathscr G$ is identified with a tractor of weight $w$, and we have
$$DV = (n+2w-2)\tilde{\nabla} V - X\otimes\tilde{\Delta}V.$$
In particular, if $\tilde{\tau}$ is a harmonic extension of $\tau$, we have $D\tau=(n+2w-2)\tilde{\nabla}\tilde{\tau}$, and
$$(D\otimes D)\tau = (n+2w-2)(n+2w-4)\tilde{\nabla}\otimes\tilde{\nabla}\tilde{\tau} + (n+2w-2)X\otimes \tilde{\Delta}(\tilde{\nabla}\tilde{\tau}).$$
The last term is zero, though, because of the Einstein vacuum equation in the ambient space and the assumption that $\tilde{\tau}$ is harmonic.
\end{proof}

\subsection{Osculation by cosmological metrics}
The previous section shows that the notion of two densities osculating can be formulated in an entirely conformally-invariant manner.  So it is very natural to try to use this information to decouple the Ricci curvature entirely from the conformal curvature.  This idea is also rather phenomenologically appealing, because of the tendency of Ricci and Weyl curvature to act on different cosmic scales: the Weyl curvature is the vehicle by which gravity propagates in free space while the Ricci curvature then determines what a clump of matter would do if it were ``left to its own devices'', forbidden from interacting with the ambient gravitational field [59].  So, we look for the simplest model in which a matter field, as represented by the Ricci curvature in an infinitesimal neighborhood of a point, can break conformal invariance.  This is a cosmology of the form of a homogeneous quadratic polynomial on the tractor space restricted to the null cone, localized at the marked point of contact.

\begin{definition}
Let $\mathbb V$ be a vector space with a non-degenerate symmetric bilinear form $k$ and let $\mathbb K$ be the projective null cone of $k$, a quadric in $\mathbb P\mathbb V$.  A {\em quadric cosmology} is a metric on $\mathbb K$ of the form
$$g_q = \frac{k(dx,dx)}{q(x, x)} $$
where $q$ is a symmetric form on $\mathcal T$, linearly independent of the form $k$.
\end{definition}
The metric $g_q$ associated to the quadratic form $q$ is non-singular away from the intersection of the quadric $\mathbb K$ with the null quadric of $q$.  We have in mind specifically the case where $\mathbb V$ is the tractor space $\mathcal T_x$ at a point $x$ of the space-time $M$ and $k$ is the tractor metric.  Then we have the following:

\begin{theorem}
Let $(M,g)$ be a space-time.  At each point $x$ of $M$, there is a unique quadric cosmology on $\mathcal T_x$ that osculates with $g$ at $x$.
\end{theorem}

The statement of the theorem requires some clarification, since the space on which the metric $g$ is defined differs from that of the quadric cosmology.  There is, however, a linear isomorphism between the tractor space $\mathcal T_x$ to $M$ at $x$ and tractor space $\mathbb T$ corresponding to the quadric $\mathbb K$ at the marked point $[\mathcal X]$.  Under this linear isomorphism, the conditions for osculation are still well-defined:
$$D_{\mathcal T}^2\tau_g^2(x)=D_{\mathbb T}^2\tau_q^2([\mathcal X]) $$
where $\tau_g$ and $\tau_q$ are the conformal scales corresponding to $g$ and $q$, respectively.  Here $D_{\mathcal T}$ refers to the $D$-operator on $M$ and $D_{\mathbb T}$ refers to the $D$-operator on the tractor space $\mathbb T$, which is just the ordinary differential in that vector space.

Concretely, the curvature conditions of Theorem \ref{curvatureconditions} can be used to define the notion of osculation without any reference to the background space, modulo an affine gauge that consists of a choice of $Y$ tractor.  The $D$ operator gives a way of fixing that gauge.

\begin{proof}
It is enough to exhibit $q$ and then show that it has the required osculation properties.  Let $q=D^2\tau_g^2(x)$, so $q\in\op{Sym}^2_0 \mathcal T_x^*$.  This defines a conformal scale $\tau_q$ on $\mathbb K$.  If we now identify $\mathcal T_x$ with the space $\mathbb T$ of tractors of $\mathbb K$ at the marked point $\mathcal X$, then the equality $D_{\mathbb T}^2\tau_q^2(x)=q=D_{\mathcal T}^2\tau^2_g(\mathcal X)$ holds.
\end{proof}
\subsection{Gauge theory and the $Q$ curvature}
We now confine attention to four dimensions.  The following density, known as the $Q$ curvature, a spectral invariant discovered by Branson, has conformal weight $-4$ which is the correct weight to integrate over the space-time manifold:
\begin{align*}
Q_\tau &= -\frac16\Delta_\tau\op{Sc}_\tau - \frac12 \op{tr}\left[ (g_\tau^{-1}\op{Ric}_\tau)\cdot(g_\tau^{-1}\op{Ric}_\tau)\right]  + \frac16 \op{Sc}_\tau^2\\
&= \frac12 \Delta_\tau \op{tr}_{g_\tau}P_\tau - \op{tr}\left[ (g_\tau^{-1}P_\tau)\cdot(g_\tau^{-1}P_\tau)\right] + \left(\op{tr}_{g_\tau}P_\tau\right)^2.
\end{align*}
Under conformal changes $\hat{\tau}=\Omega\tau$, $Q$ changes via
$$Q_{\hat{\tau}} = Q_\tau + L_\tau\Omega$$
where $L_\tau$ is the fourth-order conformally-invariant Paneitz operator [55]
\begin{align*}
L_\tau f&= \op{div} \left[\op{grad}_\tau\op{div} + 2 g^{-1}_\tau(\op{Ric}_\tau) - \frac{2}{3}\op{Sc}_\tau \op{Id}\right]\op{grad}_\tau f\\
&=G_\tau d f
\end{align*}
where $G_\tau$ is a differential operator from one-forms to weight $-4$ densities given by 
$$G_\tau\omega = \delta_\tau\left(\Delta_\tau+2\op{tr}_{g_\tau}(P_\tau) - 4g^{-1}_\tau(P_\tau)\lrcorner \right).$$
Apart from having the appropriate weight to form an invariant integral, we see that (over a compact space-time) the integral of $Q$ actually is conformally-invariant because the transformation law under conformal changes is exact.  Moreover, the ingredients of $Q$ are precisely the curvature invariants that are contained in our osculating cosmologies.

The clear question is then: if $Q$ is the Lagrangian, what are the field equations?  The $Q$ curvature is closely related to other invariants of the space time.  Specifically,
$$Q_\tau=\frac{1}{16}\op{Pfaff}_\tau - \left(\frac14 \op{tr}(g_\tau^{-1}W\cdot g_\tau^{-1}W) +\frac16\Delta R\right)$$
where $\op{Pfaff}$ denotes the Chern--Gauss--Bonnet integrand (the Pfaffian of the curvature $2$-form) and $\op{tr}(g_\tau^{-1}W\cdot g_\tau^{-1}W) $ is the complete contraction of two copies of the Weyl curvature on all indices, using the metric $g_\tau$.  On a compact manifold, integral of the Pfaffian is a global topological invariant, and the Laplacian of the scalar curvature is a divergence and therefore integrates away [13, 14, 15].  So with respect to variations of the metric, $Q$ is equivalent as a Lagrangian to the square of the Weyl curvature, a well-known primitive for the Bach equation.  With appropriate boundary conditions (see Chang and Qing [15]), this also works on a non-compact manifold, and gives a conformal primitive for the space-time to be conformal to Einstein.

The pair of operators $(L_\tau, G_\tau)$ consisting of the Paneitz operator $L_\tau$ and its ``gauge companion operator'' $G_\tau$ (see Branson and Gover) appears naturally in connection with fixing the conformal gauge for solutions of the Maxwell equations, as shown in Eastwood and Singer [23].  We recall here the construction.  Let $F$ be a two-form on $M$ that solves the Maxwell equations
$$dF=0,\qquad dF^*=0.$$
We want to write $F$ in the form $F=dA$ (locally) for some potential $A$.  There is a substantial freedom in $A$, given by gauge transformations $A\mapsto A+df$ where $f$ is a function.  The standard way of fixing this gauge freedom is to impose the Lorenz gauge condition, that $d A^*=0$ as well.  This then reduces the gauge freedom to functions such that $d(df)^*=0$, which is equivalent to the wave equation $\Delta f=0$.

However the Lorenz gauge is not a conformally-invariant condition.  We should instead demand that $G_\tau A=0$.  Although $G_\tau$ is not itself conformally-invariant, this equation actaully is, and so gives a conformally-invariant gauge fixing condition.  Then the residual gauge freedom is of the form $A\mapsto A+df$ where $f$ is a solution of the Paneitz equation $P_\tau f=0$, since $P_\tau=G_\tau d$.

So, we note that by fixing the value of the $Q$ curvature, we reduce the gauge freedom in the conformal factor to that of the conformally-invariant Maxwell gauge.  This suggests the tantalizing prospect of coupling the $Q$ curvature to a Lagrangian for the Maxwell field.  It is a similar prospect that originally led Weyl in 1918  to consider his original coupling of the Maxwell field into the conformal factor of the metric [68].  However, Weyl turned out to have the wrong gauge group [69].  It was not until 1929, when Weyl used the gauge group $U(1)$, that the appropriate gauge theory was constructed [70].  So to couple gravity and electromagnetism properly, we need a suitable generalization of $Q$ for a $U(1)$ gauge group.  This is the subject of a forthcoming article by the first author.

\section{Quadric cosmologies}
By the results of the last chapter, we may completely describe the structure of matter, as encoded in the Ricci tensor and its derivatives, in a bundle of osculating quadric cosmologies, one for each space-time point.  Thus an in-depth investigation of these quadric cosmologies is warranted.   The first main result is that the geodesics of these space-times (in any dimension) form a completely integrable Hamiltonian system, a result which effectively goes back to the work of Carl Gustav Jacob Jacobi [46], extended particularly by J\"{u}rgen Moser [51]. 
\subsection{The Jacobi-Moser integrable system:  a free particle on an ellipsoid}
Let $\mathbb{V}$ be a vector space equipped with a non-degenerate symmetric bilinear form, denoted $g$, whose corresponding flat metric is written $g(dx, dx)$, for $x \in \mathbb{V}$.  We consider a particle moving in the space $\mathbb{V}$, constrained to move on an ellipsoid, whose equation   is $h(x, x) = B$, where $h$ is a non-degenerate symmetric bilinear form on $\mathbb{V}$ and $B$ is a real constant. The particle is otherwise free. The Lagrangian  $\mathcal{L}$ for this particle is:
\[ \mathcal{L} = \frac{1}{2} (g(x', x')  +  \lambda (h(x, x) - B)).\]
Lagrange's variational equations for the stationary points of the Lagrangian give:
\[ \frac{d}{dt} g(x', \hspace{3pt})  =   \lambda  h(x, \hspace{3pt}), \hspace{10pt} H = h(x, x) = B,  \]
\[ x'' = \lambda J(x), \hspace{10pt} J  = g^{-1} \circ h.\]
Here we are regarding $g$ and $h$ as endomorphisms from $\mathbb{V}$ to $\mathbb{V}^*$, the dual space of $\mathbb{V}$.
We have:
\[ 0 = h(x, x'), \hspace{10pt} 0 = h(x', x') + h(x, x'')  = h(x', x') + \lambda b(x, x),\hspace{10pt} b = h\circ g^{-1} \circ h. \]
We also regard $b$ as giving an endomorphism from $\mathbb{V}$ to $\mathbb{V}^*$.   If now $b(x, x) =0$, we get  $h(x', x') =0$.  Here and in the following, we will assume the generic case that $b(x, x) \ne 0$.  Then we get:
\[ \lambda = - \frac{h(x', x')}{b(x, x)}, \]
\[  x'' = - \frac{h(x', x')}{b(x, x)} J(x), \hspace{10pt} h(x, x') = 0.\]
Put:
\[ p = h(x'), \hspace{10pt} x' = h^{-1}(p), \hspace{10pt} p(x) = 0.\]
Then we have:
\[ x' =   h^{-1}(p), \hspace{10pt}p'  = - \frac{h^{-1}(p, p)}{b(x, x)} b(x), \hspace{10pt} p(x) = 0.\]
If now we rescale the time variable of the motion, putting  $b(x, x) z' = \dot{z}$,  for any function $z$, we get:
\[ \dot{x} =  b(x, x) h^{-1}(p) , \hspace{10pt}  \dot{p}  =   -  h^{-1}(p, p) b(x), \hspace{10pt} p(x) = 0.\]
These are Hamilton's equations for the Hamiltonian:  $F = 2^{-1} b(x, x) h^{-1}(p, p)$.\\
If we put $S = p(x)$, we have the relations:
\[ \dot{H} = 2h(x, \dot{x}) = 2b(x, x)S, \]
\[ \dot{S} =  \dot{p}(x) + p(\dot{x}) = 0.\]
In particular the evolution is consistent with the evolution of the Hamiltonian $S$, which is $(x, q) \rightarrow (e^s x, e^{-s} q)$, for $s$ real and with the condition that $H$ be constant, with $S = 0$. Then the Hamiltonian evolution of $F$ induces the geodesic evolution on the cotangent bundle of the projective space of $\mathbb{V}$ of the metric $\displaystyle{\frac{h(dx, dx)}{b(x, x)}}$, restricted to the null cone of $h$: $h(x, x) = 0$.
\subsection{The geodesic equations of the quadric cosmologies}
Let $\mathbb{V}$ be a vector space, with dual space $\mathbb{V}^*$. \\
We work in a phase space $\mathbb{V}\times \mathbb{V}^*$, equipped with the Poisson-structure $P$ given by the formula:
\[ P  =  \partial_p.\partial_x -  \partial_x.\partial_p.\]
Here $x$ is a variable with values in $\mathbb{V}$, whereas $p$ is a variable with values in $\mathbb{V}^*$.   \\
So the Poisson bracket of functions $f$ and $g$ on phase space is given by:
\[ \{f, g\} = f_p.g_x - f_x.g_p \]
Here the derivative are denoted by subscripts, so,  for example,  $f_p = \partial_p f$ and the dot indicates the dual pairing.
Then the Hamiltonian vector field of the function $f$ is $f' =  f_p \partial_x - f_x \partial_p$ and the dynamical equations of its trajectories are:
\[   x' = f_p, \hspace{10pt} p' = - f_x.\]
Here the prime denotes the derivative with respect to "time".\\\\
Let  $h$ and $g$ be given (non-degenerate) symmetric bilinear forms, $h$ on the vector space $\mathbb{V}$ and $g$ on the dual vector space $\mathbb{V}^*$.  We also regard $h$ as an endomorphism from $\mathbb{V}$ to $\mathbb{V}^*$ and $g$ as an endomorphism from $\mathbb{V}^*$ to $\mathbb{V}$. \\Put $J = g\circ h$ so $J$ is an endomorphism from  $\mathbb{V}$ to $\mathbb{V}$.
Also put:
\[ G = g(p, p), \hspace{10pt} H = h(x, x), \hspace{10pt} Q = g^{-1}(x, x), \hspace{10pt} R = h^{-1}(p, p).\]
Our system is given by the pair of Hamiltonians on the phase space, $E$ and $S$:
\[ S = p(x), \hspace{10pt} E =  2^{-1}HG. \]
Ultimately, we wish to restrict to the subspace $S = 0$ and divide out by the flow of the Hamiltonian $S$  and then further restrict to the null cone $Q = 0$.
We have the Poisson bracket:
\[ \{S, E\} = x.E_x - p.E_p = 0.\]
This means that the Hamiltonian flows of $E$  and $S$ are compatible. 
The flow of $S$ is the scaling $(x, p) \rightarrow (e^s x, e^{-s} p)$, $s$ being the parameter along the flow.   
The Hamiltonian flow of $E$ is:
\[ x' = E_p =  H g(p, \hspace{3pt} ), \]
\[ p' = - E_x = - G h(x, \hspace{3pt}  ).\]
Note that we have then $S' = (p(x))' = 0$, as expected.\\
Also we have:
\[ (g^{-1}(x, x))'  = 2g^{-1}(x, x') = 2 H p(x).\]
In particular, when we have $S = p(x) = 0$, the evolution takes place on the quadric $g^{-1}(x, x) = A$, with $A$ constant.   For the second derivative of the position, we have:
\[ x'' = (Hg(p, \hspace{3pt}))' = 2h(x, x') g(p, \hspace{3pt}  ) + Hg(p',  \hspace{3pt} ) = 2H^{-1} h(x, x') x' -  GH  J(x).\]
Here we have:
\[ G = g(p, p) = g^{-1}(x', x') H^{-2}.\]
So finally we have:
\[ x''  = (h(x, x))^{-1}(2 h(x, x') x' -  g^{-1}(x', x')J(x)).\]
This gives a closed system of second order  equations for the positions $x$ as a function of time.\\\\
Note that we then have:
\[ g^{-1}(x, x'')  = 2H^{-1}h(x, x') g^{-1}(x, x') -  g^{-1}(x', x'), \]
\[ (g^{-1}(x, x'))'  = 2H^{-1}h(x, x') g^{-1}(x, x').\] 
So if the initial conditions are such that $g^{-1}(x, x') = 0$, then $g^{-1}(x, x') = 0$ for all time.
Note the Poisson brackets:
\[ \{ E, Q\} =  \frac{H}{2} \{ G, Q\}  = 2HS, \]
\[ \{ E, R\} =  \frac{G}{2} \{ H, R\}  = - 2GS,  \]
\[ \{ S, Q\} = 2Q, \hspace{10pt} \{S, R\} = - 2R.\]
So the conditions $S = 0$, $Q = A$, with $A$ constant are compatible with the evolution. Imposing these constraints gives the geodesic spray on the sphere $g^{-1}(x, x) = A$, for the metric $\displaystyle{\frac{g^{-1}(dx, dx)}{h(x, x)}}$.
Let us change the parameter for the evolution to a new parameter $s$ (the old parameter being the time $t$).\\
Then we have:
\[  \frac{dx}{ds} =  \frac{x'}{s'}, \hspace{10pt} ( s')^3 \frac{d^2x}{d^2s} =  (s')^2  \left(\frac{x'}{s'}\right)' =   s'x'' - s'' x' \]
\[ =  - H^{-1}(h(x, x)s'' - 2 s'  h(x, x') ) x' -   (s')^{3} H^{-1} g^{-1}\left(\frac{dx}{ds}, \frac{dx}{ds}\right)  J(x)) .\]
We want to choose $s$ so as to knock out the  $x'$ term.  \\So we need the condition:
\[ h(x, x) s'' =  s' (h(x, x))', \hspace{10pt} s' = Ch(x, x).\]
Here $C$ is a non-zero constant which we may take to be unity.\\So we take:
\[ s' = h(x, x) = H.\]
Then in terms of the parameter $s$, we have the differential equation:
\[   \frac{d^2x}{d^2s} =  - H^{-1} g^{-1}\left(\frac{dx}{ds}, \frac{dx}{ds}\right)  J(x).\]
Representing the derivative with respect to $s$ by a prime (not forgetting that this is not our original time coordinate), we have reduced to the system:
\[ x'' = - \frac{g^{-1}(x', x') J(x)}{h(x, x)}, \hspace{10pt}  0 = g^{-1}(x, x'), \hspace{10pt} J = g\circ h.\]
Here our old time coordinate $t$ is recovered from the formula:
\[  t' = \frac{1}{h(x, x)}.\]
\[ H = b(x, x), \hspace{10pt} G = h^{-1}(p, p).\]
Replace $g^{-1}$ by $h$ and $h$ by $b$.   Then we have:
\[ x'' = - \frac{h (x', x') (h^{-1}\circ b)(x)}{b(x, x)}, \hspace{10pt}  0 = h(x, x').\]
Compare with the Jacobi-Moser system of the previous section:
\[ x'' = - \frac{h(x', x')}{b(x, x)} J(x), \hspace{10pt} h(x, x') = 0, \]
\[ J = g\circ h, \hspace{10pt} b = h \circ g \circ h, \hspace{10pt} h^{-1} \circ b = J.\]
It is identical.
In this language the original time co-ordinate is given by $t' = (b(x, x))^{-1}$.
\eject\noindent
\subsection{The conserved quantities: the proof of complete integrability}
For $n$ a positive integer, we consider a phase space $(x, p)$, of $2n$ dimensions,  described by $n$ "position" variables $x = \{x^i, i =1, 2, \dots, n\}$ and their $n$ conjugate "momentum" variables $p = \{p_i, i = 1, 2, \dots, n\}$.  For $h(x, p)$ a differentiable function on the phase space, put:
\[ h_i =  \partial_i h, \hspace{10pt} h^i = \partial^i h, \hspace{10pt}  \partial_i  = \frac{\partial}{\partial x^i}, \hspace{10pt} \partial^i  = \frac{\partial}{\partial p_i}.\] 
Then the Poisson bracket $[f, g]$ of smooth functions $f(x, p)$ and $g(x, p)$ on the phase space $(x, p)$ is:
\[ [f,  \hspace{3pt} g] =  \sum_i (f_i g^i -  f^i  g_i).\]
Here and in the following,  the summations range from $1$ to  $n$.  \\The Hamiltonian of interest is:
\[ H_\infty =   \sum_{i, j}\frac{ (x^i)^2}{a_i}  p_j^2.\]
Here each of the $a_i, i = 1, 2, \dots, n$ is a non-zero constant.   This is the Hamiltonian for the geodesic flow of the metric:
\[ G = \frac{ g(dx, dx)}{h(x, x)}.\]
Here $x \in \mathbb{V}$, a vector space of dimension $n$ and $g$ and $h$ are linearly independent symmetric bilinear forms on $\mathbb{V}$. 
Also we are assuming that $g$ and $h$ are generic, so simultaneously diagonalizable, so coordinates $x^i, i = 1,2, \dots, n$ exist for which we have:
\[ g(x, x) = \sum_i  (x^i)^2, \]
\[ h(x, x) = \sum_i \frac{(x^i)^2}{a_i}.\]
The constants $a_i, i = 1, 2, \dots, n$ are non-zero.  These constants are invariantly defined as the eigen-values of the endomorphism $h^{-1} g$, where we regard $g$ and $h$ as isomorphisms from $\mathbb{V}$ to its dual $\mathbb{V}^*$.  To recover the geodesic flow of our cosmology, we need to divide out by the Hamiltonian action of the Hamiltonian $p(x) = \sum_i p_i x^i$ and to impose the constraint $g(x, x) = 0$.   
\eject\noindent
We embed the Hamiltonian $H_\infty$ in a Hamiltonian system, $H(s)$, depending on a free parameter $s$:
\[ H(s) =   \sum_{i, j} \frac{ sa_j (x^i)^2p_j^2 + x^i x^j p_i p_j}{(1 + sa_i)(1 + sa_j)} =  K(s) L(s) +  M(s)^2, \]
\[ K(s) = \sum_i\frac{ (x^i)^2}{1 + sa_i}, \hspace{10pt} L(s) = \sum_i \frac{ sa_i p_i^2}{1 + s a_i}, \hspace{10pt} M(s) = \sum_i \frac{p_i x^i}{1 + sa_i}.\]
Note that the definition of the system $H(s)$ does not require that the $a_i$ be non-zero.  By inspection, we see that we have, provided all the $a_i$ are non-zero:
\[ H_\infty = \lim_{s \rightarrow \infty} sH(s).\]
Note also the relations:
\[ H(0) =  S^2, \hspace{10pt} S =  \sum_i  p_i x^i, \hspace{10pt}  H'(0) =  \sum_{i, j} (x_i^2 a_j p_j^2) - 2 S  \sum_i a_ip_i x^i.\]
In particular, we see that when $S = 0$ the function $H(s)$ smoothly interpolates between the Hamiltonian $ \sum_{i, j} a_i^{-1}x_i^2 p_j^2$ and its dual $ \sum_{i, j} a_i p_i^2  (x^j)^2$. 
By partial fractions, we have, under the assumption that the $a_i$ parameters are pairwise distinct:
\[ H(s) = \sum_{i} \frac{H_i}{1 + sa_i}, \]
\[ H_i = (x^i)^2\sum_j p_j^2 - \sum_{j \ne i} \frac{a_i}{a_i - a_j} (x^i p_j - x^j p_i)^2.\]
Check:
\[ \sum_i \frac{H_i}{1 + sa_i} =  K(s) \sum_j p_j^2 - \sum_{i, j, i \ne j} \frac{a_i}{(a_i - a_j)( 1+ sa_i)} (x^i p_j - x^j p_i)^2\]
\[ =  K(s) \sum_j p_j^2 -\frac{1}{2} \sum_{i, j, i \ne j} \frac{a_i( 1+ sa_j) - a_j(1 + sa_i)}{(a_i - a_j)( 1+ sa_i)(1 + sa_j)} (x^i p_j - x^j p_i)^2\]
\[ =  K(s) \sum_j p_j^2 -\frac{1}{2} \sum_{i, j } \frac{1}{( 1+ sa_i)(1 + sa_j)} (x^i p_j - x^j p_i)^2\] 
\[ = K(s) \sum_j p_j^2 -   \sum_{i, j } \frac{1}{( 1+ sa_i)(1 + sa_j)} (x^i)^2 (p_j)^2  +   \sum_{i, j } \frac{1}{( 1+ sa_i)(1 + sa_j)} (x^ip_i) (x^jp_j)\] 
\[ = K(s) \sum_j \left(p_j^2 -  \frac{p_j^2}{(1 + sa_j)}\right) + M(s)^2  = K(s)L(s) + M(s)^2 = H(s).\]
In terms of the Hamiltonians, $H_i$, we have the formulas for $H_\infty, H(0)$ and $H'(0)$, valid whenever the $a_i$ parameters are non-zero and pairwise distinct:
\[ H_\infty = \sum_i \frac{H_i}{a_i}, \hspace{10pt} H(0) = \sum_i H_i, \hspace{10pt} H'(0) =  - \sum_i a_i H_i. \]
Note that we have the commutator:
\[ [S, H(s)] = 0.\]
Also put: 
\[ X = K(0) = \sum_i (x^i)^2, \hspace{10pt} Y =   L(0) =   \sum_i a_i p_i^2.\]
Then we  have the subsidiary relations:
\[   [X, S] = 2X,   \hspace{10pt} \left[ X, \hspace{7pt} H(s)\right] =  4S K(s), \]
\[  [Y, S] = - 2Y, \hspace{10pt}\left[ Y, \hspace{7pt} H(s)\right] = -4S L(s).\]
So the quantities  $X$ and $Y$ are preserved along the flows of $H(s)$, provided that $S$ is fixed to be zero.
\begin{theorem}  The Hamiltonians $H(s)$ Poisson commute:
\[ [H(s),  \hspace{7pt} H(t)] = 0.\]
When the parameters $a_i$ are pairwise distinct,  $[H_i, H_j] =0$, for all $i$ and $j$.   \\
In particular, when the parameters $a_i$ are pairwise distinct and all non-zero, the Hamiltonian evolution given by $H_\infty$ is completely integrable: the Hamiltonians $H_i$ are independent on a Zariski open subset of the phase space.
\end{theorem}
\eject\noindent
\textbf{Proof}\\
We use the formula:  
\[ \hspace{-40pt} H(s) = K(s) L(s) +  M(s)^2, \hspace{10pt} K(s) = \sum_i\frac{ (x^i)^2}{1 + sa_i}, \hspace{10pt} L(s) = \sum_j \frac{ sa_j p_j^2}{1 + s a_j}, \hspace{10pt} M(s) = \sum_i \frac{p_i x^i}{1 + sa_i}.\]
We compute the commutators:
\[ [K(s), L(t)]  =   4\sum_j \frac{t a_j p_j x^j}{(1 + a_j s)(1 + a_jt)}, \hspace{10pt} [L(s), K(t)]  =   - 4\sum_j \frac{sa_j p_j x^j}{(1 + a_j s)(1 + a_jt)}, \]
\[ [K(s), M(t)]  =   2\sum_j \frac{( x^j)^2}{(1 + a_j s)(1 + a_jt)}, \hspace{10pt}  [M(s), K(t)]  =   - 2\sum_j \frac{( x^j)^2}{(1 + a_j s)(1 + a_jt)}, \]
\[ [L(s), M(t)] =  - 2\sum_{i} \frac{sa_i p_i^2 }{(1 + a_is)(1 + a_i t)}, \hspace{10pt}  [M(s), L(t)] =   2\sum_{i} \frac{ta_i p_i^2 }{(1 + a_is)(1 + a_j t)}. \]
This gives:
\[ [H(s), H(t)]  = [K(s) L(s) + M(s)^2, K(t) L(t)  + M(t)^2]\]
\[\hspace{-45pt} =   K(s)L(t)[L(s), K(t)]  + L(s) K(t)[K(s), L(t)]   + 2L(s)M(t)[K(s), M(t)] + 2M(s) L(t)[M(s), K(t)] \]
\[+ 2K(s)M(t)[L(s), M(t)] + 2 M(s)K(t)[M(s), L(t)]\]
The first two terms of the right-hand side are:
\[  K(s)L(t)[L(s), K(t)]  + L(s) K(t)[K(s), L(t)] \]
\[ = \sum_{i, j, k} \frac{- 4sta_j a_k (x^i)^2p_j^2p_k x^k}{(1 + a_k s)(1 + a_kt)}\left(\frac{1}{ (1 + sa_i)(1 + t a_j)} - \frac{1}{ (1 + ta_i)(1 + s a_j)} \right)\]
\[ \hspace{-40pt}= \sum_{i, j, k} \frac{- 4sta_j a_k (x^i)^2p_j^2p_k x^k}{(1 + sa_i)(1 + t a_j)(1 + ta_i)(1 + s a_j)(1 + a_k s)(1 + a_kt)}((1 + ta_i)(1 + s a_j) - (1 + sa_i)(1 + ta_j))\]
\[ = \sum_{i, j, k} \frac{4st  (a_i - a_j) a_j a_k (s - t)(x^i)^2p_j^2p_k x^k}{(1 + sa_i)(1 + t a_j)(1 + ta_i)(1 + s a_j)(1 + a_k s)(1 + a_kt)}.\]
\eject\noindent
The next two terms of the right-hand side are:
\[  2L(s)M(t)[K(s), M(t)] + 2M(s) L(t)[M(s), K(t)] \]
\[ =     \sum_{i, j, k}  \frac{4a_j  (x^i)^2p_j^2p_k x^k}{(1 + a_is)(1 + a_i t)}\left(\frac{s}{(1 + s a_j)(1 + ta_k)}  - \frac{t}{(1 + t a_j)(1 + sa_k)}\right)\]
\[ \hspace{-40pt}=     \sum_{i, j, k}  \frac{4a_j  (x^i)^2p_j^2p_k x^k}{(1 + sa_i)(1 + t a_j)(1 + ta_i)(1 + s a_j)(1 + a_k s)(1 + a_kt)}(s(1 + t a_j)(1 + sa_k) - t(1 + s a_j)(1 + ta_k))\]
\[ =     \sum_{i, j, k}  \frac{4a_j (1 + (s + t) a_k + st a_j a_k)(s - t) (x^i)^2p_j^2p_k x^k}{(1 + sa_i)(1 + t a_j)(1 + ta_i)(1 + s a_j)(1 + a_k s)(1 + a_kt)}\]
The last two terms of the right-hand side are:
\[  2K(s)M(t)[L(s), M(t)] + 2 M(s)K(t)[M(s), L(t)]\]
\[ = \sum_{i, j, k}  \frac{ -4a_j(x^i)^2p_j^2p_k x^k}{(1 + a_js)(1 + a_j t)}\left(  \frac{s}{(1 + sa_i)(1 + ta_k)} -  \frac{t}{(1 + ta_i)(1 + sa_k)}\right)\]
\[ \hspace{-40pt}= \sum_{i, j, k}  \frac{ -4a_j(x^i)^2p_j^2p_k x^k}{(1 + sa_i)(1 + t a_j)(1 + ta_i)(1 + s a_j)(1 + a_k s)(1 + a_kt)}(s(1 + ta_i)(1 + sa_k) - t(1 + sa_i)(1 + t a_k))\]
\[ = \sum_{i, j, k}  \frac{ -4a_j( 1 + (s + t)a_k  + st a_ia_k)(s - t)(x^i)^2p_j^2p_k x^k}{(1 + sa_i)(1 + t a_j)(1 + ta_i)(1 + s a_j)(1 + a_k s)(1 + a_kt)}.\]
So the last four terms of the right-hand side, put together, are:
\[  \hspace{-40pt}2L(s)M(t)[K(s), M(t)] + 2M(s) L(t)[M(s), K(t)] +  2K(s)M(t)[L(s), M(t)] + 2 M(s)K(t)[M(s), L(t)]\]
\[ \hspace{-57pt} = \sum_{i, j, k} \frac{4a_j (s - t)(x^i)^2p_j^2p_k x^k}{(1 + sa_i)(1 + t a_j)(1 + ta_i)(1 + s a_j)(1 + a_k s)(1 + a_kt)} (1 + (s + t) a_k + st a_j a_k - (1 + (s + t)a_k  + st a_ia_k))\]
\[  =  \sum_{i, j, k}\frac{4st(a_j - a_i)a_j a_k(s - t)(x^i)^2p_j^2p_k x^k}{(1 + sa_i)(1 + t a_j)(1 + ta_i)(1 + s a_j)(1 + a_k s)(1 + a_kt)}.\]
By inspection, this is the negative of the first two terms, $K(s)L(t)[L(s), K(t)]  + L(s) K(t)[K(s), L(t)]$, so we have the required relation:
\[ [H(s), H(t)]  = 0.\]
The rest of the theorem follows immediately, provided we can establish that the Hamiltonians $H_i$ are independent.  To this end, when the $a_i$ are pairwise distinct,  without loss of generality, at worst after a permutation of the phase space variables, we may assume that $a_1 \ne 0$.  Then  consider the specialization with $p_1 = 1$ and $p_i = 0, i = 2, 3, \dots, n$, in which case we have:
\[ K(s) = \sum_i\frac{ (x^i)^2}{1 + sa_i}, \hspace{10pt} L(s) =  \frac{sa_1}{1 + s a_1}, \hspace{10pt} M(s) = \frac{ x^1}{1 + sa_1}.\]
This gives:
\[ (1 + sa_1) H(s) = (x^1)^2 +  sa_1 \sum_{i = 2}^n \frac{ (x^i)^2}{1 + sa_i}.\]
 As $s$ varies, we see that the right-hand-side of this expression gives $n$-independent quantities, provided only that $x^1 x^2 \dots x^n \ne 0$.
 If also we first impose that $S = 0$, so here $x^1 = 0$, we reduce instead to:
 \[   \frac{(1 + sa_1) }{sa_1} H(s)  =  \sum_{i = 2}^n\frac{(x^i)^2}{(1 + sa_i)}.\]
 As $s$ varies, we see that the right-hand-side of this expression gives $(n - 1)$-independent quantities, provided only that $x^2 \dots x^n \ne 0$.
 Finally if also we impose that $\sum_j (x^j)^2  = 0$, we reduce to:
 \[  \frac{(1 + sa_1)(1 + sa_2)}{s^2a_1} H(s)  =         \sum_{i = 3}^n \frac{(x^i)^2(a_2 - a_i)}{1 + sa_i}.\]
  As $s$ varies, we see that the right-hand-side of this expression gives $(n - 2)$-independent quantities, provided only that $x^3 \dots x^n \ne 0$.   In all these cases, the condition that the Hamiltonians have fewer than the maximum number of degrees of freedom, is given by the vanishing of certain determinants (of appropriate minors of the Jacobian matrix, with respect to the phase space variables, of the Hamiltonians $H_i$), so by the vanishing of a system of polynomials in the phase space variables.  The special cases just  given above do have the maximum number of degrees of freedom, so the polynomial relations cannot be identities.   This proves that the Hamiltonians are independent on a Zariski open subset of the phase space and are still independent when we restrict to $S =0$ and still independent when we further restrict to $\sum_{i} (x^i)^2 = 0$ and we are done. Finally note that it would appear that this system is still integrable in the limit of a countably infinite number of variables, if appropriately formulated.
\eject\noindent
\subsection{Hyper-elliptic cosmologies in dimension four}
We now specialize to the case when the space-time is of dimension four.   Denote by $G$ the symmetric bilinear form on $\Omega^2(\mathbb{T})$, with values in $\mathbb{L} = \Omega^4(\mathbb{T})$, such that $G(X, X ) = X \wedge X$, for any $X \in \Omega^2(\mathbb{T})$.  By definition, the four dimensional hyper-elliptic cosmologies, are those Klein cosmologies, $\mathcal{K}_{Q}$, for which the metric, $g_Q$ takes the following form:
\[ g_Q  =  \frac{dX \wedge dX}{Q(X, X)}  = \frac{G(dX, dX)}{Q(X, X)}, \hspace{10pt} G(X, X) = X\wedge X = 0.\]
Here $X \in \Omega^2(\mathbb{T})$ and $Q$ is a non-zero symmetric bilinear form on $\Omega^2(\mathbb{T})$, with values in $\mathbb{L}$, such that $Q$ and $G$ are linearly independent tensors.    Note that the metric $g_Q$ only depends on $Q$ modulo multiples of $G$, so without loss of generality, if desired, we may take $Q$ to be a non-zero tensor, trace-free with respect to $G$:  $\textrm{tr}(G^{-1}Q) = 0$.  The space of symmetric bilinear forms on the six-dimensional vector space $\Omega^2(\mathbb{T})$ has dimension twenty-one,  so the space of metrics $g_Q$ has dimension  twenty.  Then the metric $g_Q$ is singular on the intersection of the quadrics: $Q(X, X) = G(X, X) = 0$.  Although to describe the simplest cosmological models one typically takes $Q$ to be of low rank, we will usually concentrate on the generic case, where the linear pencil of quadrics defined by $Q$ and $G$ has no members that that have more than one degeneracy: indeed we usually require that all the six roots of the characteristic polynomial $\det(xQ + yG)$ are unequal.  In particular, when a  co-ordinate description is desired this allows us to simultaneously diagonalize the quadrics $Q$ and $G$ and therefore all elements of the pencil that they define.  \\\\Naturally associated to the pair of quadrics $Q$ and $G$ is the curve:
\[ z^2 = \det(xQ + yG), \hspace{10pt} (x, y, z) \ne (0, 0, 0).\]
Here $x$ and $y$ lie in $\mathbb{C}$ and $z\in \mathbb{L}^6$.  The right-hand side  of this equation is an homogeneous polynomial of total degree six in the pair $(x, y)$ (where $x$ and $y$ are not both zero) and is also homogeneous of total degree six in the entries of the pair $(Q, G)$.  The resulting Riemann surface is generically an hyper-elliptic curve of genus two.  In the present context, since the quadric $G$ is preferred, this curve has the extra structure of a natural pair of base points corresponding to taking the variable $x$ to vanish.
\eject\noindent
The pencil defined by $Q$ and $G$ is the ensemble of quadrics $xQ + yG$, where $x$ and $y$ are complex numbers not both zero.  Geometrically, consider, for a quadric of the pencil,   so a symmetric bilinear form on $\mathbb{T}$, the space of  maximal isotropic subspaces of $\mathbb{T}$, defined by the chosen quadric.  Each maximal  isotropic subspace has dimension three.   The  space of all the isotropic subspaces also has dimension three and consists of  two disjoint copies of complex projective three-space when the quadric is non-singular and just one copy when the quadric has one degeneracy.   So as the pair $(x, y)$ varies, the ensemble of these spaces gives a two-fold cover of the Riemann sphere of the variables $(x,y)$, branched at the roots of the characteristic polynomial $\det(xQ + yG)$.   This gives the hyper-elliptic curve in question. By its construction it comes naturally equipped with 
with a tautological holomorphic bundle of complex three-dimensional projective spaces. 
\\\\
Now consider the intersection $F$ of the quadrics $Q$ and $G$: this constitutes the conformal infinity of the cosmology.  Note that the cosmology is only directly aware of $G$ and of the intersection $F$, not the specific quadric $Q$ that defines the intersection.   A basic fact is that the space of complex projective lines lying entirely in $F$ is a two-dimensional abelian variety and is exactly a model for the Jacobian of the hyper-elliptic curve of the pencil: the space of line bundles of Chern class zero over the curve, with addition given by the tensor product.
\\\\
The hypersurface $F$ of the Klein space-time where the metric of the cosmology is undefined, so where $Q = G = 0$, is the intersection of two quadrics in the five-dimensional projective space $\mathbb P(\Omega^2\mathbb T)$.  As such it has an intrinsic conformal metric, which is just the restriction to $F$ of the Klein conformal structure and a conformal second fundamental form (this is the ordinary second fundamental form with its trace removed).   The intrinsic conformal metric is non-degenerate provided the hypersurface is non-null.  A direct calculation shows that the points where the hypersurface is null are exactly those lying in the intersection $\Sigma$ of the three quadrics, $G(X, X) = 0$, $Q(X, X) = 0$ and $(QG^{-1} Q)(X, X) = 0$.  Also the conformal second fundamental form is just the restriction of the form $Q(dX, dX)$ to the hypersurface.
\eject\noindent
These algebraic varieties are of a type that is well-known in algebraic geometry [22, 45]:
\begin{itemize}
\item The three-dimensional variety $S$, the intersection of the Klein quadric with the quadric $ Q$, is a Fano manifold, which is by definition a complex manifold with an ample tangent bundle.  It carries a canonical family of  one-forms,  represented by the closed projective three-form  $\displaystyle{\frac{X\wedge dX \wedge dX \wedge dX}{G(X, X) Q(X, X)}}$, which is integrated over a two-torus surrounding the Fano manifold to produce the one-form (here we regard $X$ as lying in a six-dimensional vector space and the wedge refers to the exterior product in six dimensions).   The space of projective lines on $S$ is two dimensional and is a model of the Jacobian of the hyperelliptic curve.
\item The two-dimensional variety $\Sigma$, the totally null surface in $S$, is a $K3$ surface.  It carries a canonical closed two-form $\sigma$ represented by the closed projective five-form  $\displaystyle{\frac{X\wedge dX \wedge dX \wedge dX \wedge dX \wedge dX}{G(X, X) Q(X, X)(QG^{-1} Q)(X, X)}}$, which is integrated over a three-torus surrounding  the $K3$-surface to produce the two-form.   Then the two-form $\sigma$ renders the $K3$ surface $\Sigma$ a Calabi-Yau manifold.
\end{itemize}
A tangent plane to $Q$ at a point $s$ of $S = Q\cap G$ has polar $s'$ with respect to $G$, a point  in $\mathbb{P}(\Omega^2(\mathbb{T}))$.   The tangent plane is null if and only if the polar point, $s'$ lies on $G$.   The collection of points of $S$ with null tangent planes is the $K3$ manifold $\Sigma$.  .
Next consider the collection of lines that lie in $S$.   Each such line $l$ is a fortiori a line in $G$, so a null geodesic, so gives rise to a projective twistor $Z(l)$ and a dual projective twistor $W(l)$.     If we consider a general projective twistor $Z$, its $\alpha$-plane intersects $S$ in the intersection of the $\alpha$-plane and the quadric $Q$, so in a conic $Q_Z$.  Generically this is a nonsingular conic; then the space of projective twistors for which the conic $Q_Z$ is singular is a hypersurface, $\alpha_Q$ in $\mathbb{PT}$.  For a projective twistor $Z$, for which $Q_Z$ is singular, we obtain a pair of lines (possibly coincident) in $S$.  Then the ensemble of lines of  $S$ is a double cover of the space $\alpha_Q$.  It is branched at sixteen points, corresponding to those $Z$ for which $Q_Z$ is a repeated pair of lines.

If $Z$ is a projective twistor, it gives rise to an  $\alpha$-plane in $G$,  which intersects $Q$ in a conic, $Q_Z$.  This conic is degenerate for $Z$ lying in an hypersurface $\mathcal{H}_Q$ in $\mathbb{PT}$.   The hypersurface $\mathcal{H}_Q$ is then a Kummer surface.   The corresponding conics in $Q\cap G$ are pairs of lines (usually distinct).  The variety of lines in $Q\cap G$ is then a double cover of the twistor hypersurface, ramified over those points $Z$ where the conics are a repeated line.  There are sixteen such points, the sixteen singularities of the Kummer surface.   For a detailed treatment of the equations of the Kummer surface (in the Battaglini case), see the Appendix.  
\eject\noindent
\subsection{Calabi-Yau and   K\"{a}hler-Einstein metrics for the cosmologies}
Calabi-Yau metrics occur in at least two places:
\begin{itemize}\item The caustic surface of the conformal infinity of the hyper-elliptic cosmologies is a $K3$ surface which possesses a unique Calabi-Yau metric.   Under small perturbations, the Calabi-Yau metric will survive, but in general the perturbed surfaces are not algebraic.  
\item  The generalized Battaglini cosmologies discussed below are singular Calabi-Yau cosmologies and will become ordinary Calabi-Yau metrics after the resolution of their singularities.  These are not quadric cosmologies:  they are given by the intersection of a quartic $Q(X, X, X, X) = 0$ with the Klein quadric, so have metrics of the form:
\[   z^{-1} (dX\wedge dX), \hspace{10pt}  X\wedge X = 0, \hspace{10pt} z^2 = Q(X, X, X, X).\]
Thus the space-time is branched over its conformal infinity.
\end{itemize}
Deep modern analysis, due particularly to Simon Donaldson and his co-workers and to Robert Berman, is needed to understand the  Riemannian geometry of the Fano manifolds   constituting the conformal infinity of the hyper-elliptic cosmologies [6, 16, 17, 18].    These results,  finalized after an effort lasting forty years or more, involving many distinguished mathematicians, may be summarized for our purposes as follows: 
\begin{itemize}\item Not every Fano manfold is K\"{a}hler-Einstein.   There is an obstruction which needs to vanish.   A notion of $K$-polystability arose, due mainly to Tian, to organize this obstruction.  
\item Berman proved that if a Fano variety (possibly singular) admits a K\"{a}hler-Einstein metric, then it must be $K$-polystable.    
\item Donaldson and his co-authors proved that  if the variety is $K$-polystable, then it has a unique K\"{a}hler-Einstein metric.
\item  \emph{Further the condition of $K$-polystability is preserved under small deformations}.
\item 
More recently, Yuji  Odaka proved that smooth $K$-polystable Fano varieties form separated moduli algebraic space and moreover can have at worst only quotient singularities.  In particular, in our context, the space of quadric intersections has a Zariski open subspace, such that each quadric intersection has a unique K\"{a}hler-Einstein metric, provided only that  it can be proved that it is not empty.
\end{itemize}
The space is indeed not empty!  In fact Alan Nadel constructed an explicit example of an intersection of pair of quadrics in complex projective five-space, whose associated hyper-elliptic curve has the equation $y^2 = x^6 - z^6$, for which he was able to show that the obstruction vanished,  so it at least has a unique K\"{a}hler-Einstein metric [52].  See below for the details of the Nadel space-time.   It then follows from the above discussion that the generic quadric pair intersection in complex projective five-space carries a unique K\"{a}hler-Einstein metric.  Very fortunately, the high degree of symmetry of the Nadel  example (which he needed to carry out his existence proof) also guarantees that the Nadel space-time is a Battaglini cosmology (see below).     So we have the theorem:
\begin{theorem}
The generic quadric cosmology has its conformal infinity carrying a unique K\"{a}hler-Einstein metric.   \\The generic Battaglini cosmology has its conformal infinity carrying a unique K\"{a}hler-Einstein metric.   \\Small perturbations of either of these cosmologies  have their conformal infinity carrying a unique K\"{a}hler-Einstein metric.  
\end{theorem}
\noindent Notice that these perturbations need not be algebraic, \emph{nor need they keep the Weyl curvature zero}.  So they would persist even in non-conformally flat small perturbations of our cosmologies, so are apparently a general feature of cosmologies.  
\eject\noindent
\section{The Battaglini cosmologies}
The general hyper-elliptic cosmologies in four dimensions naturally fall into two broad categories: those for which the associated hyper-elliptic curve has an extra involution and those for which it does not.   We call those hyper-elliptic cosmologies with an extra involution the Battaglini cosmologies.   They are distinguished by the fact that the sextic which defines the hyper-elliptic curve of the cosmology becomes an even polynomial, at worst after  a projective transformation of its variables.  It is relatively straightforward to derive the condition on the curve that it be of Battaglini type, either by brute force, using Maple, where it comes down to determining a resultant of two complicated polynomials in one variable, one of degree twenty and the other of degree twelve.  Alternatively we can apply some results given in the masterwork of Igor Dolgachev and express the required condition as the product of fifteen polynomials, each homogeneous cubics in  the roots of the sextic.    The details are given in the appendix, together with the relation to the Igusa invariants.  We first briefly discuss the cosmologies arising in the limiting case where the two quadrics of the Battaglini construction become the same.  
\subsection{Battaglini cosmologies defined by a single quadric}
Let $A_{\alpha\beta} = A_{\beta\alpha}$ be a non-zero symmetric tensor, so $A$ represents the quadric with equation $A_{\alpha\beta}Z^\alpha Z^\beta =0$ in twistor space.  Up to equivalence under the general linear group of the twistor space, there are only four such quadrics, classified by the rank of the transformation from $\mathbb{T}$ to $\mathbb{T}^*$ represented by $A$.  Given $A_{\alpha\beta}$, put $X(A)_\alpha^\beta = A_{\alpha\gamma}X^{\beta\gamma}$ and $Q_A = \textrm{tr}(X(A)^2)$.   Then $Q_A$ vanishes identically if $A$ has rank one, leaving three cases:
\begin{itemize}\item 
When $A$ has rank two, we have $A_{\alpha\beta} = 2B_{(\alpha}C_{\beta)}$ for linearly independent dual twistors $B$ and $C$.  Then we get:
\[ X(A) = B_\alpha C_\gamma X^{\beta\gamma} + C_\alpha B_\gamma X^{\beta\gamma} , \]
\[ Q_A = 2 (B_\alpha  C_\beta X^{\alpha\beta})^2.\] 
So such quadrics give us the flat Klein cosmology, singular on the null cone of the projective line defined by the intersection of the planes where the planes represented by $B$ and $C$.
\item When $A$ has rank three, we have $A_{\alpha\beta} = 2B_{(\alpha}C_{\beta)} + D_\alpha D_\beta$ for linearly independent dual twistors $B$, $C$ and $D$.  Then we get:
\[ X(A) = B_\alpha C_\gamma X^{\beta\gamma} + C_\alpha B_\gamma X^{\beta\gamma} + D_\alpha D_\gamma X^{\beta\gamma} , \]
\[ Q_A = 2 (B_\alpha  C_\beta X^{\alpha\beta})^2 - 2B_\alpha C_\beta X^{\beta\gamma} X^{\alpha\delta} D_\gamma D_\delta.\] 
In appropriate co-ordinates,  $(t, x, y, z)$, the metric is:
\[   \frac{dt^2 - dx^2 + dy^2 - dz^2}{b(a + (t + z)(x + y))}.\]
Here $a$ and $b$ are non-zero constants.\\\\
The Einstein tensor is $\displaystyle{\frac{4aduds  -  (s du - u ds)^2}{2(a + us)^2}}$, where $u = t + z$ and $s = x + y$.    As an endomorphism, the Einstein tensor has square zero and rank two.   Geometrically, there is preferred completely null two-plane given by $t + z = x + y =0$, corresponding to the twistor that is the vertex of the quadric cone given by $A_{\alpha\beta}$.     Also the metric is singular when $ (t + z)(x + y) = -a$, a null hypersurface.  There is a preferred conic in the space-time.
\item When the quadric has rank four.    In suitable coordinates the metric can be taken to be the spherically symmetric metric:
\[ g = \frac{dt^2 - dx^2 - dy^2 - dz^2}{x^2 + y^2 + z^2}.\]
The Einstein tensor $G$ is:
\[ g + \frac{2}{(x^2+ y^2 + z^2)^2}((ydz - zdy)^2 + (zdx - xdz)^2 + (xdy - ydx)^2).\]
The Einstein endomorphism has minimal polynomial $\lambda^2 -1$.
\end{itemize}
\eject\noindent
\subsection{The Battaglini space-times of a pair of quadrics in twistor space}
The general hyper-elliptic cosmology possessing an extra involution is given by a beautiful construction due to Guiseppe Battaglini [3, 4, 5].   Fix a pair of quadrics in twistor space, $\mathbb{T}$, so generically these correspond to a quartet of conformal geodesics in space-time, two for each quadric.  Also these quadrics generically  intersect in an elliptic curve.    Now consider a point of the Klein quadric, so a complex projective line in the twistor space.  This line generically intersects each quadric in a pair of points.   Now impose the condition that each pair of points is harmonically separated with respect to the other pair.   This determines a three dimensional family of lines: an hyper-surface in the Klein quadric. 
\begin{theorem} The completion of the space of such lines is the intersection with the Klein quadric of a quadric in the projective space of $\Omega^2(\mathbb{T})$ and the quadric pencil formed by this quadric together with the Klein quadric gives an hyper-elliptic curve of genus two with an extra involution.   Further all such quadric pencils and all hyper-elliptic curves with an extra involution are obtained in this way.   The corresponding quadric cosmologies with  the quadric intersection as their conformal infinities are called the Battaglini cosmologies.   
\end{theorem}
If the two quadrics are represented by symmetric  tensors $A_{\alpha\beta}$ and $B_{\alpha\beta}$,  then the equation of the Battaglini quadric, denoted $AB(X, X)$ in complex projective five-space is
\[ A_{\alpha \gamma}B_{\beta\delta} X^{\beta \gamma} X^{\alpha\delta} = 0, \hspace{10pt} X^{\alpha\beta} = - X^{\beta\alpha}.\]
\noindent 
Note that the Battaglini quadric  depends on the specific quadrics, not just on the pencil that they define.   Also note that we can repeat this construction, dropping the condition that the cross-ratio of the four intersection points be $-1$ and replacing it by the requirement that the cross-ratio be a fixed complex number.   We call the resulting cosmologies generalized Battaglini cosmologies.  
\begin{theorem} The generalized Battaglini cosmologies are the intersection with the Klein quadric of a singular quartic hypersurface.
\end{theorem}
\noindent \textbf{Proof}  We just calculate the condition directly,   giving the relation:
\[   AB(X, X)^2 = t AA(X, X) BB(X, X).\]
Here $AB(X, X)$ is the usual Battaglini quadric, defined by quadrics $A$ and $B$ in twistor space, whereas $AA(X, X)$ and $BB(X, X)$ are the Battaglini quadrics, defined by putting $B$ equal to $A$ and $A$ equal to $B$, respectively.  Also $t - 1$ is the cross-ratio in question, so $t$ vanishes in the standard Battaglini case.

The intersection $E$ of the two quadrics $A$ and $B$ in $\mathbb{PT}$ is an elliptic curve.   Denote by $C$ the elliptic curve with equation $z^2 = \det(xA + yB)$.    Geometrically $C$ is the space of pairs consisting of a quadric of the pencil defined by $A$ and $B$, together with a choice of one family of lines lying on that quadric.   Define a map $\rho:C\to\op{Aut}(X)$ from $C$ into the automorphism group of $E$ as follows.  Fix a point  $c\in C$, so a quadric and a choice of one of the (at most) two families of lines on the quadric.  Now let $x \in E$ be given.  Then the point $x$ lies on a unique line $\ell_x$ in this family.  Then $\ell_x$ intersects $E$ in another point, denoted $\rho(c)(x)$.  Then $\rho(c)$ is an involution for all $c\in C$, so the image of $\rho$ does not lie in the connected component of the identity.  Moreover, $\rho$ is one-to-one, so it gives $C$ the structure of an $\op{Aut}_0(X)$-torsor.


The tangent map $\tau:E\to\mathbb K$ associates to each point $x$ of $E$ the line tangent to $E$ at $x$ gives a mapping from $E$ into the Grassmannian $\op{Gr}(2,\mathbb T)$ which is identified with the Klein quadric $\mathbb K$ under the Pl\"ucker embedding.

Any quadric $\mathbb Q$ in $\mathbb{PT}$ is ruled by two families of lines.  A point $X\in\mathbb K\subset\mathbb P(\Omega^2\mathbb T)$ determines a line on the quadric $q$ if the endomorphism $X.q: \mathbb T\to\mathbb T$ is nilpotent of order two.  On $\mathbb K$, this quadratic condition is reducible, having two irreducible components $\alpha(\mathbb Q),\beta(\mathbb Q)\subset\mathbb K$ consisting of the $\alpha$ and $\beta$ lines on $\mathbb Q$.  Each of $\alpha(\mathbb Q)$ and $\beta(\mathbb Q)$ is the intersection with $\mathbb K$ of an eigenspace of the Hodge duality endomorphism of $\Omega^2\mathbb T$ associated to the quadratic form $q$ (projectively a $2$-plane).  In other words, each of the two families of lines on the quadric defines a conic in $\mathbb K$.  At each point $x$ of $\mathbb Q$, the tangent $2$-plane to $\mathbb Q$ at $x$ determines a dual twistor.  So each point of $\mathbb Q$ gives both a twistor and an incident dual twistor: thus a null geodesic.  This null geodesic contains one line from each of the two families of lines ruling the conic, and so connects a point of $\alpha(\mathbb Q)$ and a point of $\beta(\mathbb Q)$.  Conversely, each point of $\alpha(\mathbb Q)\times\beta(\mathbb Q)$ determines a null geodesic: every point of $\alpha(\mathbb Q)$ is null related to every point of $\beta(\mathbb Q)$.  Each such null geodesic determines a unique twistor of the quadric $\mathbb Q$ itself.  The correspondence $\alpha(\mathbb Q)\times\beta(\mathbb Q)\to\mathbb Q$ is an isomorphism.

Now suppose we are in the situation of two quadrics $\mathbb Q_0$ and $\mathbb Q_1$ in $\mathbb{PT}$.  Let $\alpha_0:\mathbb Q_0\to\mathbb K$ be the map that associates to each $x\in\mathbb Q_0$ the $\alpha$-line through $x$, and $\beta_0:\mathbb Q_0\to\mathbb K$ to map that associates the $\beta$-line through $x$.  Also define $\alpha_1,\beta_1:\mathbb Q_1\to\mathbb K$.  
Each point $x$ of their intersection $E$ determines a pair of tangent planes $T_x\mathbb Q_0$ and $T_x\mathbb Q_1$ to each of the two quadrics.  So $(x,T_x\mathbb Q_0)_{\mathbb K}$ is a null geodesic in $\mathbb K$ that contains the three points $\tau(x)_{\mathbb K}$, $\alpha_0(x)$, and $\beta_0(x)$.  Similarly, $(x,T_x\mathbb Q_1)_{\mathbb K}$ is a null geodesic that contains the points $\tau(x)_{\mathbb K}$, $\alpha_1(x)$, and $\beta_1(x)$.

\begin{theorem}
For $Z\in E$, the null geodesics $(Z,T_Z\mathbb Q_i)_{\mathbb K}$, $i=0,1$, lie on the Fano variety associated to the Battaglini quadric in $\mathbb P(\Omega^2\mathbb T)$.
\end{theorem}

\begin{proof}
The points of $(Z,T_Z\mathbb Q_i)_{\mathbb K}$ are of the form $Z\wedge Z'$ where $Z'\in\mathbb T$ is such that $q_i(Z,Z')=0$.  So $(q_0\varowedge q_1)(Z\wedge Z',Z\wedge Z') = 0$ as well.
\end{proof}
\subsection{The cosmology of Alan Nadel}
If $q_0(Z)$ and $q_1(Z)$ are generic quadrics in projective twistor space, with $q_0$ non-degenerate, then there exist co-ordinates  $(Z^1, Z^2, Z^3, Z^4) \in \mathbb{C}^4$, for the points $Z$ of twistor space, such that the quadrics take the diagonal form:
\[ q_0(Z) = (Z^1)^2+ (Z^2)^2+ (Z^3)^2+ (Z^4)^2, \]
\[ q_1(Z) = a_1(Z^1)^2+a_2(Z^2)^2+a_3(Z^3)^2+a_4(Z^4)^2.\]
Here the parameters $(a_1, a_2, a_3, a_4) \in \mathbb{C}^4$ are pairwise distinct and give the branch points $x:y$ of the elliptic curve:
\[ z^2 = \det(xq_0 - y q_1) = \det(q_0)(x^4 - s_1 x^3 y + s_2 x^2y^2 - s_3xy^3 + s_4 y^4).\]
The $j$-invariant of the elliptic curve is determined   by the cross ratio $ \displaystyle{\alpha = \frac{(a_1 - a_3)(a_2 - a_4)}{(a_1 - a_4)(a_2 - a_3)}}$ and is invariant under permutations of the roots (see Appendix 6.4).    This elliptic curve is isomorphic to the curve of intersection of the two quadrics.   It is shown in the Appendix 6.3 that the Battaglini quadric,  $Q(X)$ in $\mathbb P(\Omega^2\mathbb T)$ associated to the pair $(q_0,q_1)$ has the equation, $Q(X) = 0$, where we have:
\[ \hspace{-63pt} Q(X) = (a_1+a_2)(X^{12})^2+(a_1+a_3)(X^{13})^2+(a_1+a_4)(X^{14})^2+(a_3+a_4)(X^{34})^2+(a_4+a_2)(X^{42})^2+(a_2+a_3)(X^{23})^2.\]
Here we are using standard co-ordinates  $X^{ij} = -X^{ji}$ to represent a point $X$  for $\Omega^2(\mathbb{T})$, for which the Klein quadric takes the customary form:
\[ G(X, X) = 2(X^{12} X^{34} + X^{13}X^{42} + X^{14}X^{23}) = 0.\]
Then the characteristic polynomial of endomorphism $G^{-1}Q$ is:
\[ (t^2 - \alpha)(t^2 - \beta) (t^2 - \gamma)  = 0, \]
\[ (\alpha, \beta, \gamma)  = ((a_1 + a_2)(a_3 + a_4), \hspace{7pt} (a_1 + a_3)(a_4 + a_2), \hspace{7pt} \gamma = (a_1 + a_4)(a_2 + a_3)).    \]
Note that this characteristic polynomial can be written directly in terms of the coefficients of the characteristic polynomial $\det(x - y q_0^{-1}q_1)$, given above and and so directly in   terms of traces of powers of the endomorphism $q_0^{-1} q_1$:
\[  (t^2 - \alpha)(t^2 - \beta) (t^2 - \gamma)  = t^6-2s_2t^4- ( 4s_4 - s_3s_1 - s_2^2)t^2+ s_4s_1^2 + s_3^2  - s_3s_2s_1.\]
The fact that this polynomial only depends on the square of the variable $t$ shows that the hyper-elliptic curve of the quadric pencil defined by $G$ and $Q$ has an extra involution.  Note that the elliptic curve $y^2 =  (u - \alpha)(u - \beta) (u - \gamma)  $ has the same $j$-invariant as does the intersection $q_0 \cap q_1$.

The Nadel cosmology is the special case:
\[ q_1 = \omega(Z^1)^2+(Z^3)^2+\omega^2(Z^4)^2,   \hspace{10pt} \omega^2 + \omega + 1 = 0, \hspace{10pt} \omega^3 = 1.\]
So here we have $(a_1, a_2, a_3, a_4) = (\omega, 0, 1, \omega^2)$ and the cross-ratio is:
\[  \frac{(a_1 - a_3)(a_2 - a_4)}{(a_1 - a_4)(a_2 - a_3)}  = \frac{(\omega - 1)(0 - \omega^2)}{(\omega - \omega^2)(0 - 1)} =  - \omega. \]
Also we have: $(\alpha, \beta, \gamma)  = (- \omega^2, \hspace{7pt} - \omega, \hspace{7pt}  -1)$, so the characteristic polynomial of the endomorphism $q_0^{-1} q_1$ is $t^6 + 1$.
So in this case the elliptic curve of intersection of the two quadrics in twister space corresponds to the equianharmonic cross-ratio, so has $j$ invariant zero.  This means that the elliptic curve corresponds to the quotient of $\mathbb{C} = \mathbb{R}^2$ by a regular hexagonal lattice.  The Battaglini quadric is then:
\[ Q(X) =  \omega (X^{12})^2  - (X^{34})^2 +  \omega^2(X^{42})^2 - (X^{13})^2) + (X^{23})^2 - (X^{14})^2.\]  
\[ X^{12}X^{34} + X^{13} X^{42}+ X^{14} X^{23} = 0.\] 
Put $X^{12} = P + iQ$, $X^{34} = iP  + Q$,  $X^{13} = R + iS$, $X^{42} = iR + S$,  $X^{14} = T + iU$, $X^{23} = iT  + U$.  
Then we have:
\[ P^2 + Q^2 + R^2 + S^2 + T^2 + U^2 = 0, \]
\[ \omega(P^2 - Q^2) + \omega^2(R^2 - S^2) + T^2 - U^2 = 0.\]
These equations define precisely the Fano projective three-manifold studied by Nadel.  So we have the pivotal theorem of Nadel [52]:
\begin{theorem}  The Nadel cosmology has for its conformal infinity the Battaglini manifold associated to the intersection of quadrics in complex projective three-space, whose elliptic curve of intersection has $j$-invariant zero.  The conformal infinity is a Fano manifold which has a unique K\"{a}hler-Einstein metric. 
\end{theorem}
\eject\noindent

\subsection{Battaglini cosmologies}
If $A,B\in \op{Sym}^2\mathbb T^*$ be two non-zero elements of $\op{Sym}^2\mathbb T^*$.

The intersection $E$ of the two quadrics in $\mathbb{PT}$ is an elliptic curve.  Define a map $\rho:C\to\op{Aut}(X)$ from $C$ into the automorphism group of $E$ as follows.  Each $c\in C$ determines a quadric and a choice of one of the two families of lines on the quadric.  So the point $x$ lies on a unique line $\ell$ in this family.  Then $\ell$ intersects $c$ in another point, denoted $\rho(c)x$.  Then $\rho(c)$ is an involution for all $c\in C$, so the image of $\rho$ does not lie in the connected component of the identity.  Moreover, $\rho$ is one-to-one, so it gives $C$ the structure of an $\op{Aut}_0(X)$-torsor.


The tangent map $\tau:E\to\mathbb K$ associates to each point $x$ of $E$ the line tangent to $E$ at $x$ gives a mapping from $E$ into the Grassmannian $\op{Gr}(2,\mathbb T)$ which is identified with the Klein quadric $\mathbb K$ under the Pl\"ucker embedding.

Any quadric $\mathbb Q$ in $\mathbb{PT}$ is ruled by two families of lines.  A point $X\in\mathbb K\subset\mathbb P(\Omega^2\mathbb T)$ determines a line on the quadric $q$ if the endomorphism $X.q: \mathbb T\to\mathbb T$ is nilpotent of order two.  On $\mathbb K$, this quadratic condition is reducible, having two irreducible components $\alpha(\mathbb Q),\beta(\mathbb Q)\subset\mathbb K$ consisting of the $\alpha$ and $\beta$ lines on $\mathbb Q$.  Each of $\alpha(\mathbb Q)$ and $\beta(\mathbb Q)$ is the intersection with $\mathbb K$ of an eigenspace of the Hodge duality endomorphism of $\Omega^2\mathbb T$ associated to the quadratic form $q$ (projectively a $2$-plane).  In other words, each of the two families of lines on the quadric defines a conic in $\mathbb K$.  At each point $x$ of $\mathbb Q$, the tangent $2$-plane to $\mathbb Q$ at $x$ determines a dual twistor.  So each point of $\mathbb Q$ gives both a twistor and an incident dual twistor: thus a null geodesic.  This null geodesic contains one line from each of the two families of lines ruling the conic, and so connects a point of $\alpha(\mathbb Q)$ and a point of $\beta(\mathbb Q)$.  Conversely, each point of $\alpha(\mathbb Q)\times\beta(\mathbb Q)$ determines a null geodesic: every point of $\alpha(\mathbb Q)$ is null related to every point of $\beta(\mathbb Q)$.  Each such null geodesic determines a unique twistor of the quadric $\mathbb Q$ itself.  The correspondence $\alpha(\mathbb Q)\times\beta(\mathbb Q)\to\mathbb Q$ is an isomorphism.

Now suppose we are in the situation of two quadrics $\mathbb Q_0$ and $\mathbb Q_1$ in $\mathbb{PT}$.  Let $\alpha_0:\mathbb Q_0\to\mathbb K$ be the map that associates to each $x\in\mathbb Q_0$ the $\alpha$-line through $x$, and $\beta_0:\mathbb Q_0\to\mathbb K$ to map that associates the $\beta$-line through $x$.  Also define $\alpha_1,\beta_1:\mathbb Q_1\to\mathbb K$.  
Each point $x$ of their intersection $E$ determines a pair of tangent planes $T_x\mathbb Q_0$ and $T_x\mathbb Q_1$ to each of the two quadrics.  So $(x,T_x\mathbb Q_0)_{\mathbb K}$ is a null geodesic in $\mathbb K$ that contains the three points $\tau(x)_{\mathbb K}$, $\alpha_0(x)$, and $\beta_0(x)$.  Similarly, $(x,T_x\mathbb Q_1)_{\mathbb K}$ is a null geodesic that contains the points $\tau(x)_{\mathbb K}$, $\alpha_1(x)$, and $\beta_1(x)$.

\begin{theorem}
For $Z\in E$, the null geodesics $(Z,T_Z\mathbb Q_i)_{\mathbb K}$, $i=0,1$, lie on the Fano variety associated to the Battaglini quadric in $\mathbb P(\Omega^2\mathbb T)$.
\end{theorem}

\begin{proof}
The points of $(Z,T_Z\mathbb Q_i)_{\mathbb K}$ are of the form $Z\wedge Z'$ where $Z'\in\mathbb T$ is such that $q_i(Z,Z')=0$.  So $(q_0\varowedge q_1)(Z\wedge Z',Z\wedge Z') = 0$ as well.
\end{proof}

\subsection{Outlook}
The work here could lead in many directions.   We briefly mention some of these.
\begin{itemize} \item The gauge theory of the $Q$-curvature needs to be finalized.
\item We have discussed one aspect of the "multi-verse"  in connection with the osculating space-times.   Another aspect is the possibility of a space of universes.  For example in connection with one-dimensional cosmologies (in the simplest case these are by definition curves equipped with a projective structure), it is often natural to consider these cosmologies as Legendrian sub-manifolds of a contact three-manifold.  The ambient contact three manifold here forms the framework in which the various possible universes live.
\item There is apparently  a natural hierarchy of universes, wherein we start with one universe, go to its conformal infinity, go to the latter's caustic hypersurface, where the conformal infinity goes null, then go to the part of the caustic hypersurface which acquires a second null direction and so on.   This seems worthy of exploration. 
\end{itemize}

\eject\noindent
\section{Appendix: Battaglini invariants}
\subsection{The Battaglini condition and its relation to the invariants of Jun-Ichi Igusa}
Let $x_1$, $x_2$, $x_3$, $x_4$, $x_5$ and $x_6$ be indeterminants.  By definition, the Battaglini cubics are the following fifteen cubics, each labeled by a partition of the numbers $1, 2, 3, 4, 5, 6$ into three disjoint (unordered) subsets of two (unordered) elements:
\[ \hspace{-108pt} F_{\{(12), (35), (46)\}} = x_2 x_6 x_4 - x_5 x_2 x_3  - x_2 x_1 x_6 + x_5 x_1 x_2 - x_2 x_1 x_4 + x_3 x_1 x_2 - x_3 x_6 x_4 + x_4 x_1 x_6 + x_4 x_3 x_5 - x_5 x_1 x_3 + x_5 x_6 x_3 - x_4 x_6 x_5, \]
\[ \hspace{-108pt} F_{\{(13), (26), (45)\}} =  x_2 x_6 x_4 - x_4 x_2 x_5 - x_2 x_1 x_6 + x_5 x_6 x_2 + x_3 x_1 x_2 - x_3 x_6 x_2 + x_4 x_1 x_5 - x_4 x_1 x_3 + x_4 x_3 x_5 -x_5 x_1 x_3 - x_4 x_6 x_5 + x_3 x_1 x_6,\]
\[ \hspace{-108pt} F_{\{(14), (26), (35)\}} =   x_2 x_1 x_4 - x_5 x_2 x_3 - x_2 x_6 x_4 - x_2 x_1x_6 + x_5 x_6 x_2 + x_3 x_6 x_2 - x_4 x_1 x_5 - x_4 x_1 x_3 + x_4 x_3 x_5 + x_5 x_1 x_3 - x_5 x_6 x_3 + x_4 x_1 x_6, \]
\[ \hspace{-108pt} F_{\{(15), (26), (34)\}} =  x_2 x_6 x_4 - x_4 x_2 x_3 - x_2 x_1 x_6 + x_5 x_1 x_2 - x_5 x_6 x_2 + x_3 x_6 x_2 - x_4 x_1 x_5 + x_4 x_1 x_3 + x_4 x_3 x_5 - x_5 x_1 x_3 - x_3 x_6 x_4 + x_5 x_1 x_6, \]
\[ \hspace{-108pt} F_{\{(13), (24), (56)\}} =   x_4 x_2 x_5 - x_4 x_2 x_3 - x_2 x_1 x_4 - x_5 x_6 x_2 + x_3 x_1 x_2 + x_2 x_6 x_4 + x_5 x_1 x_6 + x_4 x_1 x_3 - x_4 x_6 x_5 - x_5 x_1 x_3 - x_3 x_1 x_6 + x_5 x_6 x_3, \]
\[ \hspace{-108pt} F_{\{(13), (25), (46)\}} = x_4 x_2 x_5 - x_2 x_6 x_4 - x_5 x_1 x_2 + x_5 x_6 x_2 + x_3 x_1 x_2 - x_5 x_2 x_3 - x_4 x_1 x_3 - x_4 x_6 x_5 + x_5 x_1 x_3 - x_3 x_1 x_6 + x_3  x_6  x_4 + x_4 x_1 x_6, \]
\[ \hspace{-108pt} F_{\{(14), (25), (36)\}} =  x_2 x_1 x_4 - x_4 x_2 x_5 - x_5 x_1 x_2 + x_5 x_6 x_2 - x_3 x_6 x_2 + x_5 x_2 x_3 - x_4 x_1 x_3 - x_4 x_1x_6 -x_5 x_6 x_3 + x_3 x_6 x_4 + x_3 x_1 x_6 + x_4 x_1 x_5, \]
\[  \hspace{-108pt} F_{\{(15), (24), (36)\}} =   x_4 x_2 x_3 -  x_4 x_2 x_5  + x_5 x_1 x_2 - x_2 x_1 x_4 - x_3 x_6 x_2 + x_2 x_6 x_4 - x_5 x_1 x_6 - x_5 x_1 x_3 + x_3 x_1 x_6 +  x_4 x_1 x_5 + x_5 x_6 x_3 - x_3 x_6 x_4, \]
\[  \hspace{-108pt} F_{\{(14), (23), (56)\}} =  x_2 x_1 x_4 - x_4 x_2 x_3 - x_5 x_6 x_2 - x_3 x_1 x_2 + x_3 x_6 x_2 + x_5 x_2 x_3 + x_5 x_1 x_6 + x_4 x_1 x_3 - x_4 x_1 x_6 + x_4 x_6 x_5 - x_5 x_6 x_3 - x_4 x_1 x_5, \]
\[  \hspace{-108pt} F_{\{(15), (23), (46)\}} = x_4 x_2 x_3 - x_2 x_6 x_4 + x_5 x_1 x_2 - x_3 x_1 x_2 + x_3 x_6 x_2 - x_5 x_2 x_3 + x_4 x_6 x_5 - x_5 x_1 x_6 + x_4 x_1 x_6 + x_5 x_1 x_3 - x_3 x_6 x_4 - x_4 x_1 x_5, \]
\[  \hspace{-108pt} F_{\{(16), (23), (45)\}} =  x_4 x_2 x_3 - x_4 x_2 x_5 - x_3 x_1 x_2 - x_3 x_6 x_2 + x_5 x_2 x_3 + x_2 x_1 x_6 - x_5 x_1 x_6 - x_4 x_1 x_6 + x_4 x_6 x_5 + x_3 x_1 x_6 + x_4 x_1 x_5 - x_4 x_3 x_5, \]
\[  \hspace{-108pt} F_{\{(16), (25), (34)\}} =  x_4 x_2 x_5 - x_4 x_2 x_3  - x_5 x_1 x_2 - x_5 x_6 x_2 +  x_5 x_2 x_3 + x_2 x_1 x_6 + x_4 x_1 x_3 + x_3 x_6 x_4 + x_5 x_1 x_6 - x_3 x_1 x_6 - x_4 x_1 x_6 - x_4 x_3 x_5, \]
\[  \hspace{-108pt} F_{\{(16), (24), (35)\}}  = x_4 x_2 x_3  - x_2 x_6 x_4 + x_4 x_2 x_5 -  x_2 x_1 x_4 - x_5 x_2 x_3 + x_2 x_1 x_6 - x_5 x_1 x_6 + x_4 x_1 x_6 + x_5 x_1 x_3 - x_3 x_1 x_6 - x_4 x_3 x_5 + x_5 x_6 x_3, \]
\[  \hspace{-108pt} F_{\{(12), (34), (56)\}}  = x_2 x_1 x_4  - x_4 x_2 x_3 - x_2 x_1 x_6 - x_5 x_1 x_2 + x_5 x_6 x_2 + x_3 x_1 x_2 - x_4 x_1 x_3 + x_3 x_6 x_4 - x_4 x_6 x_5 + x_5 x_1 x_6 - x_5 x_6 x_3 + x_4 x_3 x_5,\]
\[ \hspace{-108pt} F_{\{(12), (36), (45)\}} = - x_5 x_1 x_2 + x_4 x_2 x_5 + x_2 x_1 x_6 - x_2  x_1 x_4 + x_3 x_1 x_2 - x_3 x_6 x_2 + x_3 x_6 x_4 - x_4 x_6 x_5 + x_5 x_6 x_3 - x_4 x_3 x_5 - x_3 x_1 x_6 + x_4 x_1x_5.\]
The function $F_{\{(ab),(cd),(ef)\}}$ vanishes identically when $x_a = - x_b, x_c = -x_d$ and $x_e = -x_f$, as required.  In fact,  given $\{(ab), (cd), (ef)\}$, the other fourteen functions in the above list that are not $F_{\{(ab),(cd),(ef)\}}$ do not vanish identically when $x_a = - x_b, x_c = -x_d$ and $x_e = -x_f$, as is easily checked.  Note that these fifteen functions only depend on the differences $x_i - x_j$.  Also each function is invariant up to a scale under the action of the six-element symmetric group  $\mathbb{S}_3$  (acting simultaneously on each variable), generated by the transformations $x \rightarrow 1- x$ of order two  and $x\rightarrow 1 - x^{-1}$ of order three. 
\eject\noindent
Each of the cubics is non-singular except at  ten points, regarded as cubics in the projective four-space  represented by the difference of coordinates: for example the coordinates $x_2 - x_1, x_3 - x_1, x_4 - x_1, x_5 - x_1 $ and $x_6 - x_1$.   So each cubic is an example of a cubic primal introduced by Beniamino Segre (which, in particular,  parametrizes the space of Kummer surfaces).  Following Igor Dolgachev, we may write each of the fifteen functions explicitly as a sum of two products of three differences [22]:
\[ F_{\{(ab), (cd), (ef)\}} =  (x_a - x_d)(x_b - x_e)(x_c - x_f)  + (x_a - x_f)(x_b - x_c)(x_d - x_e).\]
Then in suitable co-ordinates $(z_1, z_2, z_3, z_4, z_5, z_6)$ for complex projective  five-space, on the four-plane where $z_1 + z_2 + z_3 + z_4 + z_5 + z_6 = 0$,   we may write the equation of one of these cubics as $z_1 z_2 z_3 + z_4 z_5 z_6 = 0$.   In these co-ordinates, one singular point is $a(1,  1, 1, -1, -1, -1)$, with $a \ne 0$ and then the others have exactly  one of the three co-ordinates $(z_1, z_2, z_3)$ a non-zero number $b$, say, and then exactly one of the three co-ordinates  $(z_4, z_5, z_6)$ non-zero, the number $-b$, giving the remaining nine singular points.\\\\
The product $F$ of all the fifteen Battaglini cubics is a symmetric function, homogeneous of degree forty-five in the variables, $(x_1, x_2, x_3, x_4, x_5, x_6)$ and has five hundred and eight thousand,  three hundred and ninety-two terms, when written out.  Then $F$ can be re-written as a polynomial $F(s_1, s_2, s_3, s_4, s_5, s_6)$  of the coefficients of the homogeneous polynomial $f(t, u)$, whose roots are $(x_1, x_2, x_3, x_4, x_5, x_6)$:
\[ \hspace{-77pt} f(t, u) = (t - ux_1)(t - ux_2)(t - ux_3)(t - ux_4)(t - ux_5)(t - ux_6) = t^6 - s_1 t^5u + s_2 t^4u^2 - s_3 t^3u^3 + s_4 t^2 u^4 - s_5 tu^5 + s_6u^6.\]
The polynomial $F(s_1, s_2, s_3, s_4, s_5, s_6)$ has one thousand three hundred and seventy terms.  When $f(t, u)$ is the characteristic polynomial of the quadric pencil the vanishing of $F$ is the condition that the quadric pencil be of the Battaglini type.\\\\
The standard invariant characterizing Battaglini pencils is perhaps that given in the paper of Tanush Shaska and Helmut V\"{o}lklein and is:
\[ G = 4743360J_{10}J_{4}^3J_{2}J_{6}-870912J_{10}J_{4}^2J_{2}^3J_{6}+507384000J_{10}^2J_{4}^2J_{2}+8748J_{10}J_{2}^4J_{6}^2\]
\[ -19245600J_{10}^2J_{4}J_{2}^3-81J_{2}^3J_{6}^4+384J_{4}^6J_{6}+41472J_{10}J_{4}^5+159J_{4}^6J_{2}^3\]
\[ +104976000J_{10}^2J_{2}^2J_{6}+6048J_{4}^4J_{2}J_{6}^2+12J_{2}^6J_{4}^3J_{6}+29376J_{2}^2J_{4}^2J_{6}^3-J_{2}^7J_{4}^4\]
\[ -9331200J_{10}J_{4}^2J_{6}^2-54J_{2}^5J_{4}^2J_{6}^2+108J_{2}^4J_{4}J_{6}^3-1728J_{4}^5J_{2}^2J_{6}\]
\[ -592272J_{10}J_{4}^4J_{2}^2+31104J_{6}^5-2099520000J_{10}^2J_{4}J_{6}-8910J_{2}^3J_{4}^3J_{6}^2\]
\[ +972J_{10}J_{2}^6J_{4}^2+77436J_{10}J_{4}^3J_{2}^4-47952J_{2}J_{4}J_{6}^4+3090960J_{10}J_{4}J_{2}^2J_{6}^2\]
\[ -5832J_{10}J_{2}^5J_{4}J_{6}-80J_{4}^7J_{2}-3499200J_{10}J_{2}J_{6}^3+1332J_{2}^4J_{4}^4J_{6}-125971200000J_{10}^3\]
\[ -236196J_{10}^2J_{2}^5-6912J_{4}^3J_{6}^3-78J_{2}^5J_{4}^5.\]
Here $J_2$, $J_4$, $J_6$ and $J_{10}$ are the  Igusa invariants, of weights $6, 12, 18$ and $30$, respectively [44].   Explicitly we have:
\[  J_2 = -240s_6+40s_5s_1-16s_4s_2+6s_3^2,\]
\[ J_4 = 1620s_6^2+12s_6(-45s_5s_1-42s_4s_2+25s_4s_1^2+27s_3^2-15s_3s_2s_1+4s_2^3)\]
\[ \hspace{-5pt} +20s_5^2(15s_2-4s_1^2)+4s_5(-45s_4s_3+s_4s_2s_1+9s_3^2s_1-3s_3s_2^2)+48s_4^3+4s_4^2(-3s_3s_1+s_2^2),\]
\[ J_6 = -119880s_6^3\]
\[ \hspace{-10pt} +6s_6^2(9990s_5s_1+4s_4(861s_2-775s_1^2)-1674s_3^2+5s_3(102s_2s_1+75s_1^3) -16s_2^3-150s_2^2s_1^2)\]
\[\hspace{-50pt}   +s_6(-40s_5^2(465s_2+56s_1^2) +s_5(4s_4(765s_3+868s_2s_1+400s_1^3)+1818s_3^2s_1-4s_3s_2(219s_2+465s_1^2)+616s_2^3s_1)\]
\[ \hspace{-88pt} -96s_4^3-876s_4^2s_3s_1+424s_4^2s_2^2-640s_4^2s_2s_1^2-468s_4s_3^2s_2+330s_4s_3^2s_1^2+492s_4s_3s_2^2s_1  -160s_4s_2^4+162s_3^4+60s_3^2s_2^3-198s_3^3s_2s_1)\]
\[+s_5^3(2250s_3+1600s_2s_1-320s_1^3)\]
\[ +s_5^2(-900s_4^2-4s_4(465s_3s_1+160s_2^2-16s_2s_1^2)+22s_3^2(15s_2+8s_1^2)+26s_3s_2^2s_1-36s_2^4)\]
\[ +s_5(616s_4^3s_1+2s_4^2(246s_3s_2+13s_3s_1^2+14s_2^2s_1)+2s_4(-99s_3^3-119s_3^2s_2s_1+38s_3s_2^3)+72s_3^4s_1-24s_3^3s_2^2)\]
\[ -4s_4^4(40s_2+9s_1^2)+4s_4^3(15s_3^2+19s_3s_2s_1-6s_2^3)-8s_4^2s_3^2(3s_3s_1-s_2^2).\]
\[ J_{10} =  -46656s_6^5\]
\[ \hspace{-97pt} -s_6^4(-38880s_5s_1+(32400s_1^2-62208s_2)s_4-34992s_3^2+(77760s_2s_1-27000s_1^3)s_3-43200s_2^2s_1^2-3125s_1^6+22500s_2s_1^4+13824s_2^3)\]
\[\hspace{-63pt}  -((-540s_1^2+32400s_2)s_5^2+((1800s_1^3-31968s_2s_1+77760s_3)s_4-15552s_3^2s_1+(-46656s_2^2-2250s_1^4+31320s_2s_1^2)s_3\]
\[ \hspace{-30pt} +2500s_1^5s_2-15600s_2^2s_1^3+21888s_2^3s_1)s_5 +13824s_4^3+(17280s_2^2-1500s_1^4+6480s_2s_1^2-46656s_3s_1)s_4^2\]
\[ \hspace{-33pt} +((27540s_1^2-3888s_2)s_3^2+(3456s_2^2s_1+3750s_1^5-19800s_2s_1^3)s_3+10560s_2^3s_1^2-2000s_2^2s_1^4-9216s_2^4)s_4\]
\[ \hspace{-40pt} +8748s_3^4+(1350s_1^3-21384s_2s_1)s_3^3+(-2250s_2s_1^4+8640s_2^3+9720s_2^2s_1^2)s_3^2+(-6912s_2^4s_1+1600s_2^3s_1^3)s_3\]
\[ -256s_2^5s_1^2+1024s_2^6)s_6^3-((-27000s_3-410s_1^3+1800s_2s_1)s_5^3\]
\[ +(-43200s_4^2+(31320s_3s_1+6480s_2^2-8748s_2s_1^2+1700s_1^4)s_4+(-15417s_1^2+27540s_2)s_3^2\]
\[ +(-16632s_2^2s_1-2000s_1^5+12330s_2s_1^3)s_3-248s_2^3s_1^2+192s_2^4+50s_2^2s_1^4)s_5^2\]
\[ +(21888s_4^3s_1+((-16632s_1^2+3456s_2)s_3-2250s_1^5-15264s_2^2s_1+13040s_2s_1^3)s_4^2\]
\[ \hspace{-38pt} +(-21384s_3^3+(22896s_2s_1-1980s_1^3)s_3^2+(-10152s_2^2s_1^2+2050s_2s_1^4+5760s_2^3)s_3-160s_2^3s_1^3+640s_2^4s_1)s_4\]
\[ \hspace{-25pt} +6318s_3^4s_1+(900s_1^4-5832s_2^2-3942s_2s_1^2)s_3^3+(4464s_2^3s_1-1020s_2^2s_1^3)s_3^2+(192s_2^4s_1^2-768s_2^5)s_3)s_5\]
\[\hspace{-15pt}  +(192s_1^2-9216s_2)s_4^4+(8640s_3^2+(5760s_2s_1+120s_1^3)s_3-4816s_2^2s_1^2+4352s_2^3+900s_2s_1^4)s_4^3\]
\[  \hspace{-25pt} +(-5832s_1s_3^3+(4536s_2s_1^2-8208s_2^2-825s_1^4)s_3^2+(-560s_2^2s_1^3+2496s_2^3s_1)s_3-512s_2^5+128s_2^4s_1^2)s_4^2\]
\[ \hspace{-40pt}  +((4860s_2-162s_1^2)s_3^4+(-2808s_2^2s_1+630s_2s_1^3)s_3^3+(576s_2^4-144s_2^3s_1^2)s_3^2)s_4-729s_3^6+(486s_2s_1-108s_1^3)s_3^5\]
\[ +(-108s_2^3+27s_2^2s_1^2)s_3^4)s_6^2-((-320s_1^4+22500s_4+1700s_2s_1^2-2250s_3s_1-1500s_2^2)s_5^4\]
\[ \hspace{-18pt}+(-15600s_1s_4^2+((-19800s_2+12330s_1^2)s_3+1600s_1^5-9768s_2s_1^3+13040s_2^2s_1)s_4+1350s_3^3\]
\[ +(-1980s_2s_1+208s_1^3)s_3^2+(120s_2^3-160s_2s_1^4+682s_2^2s_1^2)s_3+36s_2^3s_1^3-144s_2^4s_1)s_5^3\]
\[ \hspace{-25pt}+((10560s_2-248s_1^2)s_4^3+(9720s_3^2+(682s_1^3-10152s_2s_1)s_3-1020s_2s_1^4+5428s_2^2s_1^2-4816s_2^3)s_4^2\]
\[ \hspace{-15pt}+(-3942s_1s_3^3+(4536s_2^2-560s_1^4+2412s_2s_1^2)s_3^2+(746s_2^2s_1^3-3272s_2^3s_1)s_3-144s_2^4s_1^2+576s_2^5)s_4\]
\[ -162s_2s_3^4+(-24s_2s_1^3+108s_2^2s_1)s_3^3+(6s_2^3s_1^2-24s_2^4)s_3^2)s_5^2\]
\[ \hspace{-28pt}+((-144s_1^3+640s_2s_1-6912s_3)s_4^4+(4464s_3^2s_1+(630s_1^4+2496s_2^2-3272s_2s_1^2)s_3+96s_2^3s_1-24s_2^2s_1^3)s_4^3\]
\[ +((108s_1^2-2808s_2)s_3^3+(-356s_2s_1^3+1584s_2^2s_1)s_3^2+(-320s_2^4+80s_2^3s_1^2)s_3)s_4^2\]
\[  \hspace{-45pt}+(486s_3^5+(72s_1^3-324s_2s_1)s_3^4+(72s_2^3-18s_2^2s_1^2)s_3^3)s_4)s_5+1024s_4^6+(-108s_1^4-512s_2^2-768s_3s_1+576s_2s_1^2)s_4^5\]
\[ +((576s_2-24s_1^2)s_3^2+(72s_2s_1^3-320s_2^2s_1)s_3-16s_2^3s_1^2+64s_2^4)s_4^4\]
\[ +(-108s_3^4+(-16s_1^3+72s_2s_1)s_3^3+(-16s_2^3+4s_2^2s_1^2)s_3^2)s_4^3)s_6+3125s_5^6\]
\[ -(2500s_1s_4+(-2000s_1^2+3750s_2)s_3+1600s_2s_1^3-256s_1^5-2250s_2^2s_1)s_5^5\]
\[ \hspace{-25pt} -((-2000s_2+50s_1^2)s_4^2+(-2250s_3^2+(-160s_1^3+2050s_2s_1)s_3-1020s_2^2s_1^2+900s_2^3+192s_2s_1^4)s_4\]
\[  \hspace{-10pt}+900s_1s_3^3+(-825s_2^2-560s_2s_1^2+128s_1^4)s_3^2+(-144s_2^2s_1^3+630s_2^3s_1)s_3+27s_2^4s_1^2-108s_2^5)s_5^4\]
\[ \hspace{-32pt} -((36s_1^3-160s_2s_1+1600s_3)s_4^3+(-1020s_3^2s_1+(-560s_2^2+746s_2s_1^2-144s_1^4)s_3+6s_2^2s_1^3-24s_2^3s_1)s_4^2\]
\[  \hspace{-28pt}+((-24s_1^2+630s_2)s_3^3+(-356s_2^2s_1+80s_2s_1^3)s_3^2+(72s_2^4-18s_2^3s_1^2)s_3)s_4-108s_3^5+(-16s_1^3+72s_2s_1)s_3^4\]
\[ +(-16s_2^3+4s_2^2s_1^2)s_3^3)s_5^3-(-256s_4^5+(-144s_2s_1^2+27s_1^4+192s_3s_1+128s_2^2)s_4^4\]
\[ +((6s_1^2-144s_2)s_3^2+(80s_2^2s_1-18s_2s_1^3)s_3-16s_2^4+4s_2^3s_1^2)s_4^3\]
\[ +(27s_3^4+(-18s_2s_1+4s_1^3)s_3^3+(-s_2^2s_1^2+4s_2^3)s_3^2)s_4^2)s_5^2.\]
The invariant $G$ is of weight $90$, so the following algebraic identity is perhaps not too surprising:
\[ 2^{11} 3^9 F^2 + G =0.\]
So the invariant $F$ is more economical.  Also this formula shows that for the case when all roots are real, that $G$ is non-positive and is zero only in the Battaglini case.  Note that the invariant $F$ cannot belong to the polynomial ring of Igusa, which is generated by $J_2$, $J_4$, $J_6$ and $J_{10}$, because any such polynomial (or even rational function) in the Igusa variables  would  have weight an integral multiple of six.  Note that the last equation shows that the space of invariants $(F, J_2, J_4, J_6, J_{10})$  is a four-dimensional variety, in a weighted projective space with five variables, equipped with a natural involution.  
\\\\
Finally we write out the Battaglini polynomial explicitly:
\[ F = b_0 + b_1 s_6 + b_2 s_6^2 + b_3 s_6^3 + b_4 s_6^4 + b_5 s_6^5 + b_6 s^6.\]
The polynomial $ F$ has total weight $45$, where $s_i $ is assigned weight $i$, for each $i = 1, 2, 3, 4, 5$ and $6$.   Here the coefficients $b_i$ are polynomials in  the variables $(s_1, s_2, s_3, s_4, s_5)$ of total weights $45 - 6i$, for $i = 0, 1, 2, 3, 4, 5, 6$.
\begin{itemize}\item  The coefficient of $s_6^6$ is:
\end{itemize} 
\[ b_6 =  15625(5s_1^2-12s_2)^3(5s_1^3 - 18s_1s_2+27s_3).\]
\begin{itemize}\item  The coefficient of $s_6^5$ is:
\end{itemize} 
\[  b_5 = -95040000s_1s_2^5s_4-5962500s_1^4s_2^4s_3+53156250s_1^4s_2s_3^3+729000000s_3s_4^3\]
\[ -162000000s_2^5s_5+189843750s_1^4s_3^2s_5-468750s_1^6s_2^3s_3+223560000s_2^4s_3s_4\]
\[ +113906250s_1^5s_3^2s_4+27000000s_1^2s_2^5s_3-162000000s_1s_2^3s_4^2+253125000s_1^4s_2s_4s_5\]
\[ -607500000s_1^2s_2^2s_4s_5+1093500000s_1^2s_2s_3s_4^2-911250000s_1^2s_2s_3^2s_5-17625000s_1^5s_2^3s_4\]
\[ -24300000s_1s_2^4s_3^2-425250000s_1^3s_2s_3^2s_4-337500000s_1^5s_2s_3s_5+1012500000s_1^3s_2^2s_3s_5\]
\[ -972000000s_1s_2^3s_3s_5+546750000s_1^2s_3^3s_4+2343750s_1^7s_2^2s_4-11250000s_1^6s_2s_3s_4-364500000s_1s_2^2s_3^2s_4\]
\[ +75000s_1^5s_2^5+720000s_1^3s_2^6-91125000s_1^3s_3^4-43740000s_2^3s_3^3-2343750s_1^7s_4^2+135000000s_1^3s_4^3\]
\[ +5184000s_2^6s_3-16875000s_1^6s_3^3-3456000s_1s_2^7-3375000s_1^4s_2^2s_3s_4+113400000s_1^2s_2^3s_3s_4\]
\[ +23625000s_1^5s_2^2s_3^2  -89100000s_1^3s_2^3s_3^2+145800000s_1^2s_2^2s_3^3+33750000s_1^5s_2s_4^2\]
\[ +60300000s_1^3s_2^4s_4-40500000s_1^3s_2^2s_4^2  -486000000s_1s_2s_4^3+105468750s_1^6s_2^2s_5-331875000s_1^4s_2^3s_5\]
\[ +418500000s_1^2s_2^4s_5-11718750s_1^8s_2s_5+35156250s_1^7s_3s_5 -1093500000s_1s_3^2s_4^2-278437500s_1^4s_3s_4^2\]
\[ +1093500000s_2^2s_3^2s_5+486000000s_2^3s_4s_5-35156250s_1^6s_4s_5.\]
\eject\noindent
\begin{itemize}\item  The coefficient of $s_6^4$ is:
\end{itemize} 
\[ b_4 = 262500s_1^4s_2^5s_3s_4-19440000s_1s_2^3s_3^3s_5-32812500s_1^6s_2s_3s_5^2+98550000s_1^3s_2^2s_3^2s_4^2\]
\[ -1162500s_1^5s_2^3s_3^2s_4-17212500s_1^4s_2^2s_3^3s_4+13320000s_1^3s_2^4s_3^2s_4-459000000s_1^2s_2s_4^3s_5\]
\[ -10125000s_1^3s_2^2s_3^3s_5-3234375s_1^6s_2^2s_3^2s_5+238140000s_1s_2s_3^2s_4^3-648000000s_1^2s_2^3s_3s_5^2\]
\[ -253800000s_1^2s_2^2s_3s_4^3+168750s_1^5s_2s_3^2s_4^2+91125000s_1^3s_3^3s_4s_5+278437500s_1^4s_2^2s_3s_5^2\]
\[ -30780000s_1^2s_2^5s_4s_5+364500000s_2s_3^2s_4^2s_5-238140000s_2^3s_3^2s_4s_5-3210000s_1^3s_2^5s_3s_5\]
\[ +7149600s_1s_2^5s_3^2s_4-7593750s_1^5s_2s_3^3s_5+13250000s_1^4s_2^4s_4s_5-2218750s_1^6s_2^3s_4s_5\]
\[ -33750000s_1^4s_2^2s_4^2s_5-729000000s_1^2s_3^2s_4^2s_5-8496000s_1s_2^6s_3s_5-21870000s_1s_2^2s_3^4s_4\]
\[ -54675000s_1^2s_2s_3^4s_5-151875000s_1^3s_2s_3^2s_5^2-4912500s_1^4s_2^3s_3s_4^2+607500000s_1s_2s_4^2s_5^2\]
\[ +178200s_1^2s_2^5s_3^3+729000000s_1s_2^2s_3^2s_5^2-29531250s_1^5s_3s_4^2s_5+33750000s_1^2s_2^4s_3^2s_5\]
\[ +1406250s_1^6s_3^2s_4s_5+1734375s_1^6s_2^2s_3s_4^2+56250000s_1^5s_2s_4s_5^2-9562500s_1^4s_2s_3s_4^3\]
\[ -3523200s_1^2s_2^6s_3s_4-91125000s_1^2s_2s_3^3s_4^2+72900000s_1^2s_2^3s_4^2s_5-34020000s_1^2s_2^4s_3s_4^2\]
\[ -270000000s_1^3s_2^2s_4s_5^2+459000000s_1s_2^3s_4s_5^2+42187500s_1^4s_3s_4s_5^2+1743750s_1^5s_2^4s_3s_5\]
\[ -23490000s_1^2s_2^3s_3^3s_4-135000000s_2^3s_5^3+43740000s_3^3s_4^3-486000000s_4^4s_5+1152000s_2^8s_5\]
\[ +162000000s_1s_4^5-16125000s_1^5s_4^4+390000s_1^3s_2^6s_3^2-1040625s_1^4s_2^4s_3^3+40320s_1s_2^7s_3^2\]
\[ +6561000s_1^2s_2^2s_3^5+1518750s_1^5s_2^2s_3^4-76800s_1^2s_2^8s_3+337920s_1s_2^8s_4-16000s_1^3s_2^7s_4\]
\[ -1944000s_1s_2^4s_3^4-1215000s_1^3s_2^3s_3^4+2250000s_1^4s_2^3s_3^2s_5+1718750s_1^6s_2s_4^2s_5\]
\[ +79380000s_1s_2^3s_3^2s_4^2+12150000s_1^3s_2s_3^4s_4+1125000s_1^6s_2s_3^3s_4+250560s_2^6s_3^3\]
\[ -9216s_2^9s_3-874800s_2^3s_3^5+6144s_1s_2^{10}-928125s_1^6s_3^5+1215000s_1^4s_2s_3^5-124000s_1^3s_2^5s_4^2\]
\[ -1875000s_1^7s_3^2s_4^2-65610000s_1s_3^4s_4^2+8343750s_1^6s_3s_4^3+18984375s_1^4s_3^4s_5\]
\[ +162000000s_2^2s_4^3s_5+195750000s_1^3s_3^2s_4^3-271350000s_1^2s_3s_4^4+2812500s_1^7s_3^3s_5\]
\[ -10546875s_1^5s_3^2s_5^2-125000s_1^5s_2^4s_4^2+58500000s_1^3s_2s_4^4+91500000s_1^3s_2^4s_5^2\]
\[ -156250s_1^7s_2s_4^3+9765625s_1^6s_5^3-1093500000s_2^2s_3s_4s_5^2+911250000s_1s_3^2s_4s_5^2\]
\[ +168750000s_1^2s_2^2s_5^3-70312500s_1^4s_2s_5^3-168750000s_1^3s_4^2s_5^2-911250000s_3s_4^2s_5^2\]
\[ +206550000s_1^2s_2^2s_3^2s_4s_5-80156250s_1^4s_2s_3^2s_4s_5+18562500s_1^5s_2^2s_3s_4s_5\]
\[ +45360000s_1s_2^4s_3s_4s_5-54000000s_1^3s_2^3s_3s_4s_5-468750s_1^7s_2s_3s_4s_5+249750000s_1^3s_2s_3s_4^2s_5\]
\[ +972000000s_1s_3s_4^3s_5-5467500s_1^3s_3^6-1082880s_2^7s_3s_4-223560000s_2s_3s_4^4+271350000s_2^4s_3s_5^2\]
\[ +31680000s_1s_2^4s_4^3-7452000s_2^5s_3^2s_5+13100000s_1^3s_2^3s_4^3+65610000s_2^2s_3^4s_5\]
\[ +5859375s_1^7s_2^2s_5^2+95040000s_1s_2^2s_4^4-3906250s_1^7s_4s_5^2-95040000s_2^4s_4^2s_5\]
\[ +390625s_1^8s_4^2s_5-546750000s_2s_3^3s_5^2-3025000s_1^5s_2^2s_4^3-245000s_1^4s_2^6s_5-42812500s_1^5s_2^3s_5^2\]
\[ +28224000s_2^6s_4s_5+32805000s_1^2s_3^5s_4-45900000s_1s_2^5s_5^2+95625000s_1^4s_4^3s_5-24105600s_2^5s_3s_4^2\]
\[ +9720000s_2^4s_3^3s_4+6662400s_1s_2^6s_4^2+432000s_1^2s_2^7s_5+9112500s_1^5s_3^4s_4-62015625s_1^4s_3^3s_4^2.
\]
\begin{itemize}\item  The coefficient of $s_6^3$ is:
\end{itemize} 
\[  b_3 = 484000s_1s_2^6s_4s_5^2-6875000s_1^6s_3s_4^2s_5^2+1500000s_1^5s_2^2s_4^2s_5^2+91125000s_2^2s_3^3s_4s_5^2\]
\[ -72900000s_1s_2^2s_4^3s_5^2+20047500s_1s_2^2s_3^4s_5^2+3950000s_1^4s_2^3s_4^3s_5+5900000s_1^3s_2^3s_4^2s_5^2\]
\[ -14040000s_1^2s_2^3s_3^3s_5^2+270000000s_1^2s_2s_4^2s_5^3+648000000s_1^2s_3s_4^3s_5^2-6260000s_1^2s_2^4s_4^3s_5\]
\[ +3125000s_1^6s_2s_4s_5^3+1012500s_1^3s_2s_3^4s_5^2-1250000s_1^7s_3^2s_4s_5^2-42187500s_1^2s_2s_3s_5^4\]
\[ +23625000s_1^4s_3^3s_4s_5^2+425250000s_2s_3^2s_4s_5^3+91500000s_1^2s_2^3s_4s_5^3-39375000s_1^4s_2s_3^2s_5^3\]
\[ -113400000s_2s_3s_4^3s_5^2-24840000s_2^5s_3s_4s_5^2+3733000s_1^2s_2^6s_3s_5^2+109350000s_1s_2^4s_3s_5^3\]
\[ -91500000s_1^3s_2s_4^3s_5^2+30937500s_1^4s_3^2s_4^3s_5+19440000s_1s_3^3s_4^3s_5+905760s_2^6s_3^2s_4s_5\]
\[ -79380000s_2^2s_3^2s_4^3s_5+13500000s_1^4s_2^2s_3^3s_5^2+10880s_1^2s_2^8s_4s_5-2530000s_1^3s_2^5s_4s_5^2\]
\[ +151875000s_1^2s_3^2s_4s_5^3-312500s_1^7s_2s_4^2s_5^2+25000s_1^5s_2^3s_4^4+116640s_1^2s_2^2s_3^7-5120s_1s_2^9s_4^2\]
\[ +60928s_2^8s_3s_4^2+24105600s_2^2s_3s_4^5-4788000s_1s_2^4s_3^2s_4^3-162000s_1^3s_2^2s_3^4s_4^2-871560s_1^2s_2^3s_3^5s_4\]
\[ +16464000s_1^2s_2^3s_3s_4^4-102500s_1^4s_2^2s_3s_4^4+137500s_1^5s_2^2s_3^2s_4^3+15000s_1^4s_2^4s_3s_4^3\]
\[ -287500s_1^6s_2^2s_4^3s_5-109350000s_1^3s_3s_4^4s_5-155250000s_1^3s_3^2s_4^2s_5^2+129276s_2^4s_3^5s_4\]
\[ -1836800s_1s_2^5s_4^4+5332500s_1^3s_3^4s_4^3-31104s_2^7s_3^3s_4-29375000s_1^4s_2^2s_4s_5^3-607500s_1^3s_3^5s_4s_5\]
\[ +142500s_1^4s_2^3s_3^3s_4^2+52160s_1^2s_2^7s_3s_4^2-165200s_1^3s_2^5s_3^2s_4^2-4301100s_1^2s_2s_3^5s_4^2-2048s_2^{11}s_5\]
\[ -5184000s_3s_4^6+91125000s_3^4s_5^3-2740000s_2^6s_5^3+742500000s_4^3s_5^3-16110000s_1^3s_2s_3^2s_4^4\]
\[ +1220400s_1^2s_2^4s_3^3s_4^2-680832s_2^5s_3^3s_4^2+2232s_1s_2^7s_3^4+1787500s_1^5s_2^3s_3^2s_5^2\]
\[ +1012500s_1^6s_3^4s_4s_5+54675000s_1s_3^4s_4s_5^2-61750000s_1^3s_2^3s_3s_5^3+874800s_3^5s_4^3\]
\[ +2740000s_1^3s_4^6-256s_2^9s_3^3-5832s_2^3s_3^7-175000s_1^7s_4^5-109350s_1^3s_3^8+3996s_2^6s_3^5\]
\[ -1012500000s_1s_3s_4^2s_5^3+1882500s_1^4s_3s_4^5-1053000s_1^4s_3^5s_4^2-5600s_1^3s_2^6s_4^3-318750s_1^5s_3^2s_4^4\]
\[ -5510400s_1s_2^3s_4^5+59985000s_1^2s_2^4s_3s_4s_5^2-29700000s_1^3s_2^2s_3^2s_4s_5^2+1375000s_1^6s_2^2s_3s_4s_5^2\]
\[ +1142500s_1^4s_2^4s_3^2s_4s_5-5490000s_1^2s_2^5s_3^2s_4s_5-1687500s_1^5s_2^2s_3^3s_4s_5+7380000s_1^3s_2^3s_3^3s_4s_5\]
\[ -1890000s_1s_2^4s_3^3s_4s_5-6075000s_1^4s_2s_3^4s_4s_5+12150000s_1^2s_2^2s_3^4s_4s_5-10700000s_1^5s_2s_3s_4^3s_5\]
\[ -11400000s_1^3s_2^2s_3s_4^3s_5+1304320s_1s_2^7s_3s_4s_5-226800s_1^3s_2^6s_3s_4s_5-17125000s_1^4s_2^3s_3s_4s_5^2\]
\[ -2700000s_1^3s_2s_3^3s_4^2s_5+8662500s_1^4s_2^2s_3^2s_4^2s_5-23220000s_1^2s_2^3s_3^2s_4^2s_5-81540000s_1s_2^3s_3^2s_4s_5^2\]
\[ -45360000s_1s_2s_3s_4^4s_5+10125000s_1^5s_2s_3^2s_4s_5^2-249750000s_1s_2^2s_3s_4s_5^3+81540000s_1^2s_2s_3^2s_4^3s_5\]
\[ -206550000s_1s_2s_3^2s_4^2s_5^2+44437500s_1^4s_2s_3s_4^2s_5^2-850000s_1^5s_2^3s_3s_4^2s_5+440000s_1^3s_2^4s_3s_4^2s_5\]
\[ +5212800s_1s_2^5s_3s_4^2s_5+362500s_1^6s_2s_3^2s_4^2s_5+1024s_2^{10}s_3s_4-8505000s_1^2s_3^3s_4^4-225000s_1^6s_3^3s_4^3\]
\[ -1312200s_1s_3^6s_4^2-6264s_1^2s_2^5s_3^5-9720000s_2s_3^3s_4^4-37422s_1s_2^4s_3^6+1545000s_1^5s_2s_4^5+656100s_1^2s_3^7s_4\]
\[ +7452000s_1s_3^2s_4^5+1035904s_2^6s_3s_4^3+9234000s_1s_2s_3^4s_4^3-437400s_1s_2^2s_3^6s_4+155250000s_1^2s_2^2s_3^2s_5^3\]
\[ +1187500s_1^7s_3s_4^3s_5+378000s_1s_2^5s_3^2s_5^2-9234000s_2^3s_3^4s_4s_5-1093500s_1^2s_2s_3^6s_5-1984500s_1^3s_2^2s_3^5s_5\]
\[ -48600s_1s_2^3s_3^5s_5-8662500s_1^3s_2^4s_3^2s_5^2+252000s_1^2s_2^6s_4^2s_5+637500s_1^5s_2^4s_4s_5^2-927500s_1^4s_2^5s_3s_5^2\]
\[-20047500s_1^2s_3^4s_4^2s_5+21870000s_2s_3^4s_4^2s_5+1417500s_1^2s_2^4s_3^4s_5+47340000s_1^2s_2^2s_4^4s_5-1500000s_1^6s_2s_3^3s_5^2\]
\[ +284472s_1s_2^5s_3^4s_4-494112s_1s_2^6s_3^2s_4^2-4444000s_1^3s_2^3s_3^2s_4^3-8064s_1s_2^8s_3^2s_4+3870000s_1^4s_2s_3^3s_4^3\]
\[ -27000s_1^4s_2^2s_3^5s_4+9000s_1^3s_2^4s_3^4s_4+50000s_1^6s_2s_3s_4^4+2052000s_1^2s_2^2s_3^3s_4^3-14364000s_1s_2^2s_3^2s_4^4\]
\[ +15720s_1^2s_2^6s_3^3s_4+5400s_1^3s_2^3s_3^6-6188000s_1^3s_2^2s_4^5-190976s_1s_2^7s_4^3-412000s_1^3s_2^4s_4^4\]
\[ +1570240s_1^2s_2^5s_3s_4^3+1215000s_1^4s_3^6s_5+134000s_1^3s_2^7s_5^2-5062500s_1^5s_3^4s_5^2+8505000s_2^4s_3^3s_5^2\]
\[ -32805000s_2s_3^5s_5^2-1498880s_2^7s_4^2s_5-103680s_2^5s_3^4s_5+45900000s_1^2s_4^5s_5+9937500s_1^4s_2^4s_5^3\]
\[ -145800000s_3^3s_4^2s_5^2+28032s_2^8s_3^2s_5-14300000s_1^2s_2^5s_5^3+95040000s_2s_4^5s_5-31680000s_2^3s_4^4s_5\]
\[ -95625000s_1s_2^3s_5^4-418500000s_1s_4^4s_5^2+379687500s_3s_4s_5^4-7812500s_1^5s_2s_5^4+12812500s_1^5s_4^3s_5^2\]
\[ +70312500s_1^3s_4s_5^4+40500000s_2^2s_4^2s_5^3+51562500s_1^3s_2^2s_5^4-1562500s_1^6s_2^3s_5^3+6250000s_1^6s_3^2s_5^3\]
\[ -187500s_1^6s_4^4s_5-476800s_1s_2^8s_5^2-436800s_2^7s_3s_5^2+1312200s_2^2s_3^6s_5-58500000s_2^4s_4s_5^3+5510400s_2^5s_4^3s_5\]
\[ +24300000s_3^2s_4^4s_5-106496s_2^9s_4s_5+278437500s_2^2s_3s_5^4-195750000s_2^3s_3^2s_5^3-51562500s_1^4s_4^2s_5^3\]
\[ -189843750s_1s_3^2s_5^4-28224000s_1s_2s_4^6-247500s_1^4s_2^3s_3^4s_5+54000s_1^4s_2^5s_4^2s_5-47340000s_1s_2^4s_4^2s_5^2\]
\[ -253125000s_1s_2s_4s_5^4+4687500s_1^5s_3s_4s_5^3+22016s_1s_2^9s_3s_5-89520s_1^2s_2^7s_3^2s_5+4000000s_1^4s_2s_4^4s_5\]
\[ -91125000s_1s_2s_3^3s_5^3+7187500s_1^5s_2^2s_3s_5^3+253800000s_2^3s_3s_4^2s_5^2+135000s_1^5s_2s_3^5s_5\]
\[ +14364000s_2^4s_3^2s_4^2s_5-3712500s_1^5s_3^3s_4^2s_5+187800s_1^3s_2^5s_3^3s_5-343440s_1s_2^6s_3^3s_5\]
\[ +24840000s_1^2s_2s_3s_4^5+777600s_1^3s_2s_3^6s_4+3078000s_1s_2^3s_3^4s_4^2.\]
\begin{itemize}\item  The coefficient of $s_6^2$ is:
\end{itemize} 
\[ b_2 = -278437500s_1^2s_3s_4^2s_5^4-5508000s_1^2s_2^2s_3^2s_4^4s_5-4000000s_1s_2^4s_4s_5^4-175000s_1^5s_2^2s_3^3s_5^3\]
\[ -14100s_1^3s_2^5s_3^2s_4s_5^2+5760000s_1^2s_2^3s_3^2s_4s_5^3+1912500s_1s_2^4s_3^3s_5^3-326400s_1^2s_2^4s_3^2s_4^3s_5\]
\[ +1890000s_1s_2s_3^3s_4^4s_5-712500s_1^5s_2^3s_3s_4s_5^3+80156250s_1s_2s_3^2s_4s_5^4+16497s_1^2s_2^4s_3^6s_5\]
\[ +55200s_1^3s_2^4s_3^2s_4^4+2190000s_1^3s_2^2s_4^4s_5^2+337500000s_1s_3s_4s_5^5-7575000s_1^4s_2^2s_3^2s_4s_5^3\]
\[ +1254000s_1^3s_2^2s_3^3s_4^3s_5+85536s_1s_2s_3^6s_4^3+48600s_1s_3^5s_4^3s_5+23490000s_2s_3^3s_4^3s_5^2\]
\[ -40824s_1^2s_2s_3^7s_4^2-12738000s_1s_2^5s_3s_4s_5^3+5508000s_1s_2^4s_3^2s_4^2s_5^2+580500s_1^2s_2^4s_3^3s_4s_5^2\]
\[ +67500s_1^4s_3^5s_4s_5^2-332500s_1^5s_2s_3^3s_4^3s_5-131250s_1^5s_2s_4^4s_5^2-283200s_1^4s_2s_3s_4^6-3750000s_1^5s_2s_3^2s_5^4\]
\[ -33750s_1^2s_2^2s_3^4s_5^3-96750s_1^4s_2^2s_3^4s_4^2s_5-255150s_1^2s_3^6s_4^2s_5+245160s_2^4s_3^4s_4^2s_5\]
\[ -2335500s_1s_2^3s_3^4s_4s_5^2-33750000s_1s_3^2s_4^4s_5^2-191625s_1^3s_2^4s_3^4s_5^2+6660000s_1^3s_2^4s_3s_4s_5^3\]
\[ +200000s_1^5s_2^3s_4^3s_5^2-4302s_1s_2^6s_3^5s_5-4687500s_1^3s_2s_3s_5^5-40000s_1^3s_2^3s_3^4s_4^3-2916s_1s_2^2s_3^8s_4\]
\[ +196500s_1^4s_2^3s_3^2s_4^3s_5+33750000s_1s_2^2s_4^2s_5^4+2475000s_1^4s_2s_3^3s_4^2s_5^2+23062500s_1^3s_2^2s_3^2s_5^4\]
\[ +4301100s_2^2s_3^5s_4s_5^2+9562500s_2^3s_3s_4s_5^4-507600s_1s_2^7s_3s_5^3+478125s_1^6s_3^2s_4^4s_5+40500s_1^3s_2s_3^6s_5^2\]
\[ +3456000s_4^7s_5+5467500s_3^6s_5^3-421875000s_4^2s_5^5+5832s_3^7s_4^3+82560s_2^9s_5^3-9765625s_1^3s_5^6+321000s_1^5s_4^7\]
\[ -1152000s_1s_4^8-729s_1^3s_3^{10}+27s_2^6s_3^7+16125000s_2^4s_5^5-250560s_3^3s_4^6-52734375s_3s_5^6-180000s_1^2s_2^6s_4s_5^3\]
\[ -30240s_1s_2^7s_4^2s_5^2+117120s_1^3s_2^2s_3^2s_4^5+139320s_1^2s_2^3s_3^3s_4^4-338040s_2^5s_3^3s_4s_5^2+720s_2^6s_3^4s_4s_5\]
\[ +2322s_1s_2^5s_3^6s_4-7128s_1^2s_2^3s_3^7s_4+3552s_1s_2^7s_3^2s_4^3-150048s_2^5s_3^2s_4^3s_5-5900000s_1^2s_2^2s_4^3s_5^3\]
\[ -15408s_1^2s_2^6s_3s_4^4+150048s_1s_2^3s_3^2s_4^5+29375000s_1^3s_2s_4^2s_5^4+4788000s_2^3s_3^2s_4^4s_5-85536s_2^3s_3^6s_4s_5\]
\[ -245160s_1s_2^2s_3^4s_4^4+16110000s_2^4s_3^2s_4s_5^3+61750000s_1^3s_3s_4^3s_5^3-144480s_2^6s_3s_4^2s_5^2+93750s_1^6s_2^2s_4^2s_5^3\]
\[ +6187500s_1^3s_2^3s_4s_5^4-1790625s_1^2s_2^4s_3s_5^4+750000s_1^4s_2^3s_3s_5^4+7290s_1^3s_2s_3^8s_4+12738000s_1^3s_2s_3s_4^5s_5\]
\[ -5212800s_1s_2^2s_3s_4^5s_5-840s_1^2s_2^7s_3s_4s_5^2-44640s_1^3s_2^5s_3s_4^3s_5-110700s_1^3s_2s_3^5s_4^2s_5+54000000s_1s_2s_3s_4^3s_5^3\]
\[ +33552s_1^2s_2^6s_3^2s_4^2s_5+131220s_1^2s_2^2s_3^6s_4s_5+2250000s_1^4s_2^2s_3s_4^3s_5^2+151680s_1s_2^6s_3^2s_4s_5^2\]
\[ -11136s_1s_2^8s_3s_4^2s_5-140000s_1^4s_2^3s_3^3s_4s_5^2-5760000s_1^3s_2s_3^2s_4^3s_5^2-12150000s_1s_2s_3^4s_4^2s_5^2\]
\[ -4170000s_1^3s_2^3s_3^2s_4^2s_5^2+141660s_1^3s_2^3s_3^5s_4s_5-102600s_1^4s_2s_3^6s_4s_5-23544s_1s_2^4s_3^5s_4s_5\]
\[ -67500s_1^3s_2^2s_3^4s_4s_5^2+21008s_1s_2^7s_3^3s_4s_5-82692s_1^2s_2^5s_3^4s_4s_5+137088s_1s_2^5s_3^3s_4^2s_5\]
\[ +3562500s_1^5s_2s_3s_4^2s_5^3+29700000s_1^2s_2s_3^2s_4^2s_5^3-4400s_1^3s_2^4s_3^3s_4^2s_5+750000s_1^6s_2s_3^2s_4s_5^3\]
\[ +2700000s_1s_2^2s_3^3s_4s_5^3-44437500s_1^2s_2^2s_3s_4s_5^4-12150000s_2s_3^4s_4s_5^3+1762500s_1^5s_3^2s_4^3s_5^2\]
\[ -905760s_1s_2s_3^2s_4^6-311850s_1^3s_2s_3^4s_4^4+2176s_1s_2^9s_4s_5^2-54000s_1^2s_2^3s_3^5s_5^2+437400s_2s_3^6s_4^2s_5\]
\[ +175500s_1^5s_3^5s_4^2s_5-937500s_1^5s_2^2s_4s_5^4-2032800s_1^2s_2^3s_4^5s_5+34020000s_2^2s_3s_4^4s_5^2+6260000s_1s_2^3s_4^4s_5^2\]
\[ -544s_1^2s_2^5s_3^3s_4^3-56875s_1^4s_2^2s_3^3s_4^4-444000s_1^4s_2^2s_4^5s_5+14040000s_1^2s_3^3s_4^3s_5^2+33750s_1^3s_3^4s_4^2s_5^2\]
\[ +3375000s_2s_3s_4^2s_5^4+29531250s_1s_2^2s_3s_5^5-6187500s_1^4s_2s_4^3s_5^3-1050000s_1^3s_2^4s_4^3s_5^2+39375000s_1^3s_3^2s_4s_5^4\]
\[ -484000s_1^2s_2s_4^6s_5-23625000s_1^2s_2s_3^3s_5^4-30937500s_1s_2^3s_3^2s_5^4-1912500s_1^3s_3^3s_4^4s_5+2250000s_1^5s_3^3s_4s_5^3\]
\[ +2032800s_1s_2^5s_4^3s_5^2+8496000s_1s_3s_4^6s_5-6384s_1s_2^6s_3^4s_4^2-11880s_1^3s_2^2s_3^6s_4^2-2052000s_2^3s_3^3s_4^2s_5^2\]
\[ -16800s_2^7s_3^2s_4^2s_5+1790625s_1^4s_3s_4^4s_5^2-54300s_1^4s_2^3s_3s_4^5-56250s_1^4s_2^4s_3^2s_5^3-3078000s_2^2s_3^4s_4^3s_5\]
\[ -27540s_1^3s_2^2s_3^7s_5+1093500s_1s_3^6s_4s_5^2+73800s_1^4s_2s_3^5s_4^3-81720s_1s_2^4s_3^4s_4^3-112500s_1^4s_2s_3^4s_5^3\]
\[ +74088s_1^2s_2^2s_3^5s_4^3+11400000s_1s_2^3s_3s_4^2s_5^3+112500s_1^5s_2s_3^4s_4s_5^2-59985000s_1^2s_2s_3s_4^4s_5^2\]
\[ +23220000s_1s_2^2s_3^2s_4^3s_5^2-64896s_1s_2^6s_3s_4^3s_5-399600s_1^2s_2^3s_3^4s_4^2s_5-303750s_1^5s_2^2s_3s_4^4s_5\]
\[ +2335500s_1^2s_2s_3^4s_4^3s_5+1398000s_1^3s_2^3s_3s_4^4s_5-1736250s_1^4s_2s_3^2s_4^4s_5-6561000s_3^5s_4^2s_5^2-656100s_2s_3^7s_5^2\]
\[ +60480s_1^3s_3^6s_4^3-216000s_1^3s_3^2s_4^6-876000s_1^3s_2s_4^7+486s_2^5s_3^6s_5+436800s_1^2s_3s_4^7+14136s_2^7s_3^3s_5^2\]
\[ -13632s_2^7s_3s_4^4+876000s_2^7s_4s_5^3+2343750s_1^4s_2^2s_5^5-2343750s_1^5s_4^2s_5^4-21200s_1^2s_2^8s_5^3-4752s_2^5s_3^5s_4^2\]
\[ -53156250s_3^3s_4s_5^4+4374s_1^2s_3^9s_4-135000s_1^5s_3^6s_5^2+14300000s_1^3s_4^5s_5^2+78336s_1s_2^4s_4^6+1082880s_2s_3s_4^7\]
\[ +1536s_2^{10}s_3s_5^2+250950s_1^4s_3^3s_4^5+14656s_2^6s_3^3s_4^3-12150s_1^4s_3^7s_4^2-8748s_1s_3^8s_4^2-512s_2^9s_3s_4^3\]
\[ -27000000s_3s_4^5s_5^2-129276s_2s_3^5s_4^4-15625s_1^7s_4^4s_5^2+216000s_2^6s_3^2s_5^3-91500000s_1^2s_4^4s_5^3+234375s_1^5s_2^4s_5^4\]
\[ +35156250s_1s_2s_5^6+23040s_2^8s_4^3s_5+18225s_1^4s_3^8s_5+89100000s_3^2s_4^3s_5^3-1035904s_2^3s_3s_4^6-93750s_1^6s_3s_4^6\]
\[ +1068750s_1^5s_2^2s_3^2s_4^2s_5^2+230400s_1^2s_2^5s_3s_4^2s_5^2-176250s_1^4s_2^4s_3s_4^2s_5^2-675000s_1^6s_2s_3s_4^3s_5^2\]
\[ +28512s_1s_2^3s_3^6s_4^2+18090s_1^2s_2^4s_3^5s_4^2-975000s_1^5s_3s_4^5s_5+607500s_1s_2s_3^5s_5^3+25600s_1^3s_2^6s_4^2s_5^2\]
\[ -56250000s_1^2s_2s_4s_5^5+2656s_1^2s_2^7s_4^3s_5-217536s_1^2s_2^4s_3s_4^5-47232s_2^8s_3s_4s_5^2+255150s_1s_2^2s_3^6s_5^2\]
\[ +146250s_1^5s_2s_3^2s_4^5-1920s_2^9s_3^2s_4s_5-37500s_1^4s_2^5s_4s_5^3-975000s_1^4s_2^3s_4^2s_5^3+637500s_1^2s_2^5s_3^2s_5^3\]
\[ -850000s_1^6s_3^3s_4^2s_5^2+17000s_1^4s_2^4s_4^4s_5-22560s_1^2s_2^5s_4^4s_5+30780000s_1s_2s_4^5s_5^2+111500s_1^3s_2^6s_3s_5^3\]
\[ +338040s_1^2s_2s_3^3s_4^5-972s_1s_2^3s_3^7s_5+249750s_1^4s_2^2s_3^5s_5^2-58590s_1s_2^5s_3^4s_5^2-2190000s_1^2s_2^4s_4^2s_5^3\] \[ -18672s_1s_2^8s_3^2s_5^2+50016s_1s_2^5s_3^2s_4^4-7290s_1^2s_2s_3^8s_5-98550000s_2^2s_3^2s_4^2s_5^3-7149600s_2s_3^2s_4^5s_5\]
\[ +144480s_1^2s_2^2s_3s_4^6+84830s_1^2s_2^6s_3^3s_5^2-202500s_1^4s_3^4s_4^3s_5-715000s_1^3s_2^3s_3^3s_5^3-23062500s_1^4s_3^2s_4^2s_5^3\]
\[ -24300s_1^3s_3^7s_4s_5-378000s_1^2s_3^2s_4^5s_5-16464000s_2^4s_3s_4^3s_5^2+10125000s_1s_3^3s_4^2s_5^3+117500s_1^6s_2s_4^5s_5\]
\[ -1012500s_1^2s_3^4s_4s_5^3-252s_2^7s_3^5s_4+729s_1^2s_2^2s_3^9-60300000s_2s_4^4s_5^3-60000s_1^5s_3^4s_4^4-33750000s_2^2s_4s_5^5\]
\[ +1498880s_1s_2^2s_4^7+6188000s_2^5s_4^2s_5^3-5332500s_2^3s_3^4s_5^3+1408s_1s_2^8s_4^4+331875000s_1s_4^3s_5^4\]
\[ +2625000s_1s_2^6s_5^4+106720s_1^3s_2^3s_4^6+103680s_1s_3^4s_4^5-18984375s_1s_3^4s_5^4-78336s_2^6s_4^4s_5\]
\[ -6662400s_2^2s_4^6s_5+8748s_2^2s_3^8s_5+10546875s_1^2s_3^2s_5^5+408s_2^8s_3^4s_5+312500s_1^6s_4^3s_5^3\]
\[ +7812500s_1^4s_4s_5^5-13100000s_2^3s_4^3s_5^3-243s_1s_2^4s_3^8+486s_2^4s_3^7s_4+704s_2^8s_3^3s_4^2-83835s_1^2s_3^5s_4^4\]
\[ +14000s_1^5s_2^2s_4^6-2625000s_1^4s_4^6s_5-12812500s_1^2s_2^3s_5^5+1944000s_3^4s_4^4s_5-1882500s_2^5s_3s_5^4\]
\[ +26112s_1s_2^6s_4^5+7328s_1^3s_2^5s_4^5+62015625s_2^2s_3^3s_5^4-1593750s_1^3s_2^5s_5^4+680832s_2^2s_3^3s_4^5\]
\[ -113906250s_2s_3^2s_5^5+250000s_1^6s_3^4s_5^3+1024s_2^{10}s_4^2s_5+83835s_2^4s_3^5s_5^2+1836800s_2^4s_4^5s_5.\]
\begin{itemize}\item  The coefficient of $s_6$ is:
\end{itemize} 
\[ b_1 = -1215000s_1s_3^6s_5^4+4s_2^6s_3^5s_4^3-9112500s_2s_3^4s_5^5+1593750s_1^4s_4^5s_5^3-1220400s_2^2s_3^3s_4^4s_5^2\]
\[ -312500s_1^3s_2^3s_5^6-14136s_1^2s_3^3s_4^7-40740s_2^8s_3s_5^4-23625000s_3^2s_4^2s_5^5+1053000s_2^2s_3^5s_5^4\]
\[ -40320s_3^2s_4^7s_5-648s_2^4s_3^7s_5^2-14656s_2^3s_3^3s_4^6-108s_1^3s_3^8s_4^3+10182s_1^3s_3^4s_4^6-26112s_2^5s_4^6s_5\]
\[ -60928s_2^2s_3s_4^8-32s_2^7s_3^3s_4^4-23040s_1s_2^3s_4^8+476800s_1^2s_4^8s_5-128s_2^9s_4^4s_5+37422s_3^6s_4^4s_5\]
\[ -1120s_1^3s_2^4s_4^7-10182s_2^6s_3^4s_5^3+4752s_2^2s_3^5s_4^5+5062500s_1^2s_3^4s_5^5-486s_1s_3^6s_4^5-337920s_2s_4^8s_5\]
\[ +124000s_2^2s_4^5s_5^3-1152s_2^{10}s_4s_5^3+40740s_1^4s_3s_4^8-4374s_2s_3^9s_5^2-83900s_1^3s_2^3s_3^4s_4^2s_5^2+25700s_1^5s_2s_3s_4^6s_5\]
\[ +48880s_1^3s_2^3s_3^3s_4^4s_5 -62370s_1^2s_2^2s_3^4s_4^4s_5+3712500s_1s_2^2s_3^3s_5^5-576s_1s_2^7s_3^3s_5^3-777600s_2s_3^6s_4s_5^3\]
\[ -88500s_1^3s_2^5s_3^2s_5^4-2250000s_1s_3^2s_4^3s_5^4-32500s_1^4s_2^4s_4^3s_5^3-105468750s_1s_4^2s_5^6\]
\[ -35156250s_1s_3s_5^7-8880s_1^2s_2^7s_4^2s_5^3+6075000s_1s_2s_3^4s_4s_5^4+51384s_1^2s_2^5s_3^3s_4^2s_5^2\]
\[ -252000s_1s_2^2s_4^6s_5^2-5344s_2^8s_4^2s_5^3+3906250s_1^2s_2s_5^7+187500s_1s_2^4s_5^6-6250000s_1^3s_3^2s_5^6\]
\[ +3210000s_1s_3s_4^5s_5^3+26392s_1^2s_2^6s_3s_4^3s_5^2-12500s_1^5s_2^3s_4^2s_5^4-53750s_1^4s_2^2s_4^4s_5^3\]
\[ +6250s_1^6s_2s_4^4s_5^3-125000s_1^5s_2s_4^3s_5^4-13250000s_1s_2s_4^4s_5^4-9840s_1^3s_2^5s_4^4s_5^2-637500s_1^3s_3^2s_4^5s_5^2\]
\[ -10800s_1^4s_3^7s_4s_5^2+40824s_2^2s_3^7s_4s_5^2+7290s_1s_3^8s_4s_5^2+1216s_2^9s_3s_4^2s_5^2+715000s_1^3s_3^3s_4^3s_5^3\]
\[ +4444000s_2^3s_3^2s_4^3s_5^3+3750000s_1^4s_3^2s_4s_5^5-5320s_1^2s_2^4s_3^4s_4^3s_5+324s_1^2s_2^3s_3^7s_5^2-810s_1^3s_2s_3^8s_5^2\]
\[ +2916s_1s_2^2s_3^8s_5^2-312s_2^8s_3^3s_4s_5^2+5832s_2^5s_3^5s_4s_5^2+3230s_1s_2^6s_3^4s_4s_5^2-116640s_3^7s_4^2s_5^2\]
\[ -178200s_3^3s_4^5s_5^2-54s_1s_2^5s_3^6s_5^2-5700s_1^5s_3^2s_4^7+162s_1s_2^4s_3^7s_4s_5-54200s_1^4s_2s_4^7s_5\]
\[ +30240s_1^2s_2^2s_4^7s_5+507600s_1^3s_3s_4^7s_5+173750s_1^2s_2^6s_5^5+343440s_1s_3^3s_4^6s_5+16736s_1^3s_2^4s_3s_4^5s_5\]
\[ -28032s_1s_3^2s_4^8+2512s_1s_2^6s_3^3s_4^3s_5+1736250s_1s_2^4s_3^2s_4s_5^4-486s_2s_3^7s_4^4-2250000s_1^2s_2^3s_3s_4^2s_5^4\]
\[+2916s_2s_3^8s_4^2s_5-288s_2^8s_3^2s_4^3s_5-888s_2^5s_3^4s_4^3s_5+9720s_1^4s_3^6s_4^3s_5-28512s_2^2s_3^6s_4^3s_5\]
\[ +2530000s_1^2s_2s_4^5s_5^3-177000s_1^3s_2s_3^4s_4^3s_5^2-250950s_2^5s_3^3s_5^4-3562500s_1^3s_2^2s_3s_4s_5^5\]
\[ +162000s_2^2s_3^4s_4^2s_5^3-8343750s_2^3s_3s_5^6-1304320s_1s_2s_3s_4^7s_5+3523200s_2s_3s_4^6s_5^2+1050000s_1^2s_2^3s_4^4s_5^3\]
\[ +152s_2^7s_3^4s_4^2s_5+108s_2^4s_3^6s_4^2s_5-2916s_1^2s_3^8s_4^2s_5+288300s_1s_2^6s_3^2s_5^4-500000s_1^2s_2^4s_3^3s_5^4\]
\[ +7593750s_1s_3^3s_4s_5^5+937500s_1^4s_2s_4^2s_5^5+160000s_1^4s_2^3s_3^3s_5^4-12728s_2^6s_3^3s_4^2s_5^2-2475000s_1^2s_2^2s_3^3s_4s_5^4\]
\[ +1984500s_1s_3^5s_4^2s_5^3+45050s_1^5s_3^3s_4^5s_5+58590s_1^2s_3^4s_4^5s_5-284472s_2s_3^4s_4^5s_5-50016s_2^4s_3^2s_4^5s_5\]
\[ +888s_1s_2^3s_3^4s_4^5-5832s_1^2s_2s_3^5s_4^5+972s_1s_3^7s_4^3s_5+1728s_2^6s_3^2s_4^4s_5+81720s_2^3s_3^4s_4^4s_5-58050s_1^3s_3^5s_4^4s_5\]
\[ -1408s_1^2s_2^6s_4^5s_5+112500s_1^3s_3^4s_4s_5^4-19600s_1^3s_2^3s_3^5s_5^3+1408s_1^2s_2^4s_3^3s_4^5-58s_1^3s_2^2s_3^4s_4^5-432000s_1s_4^7s_5^2\]
\[ +975000s_1^3s_2^2s_4^3s_5^4-92500s_1^5s_3^4s_4^3s_5^2-100000s_1^5s_2s_3^4s_5^4-1500000s_1^2s_2^2s_4^2s_5^5-4920s_1^4s_2^3s_4^6s_5\]
\[ +12192s_1^2s_2^4s_4^6s_5-288300s_1^4s_3^2s_4^6s_5+494112s_2^2s_3^2s_4^6s_5-894s_1^2s_2^3s_3^5s_4^4+864s_1^3s_2s_3^6s_4^4\]
\[ -108s_1s_2^2s_3^6s_4^4-576s_1s_2^6s_3^2s_4^5+10700000s_1s_2^3s_3s_4s_5^5+8060s_1^4s_2^2s_3s_4^7+1344s_1^2s_2^3s_3s_4^7\]
\[ -46032s_1^3s_2s_3^2s_4^7+16800s_1s_2^2s_3^2s_4^7+47232s_1^2s_2s_3s_4^8-18s_2^6s_3^6s_4s_5+486s_1^3s_3^9s_4s_5\]
\[ +54000s_1^2s_3^5s_4^3s_5^2+871560s_2s_3^5s_4^3s_5^2-720s_1s_2s_3^4s_4^6-68070s_1^4s_2^2s_3^2s_4^5s_5+125000s_1^4s_2^3s_4s_5^5\]
\[ +46032s_2^7s_3^2s_4s_5^3+98000s_1^2s_2^5s_4^3s_5^3-3996s_1^3s_2s_3^7s_4^2s_5+13200s_1^3s_2^2s_3s_4^6s_5-13320000s_2s_3^2s_4^4s_5^3\]
\[ -1728s_1s_2^4s_3^2s_4^6+5920s_1^4s_2s_3^3s_4^6+12728s_1^2s_2^2s_3^3s_4^6-1406250s_1s_2s_3^2s_5^6+6875000s_1^2s_2^2s_3s_5^6\]
\[ +11250000s_2s_3s_4s_5^6+1215000s_3^4s_4^3s_5^3-14250s_1^5s_2^2s_4^5s_5^2+217536s_2^5s_3s_4^4s_5^2+500000s_1^4s_3^3s_4^4s_5^2\]
\[ +975000s_1s_2^5s_3s_5^5-7187500s_1^3s_3s_4^2s_5^5+8662500s_1^2s_3^2s_4^4s_5^3-3950000s_1s_2^3s_4^3s_5^4+22560s_1s_2^4s_4^5s_5^2\]
\[ -32500s_1^6s_3s_4^5s_5^2-98000s_1^3s_2^3s_4^5s_5^2-67500s_1^2s_2s_3^5s_5^4-9250s_1^3s_2^6s_4s_5^4-168750s_2^2s_3^2s_4s_5^5\]
\[ -74088s_2^3s_3^5s_4^2s_5^2+51300s_1^3s_3^6s_4^2s_5^2-750000s_1^5s_3^2s_4^2s_5^4-36s_1s_2^4s_3^6s_4^3+108s_1^2s_2^2s_3^7s_4^3\]
\[ +296s_1s_2^5s_3^4s_4^4+750000s_1^4s_2^2s_3^2s_5^5-1440s_1s_2^8s_4^3s_5^2-1344s_2^7s_3s_4^3s_5^2+102500s_2^4s_3s_4^2s_5^4\]
\[ -1762500s_1^2s_2^3s_3^2s_5^5-750000s_1^4s_3s_4^3s_5^4+444000s_1s_2^5s_4^2s_5^4-139320s_2^4s_3^3s_4^3s_5^2+1824s_1^2s_2^5s_3s_4^6\]
\[ -6944s_1^3s_2^3s_3^2s_4^6-151680s_1^2s_2s_3^2s_4^6s_5-321000s_2^7s_5^5-60480s_2^3s_3^6s_5^3-106720s_2^6s_4^3s_5^3\]
\[ -580500s_1^2s_2s_3^3s_4^4s_5^2-230400s_1^2s_2^2s_3s_4^5s_5^2+4170000s_1^2s_2^2s_3^2s_4^3s_5^3+777000s_1^4s_2s_3s_4^5s_5^2\]
\[ +24300s_1s_2s_3^7s_5^3-18562500s_1s_2s_3s_4^2s_5^5+23544s_1s_2s_3^5s_4^4s_5+24300s_1^4s_2s_3^5s_4^2s_5^2\]
\[ +5490000s_1s_2s_3^2s_4^5s_5^2+214500s_1^3s_2^4s_3^3s_4s_5^3+14032s_1s_2^8s_3s_4s_5^3-77940s_1^2s_2^6s_3^2s_4s_5^3\]
\[ +53750s_1^3s_2^4s_4^2s_5^4-13500000s_1^2s_3^3s_4^2s_5^4-7380000s_1s_2s_3^3s_4^3s_5^3+2176s_1s_2^7s_3s_4^4s_5\]
\[ +83400s_1^3s_2^5s_3s_4^2s_5^3+135000s_1^4s_2^4s_3s_4s_5^4-267200s_1s_2^5s_3^3s_4s_5^3+326400s_1s_2^3s_3^2s_4^4s_5^2\]
\[ +110700s_1s_2^2s_3^5s_4s_5^3-6660000s_1^3s_2s_3s_4^4s_5^3-384s_1s_2^5s_4^7+101750s_1^4s_2^3s_3s_4^4s_5^2-153750s_1^5s_2s_3^2s_4^4s_5^2\]
\[ +58050s_1s_2^4s_3^5s_5^3+399600s_1s_2^2s_3^4s_4^3s_5^2-471000s_1^3s_2^2s_3^2s_4^4s_5^2+250000s_1^5s_2s_3^3s_4^2s_5^3\]
\[ -1254000s_1s_2^3s_3^3s_4^2s_5^3+471000s_1^2s_2^4s_3^2s_4^2s_5^3-315000s_1^4s_2^3s_3^2s_4^2s_5^3+67500s_1^2s_2s_3^4s_4^2s_5^3\]
\[ +87500s_1^5s_2^2s_3s_4^3s_5^3-1398000s_1s_2^4s_3s_4^3s_5^3+393750s_1^5s_3s_4^4s_5^3-1215000s_3^5s_4s_5^4-3600s_1^5s_2s_4^8\]
\[ +32812500s_1^2s_3s_4s_5^6-3125000s_1^3s_2s_4s_5^6-1718750s_1s_2^2s_4s_5^6+17212500s_2s_3^3s_4^2s_5^4-12192s_1s_2^6s_4^4s_5^2\]
\[ -173750s_1^5s_4^6s_5^2-440000s_1s_2^2s_3s_4^4s_5^3+4912500s_2^2s_3s_4^3s_5^4-777000s_1^2s_2^5s_3s_4s_5^4-13200s_1s_2^6s_3s_4^2s_5^3\]
\[ +311850s_2^4s_3^4s_4s_5^3-212500s_1^4s_2^2s_3^4s_4s_5^3+177000s_1^2s_2^3s_3^4s_4s_5^3-150000s_1^5s_2^2s_3^2s_4s_5^4-393750s_1^3s_2^4s_3s_5^5\]
\[ +25380s_1^3s_2^2s_3^6s_4s_5^2+17125000s_1^2s_2s_3s_4^3s_5^4+50000s_1^6s_3^2s_4^3s_5^3+5962500s_3s_4^4s_5^4-18750s_1^4s_2^5s_5^5\]
\[ +42812500s_1^2s_4^3s_5^5+5344s_1^3s_2^2s_4^8+648s_1^2s_3^7s_4^4+31104s_2s_3^3s_4^7-1584s_1^4s_3^5s_4^5+267200s_1^3s_2s_3^3s_4^5s_5\]
\[ -51300s_1^2s_2^2s_3^6s_5^3+202500s_1s_2^3s_3^4s_5^4-135600s_1^3s_2^4s_3^2s_4^3s_5^2-10125000s_1^2s_2s_3^2s_4s_5^5+105468750s_4s_5^7\]
\[ +16875000s_3^3s_5^6+2343750s_2^2s_5^7+5270s_1^2s_2^5s_3^4s_5^3+9216s_3s_4^9-3996s_3^5s_4^6+27432s_1^2s_2s_3^6s_4^3s_5+64s_2^8s_3s_4^5\]
\[ -1417500s_1s_3^4s_4^4s_5^2-137088s_1s_2^2s_3^3s_4^5s_5+106496s_1s_2s_4^9+60000s_1^5s_3^5s_4s_5^3-117120s_2^5s_3^2s_4^2s_5^3\]
\[ +64896s_1s_2^3s_3s_4^6s_5+4212s_1^2s_2^3s_3^6s_4^2s_5-10704s_1^2s_2^3s_3^2s_4^5s_5-16224s_1^2s_2^5s_3^2s_4^4s_5+18000s_1^4s_2s_3^6s_5^3\]
\[ -40500s_1^2s_3^6s_4s_5^3-3870000s_2^3s_3^3s_4s_5^4-2250000s_1^3s_2s_3^3s_5^5+180000s_1^3s_2s_4^6s_5^2-1404s_1s_2^5s_3^5s_4^2s_5\]
\[ +322500s_1^3s_2^2s_3^4s_5^4-12960s_1s_2^7s_3^2s_4^2s_5^2+3025000s_2^3s_4^2s_5^5+318750s_2^4s_3^2s_5^5+7575000s_1^3s_2s_3^2s_4^2s_5^4\]
\[ +283200s_2^6s_3s_4s_5^4-3228s_1^3s_2^2s_3^5s_4^3s_5+14650s_1^2s_2^7s_3s_5^4-27432s_1s_2^3s_3^6s_4s_5^2+131250s_1^2s_2^4s_4s_5^5\]
\[ +810000s_1^3s_2^3s_3^2s_4s_5^4-3733000s_1^2s_3s_4^6s_5^2+412000s_2^4s_4^4s_5^3+13632s_2^4s_3s_4^7+10704s_1s_2^5s_3^2s_4^3s_5^2\]
\[ -1545000s_2^5s_4s_5^5+54200s_1s_2^7s_4s_5^4-82560s_1^3s_4^9+9000s_1^6s_4^7s_5-9937500s_1^3s_4^4s_5^4-8662500s_1s_2^2s_3^2s_4^2s_5^4\]
\[ +17625000s_2s_4^3s_5^5+384s_2^7s_4^5s_5+48s_2^9s_3^2s_5^3-128s_1s_2^7s_4^6+6s_2^7s_3^5s_5^2-322500s_1^4s_3^4s_4^2s_5^3\]
\[ -810000s_1^4s_2s_3^2s_4^3s_5^3+62370s_1s_2^4s_3^4s_4^2s_5^2+109350s_3^8s_5^3-720000s_4^6s_5^3-37150s_1^4s_2s_3^4s_4^4s_5\]
\[ +249000s_1^4s_2^2s_3^3s_4^3s_5^2-1570240s_2^3s_3s_4^5s_5^2-131220s_1s_2s_3^6s_4^2s_5^2-13572s_1^2s_2^4s_3^5s_4s_5^2\]
\[ -486s_1^2s_2^2s_3^8s_4s_5-360s_1s_2^9s_5^4+190976s_2^3s_4^7s_5.\]
\eject\noindent
\begin{itemize}\item  The coefficient independent of $s_6$ is:
\end{itemize}
\[ b_0 = 5700s_2^7s_3^2s_5^5+165200s_2^2s_3^2s_4^5s_5^3+102600s_1s_2s_3^6s_4s_5^4-390000s_3^2s_4^6s_5^3+13572s_1^2s_2s_3^5s_4^4s_5^2\]
\[ +287500s_1s_2^3s_4^2s_5^6-864s_2^4s_3^6s_4s_5^3+1162500s_2s_3^2s_4^3s_5^5+314s_1s_2^5s_3^5s_4s_5^3+894s_2^4s_3^5s_4^3s_5^2\]
\[ -1743750s_1s_3s_4^4s_5^5-52160s_2^2s_3s_4^7s_5^2-637500s_1^2s_2s_4^4s_5^5+11880s_2^2s_3^6s_4^2s_5^3+625s_1^4s_2^4s_4^2s_5^5\]
\[ +2015s_1^2s_2^4s_3^5s_5^4-84830s_1^2s_3^3s_4^6s_5^2-17000s_1s_2^4s_4^4s_5^4-16s_2^8s_3s_4^4s_5^2-25600s_1^2s_2^2s_4^6s_5^3\]
\[ -10825s_1^5s_3^2s_4^6s_5^2+92500s_1^2s_2^3s_3^4s_5^5-1800s_1^3s_2^2s_3^6s_5^4+18672s_1^2s_3^2s_4^8s_5+34s_1^3s_2s_3^4s_4^7\]
\[ -1408s_2^5s_3^3s_4^4s_5^2-64s_1^3s_2^2s_3^2s_4^8+10000s_1^4s_2^2s_3^4s_5^5-704s_2^2s_3^3s_4^8+3045s_1^4s_2^2s_3s_4^6s_5^2\]
\[ -55200s_2^4s_3^2s_4^4s_5^3+108s_2^3s_3^8s_5^3+486s_1s_2s_3^8s_4^2s_5^2-25700s_1s_2^6s_3s_4s_5^5+27000s_2s_3^5s_4^2s_5^4\]
\[ -142500s_2^2s_3^3s_4^3s_5^4-1187500s_1s_2^3s_3s_5^7+3996s_1s_2^2s_3^7s_4s_5^3-144s_1^2s_2^4s_3^2s_4^6s_5-296s_2^4s_3^4s_4^5s_5\]
\[ -48s_1^3s_2^4s_4^6s_5^2-249000s_1^2s_2^3s_3^3s_4^2s_5^4-1050s_1^4s_2^3s_4^5s_5^3-s_2^6s_3^5s_4^2s_5^2\]
\[ -8960s_1^2s_2^5s_3^3s_4s_5^4-18090s_2^2s_3^5s_4^4s_5^2+288s_1s_2^3s_3^2s_4^8-1142500s_1s_2s_3^2s_4^4s_5^4\]
\[ -101750s_1^2s_2^4s_3s_4^3s_5^4+1440s_1^2s_2^3s_4^8s_5-21008s_1s_2s_3^3s_4^7s_5-196500s_1s_2^3s_3^2s_4^3s_5^4\]
\[ -1216s_1^2s_2^2s_3s_4^9-10000s_1^5s_3^4s_4^2s_5^4+81s_1^2s_2^2s_3^8s_5^3+11718750s_1s_4s_5^8-81s_1^3s_3^8s_4^2s_5^2\]
\[ -27s_3^7s_4^6+100000s_1^4s_3^4s_4s_5^5+10750s_1^3s_2^4s_3^2s_4^2s_5^4+58s_2^5s_3^4s_4^2s_5^3-2610s_1s_2^8s_3s_5^5\]
\[ -45050s_1s_2^5s_3^3s_5^5-16497s_1s_3^6s_4^4s_5^2-528s_1s_2^7s_3s_4^3s_5^3+48s_1^2s_2^6s_4^4s_5^3+9840s_1^2s_2^4s_4^5s_5^3\]
\[ +89520s_1s_3^2s_4^7s_5^2-108s_2^3s_3^7s_4^2s_5^2-1734375s_2^2s_3s_4^2s_5^6+928125s_3^5s_5^6+175000s_2^5s_5^7+729s_3^{10}s_5^3\]
\[ -75000s_4^5s_5^5-9000s_2s_3^4s_4^4s_5^3+7128s_2s_3^7s_4^3s_5^2+1584s_2^5s_3^5s_5^4+12960s_1^2s_2^2s_3^2s_4^7s_5\]
\[ +6944s_2^6s_3^2s_4^3s_5^3+840s_1^2s_2s_3s_4^7s_5^2+74s_1^2s_2^3s_3^4s_4^5s_5-9000s_1s_2^7s_5^6+27540s_1s_3^7s_4^2s_5^3\]
\[ +15625s_1^2s_2^4s_5^7-1068750s_1^2s_2^2s_3^2s_4^2s_5^5-234375s_1^4s_4^4s_5^5+1195s_1^2s_2^4s_3^4s_4^2s_5^3\]
\[ -8s_1^2s_2^3s_3^3s_4^7+556s_1^3s_2^2s_3^3s_4^6s_5+37150s_1s_2^4s_3^4s_4s_5^4+88500s_1^4s_3^2s_4^5s_5^3\]
\[ -4s_1^3s_3^6s_4^6-59s_1^4s_3^3s_4^8-60000s_1^3s_2s_3^5s_5^5+8064s_2s_3^2s_4^8s_5+1800s_1^4s_3^6s_4^2s_5^3\]
\[ +360s_1^4s_4^9s_5+175000s_1^3s_3^3s_4^2s_5^5+623s_1^4s_3^4s_4^6s_5-51384s_1^2s_2^2s_3^3s_4^5s_5^2-324s_1^2s_3^7s_4^3s_5^2\]
\[ +16224s_1s_2^4s_3^2s_4^5s_5^2-1125000s_2s_3^3s_4s_5^6-87500s_1^3s_2^3s_3s_4^2s_5^5+19600s_1^3s_3^5s_4^3s_5^3\]
\[ -135000s_1s_3^5s_4s_5^5+256s_3^3s_4^9+6264s_3^5s_4^5s_5^2+1984s_1^2s_2^5s_3^2s_4^3s_5^3-750000s_1^3s_2s_3^2s_4s_5^6\]
\[ +68070s_1s_2^5s_3^2s_4^2s_5^4+8960s_1^4s_2s_3^3s_4^5s_5^2-141660s_1s_2s_3^5s_4^3s_5^3-74s_1s_2^5s_3^4s_4^3s_5^2\]
\[ +2728s_1s_2^7s_3^2s_4s_5^4+9s_1s_2^4s_3^6s_4^2s_5^2-2176s_1s_2^4s_3s_4^7s_5-112500s_1^2s_2s_3^4s_4s_5^5\]
\[ +11136s_1s_2^2s_3s_4^8s_5-2728s_1^4s_2s_3^2s_4^7s_5-1984s_1^3s_2^3s_3^2s_4^5s_5^2+144s_1s_2^6s_3^2s_4^4s_5^2\]
\[ -214500s_1^3s_2s_3^3s_4^4s_5^3+4400s_1s_2^2s_3^3s_4^4s_5^3+96750s_1s_2^2s_3^4s_4^2s_5^4+226800s_1s_2s_3s_4^6s_5^3\]
\[ +1687500s_1s_2s_3^3s_4^2s_5^5-362500s_1s_2^2s_3^2s_4s_5^6+135600s_1^2s_2^3s_3^2s_4^4s_5^3+5320s_1s_2^3s_3^4s_4^4s_5^2\]
\[ -24300s_1^2s_2^2s_3^5s_4s_5^4+675000s_1^2s_2^3s_3s_4s_5^6-14032s_1^3s_2s_3s_4^8s_5-3045s_1^2s_2^6s_3s_4^2s_5^4\]
\[ -2512s_1s_2^3s_3^3s_4^6s_5-3230s_1^2s_2s_3^4s_4^6s_5-25380s_1^2s_2s_3^6s_4^2s_5^3-5400s_3^6s_4^3s_5^3\]
\[ -8060s_2^7s_3s_4^2s_5^4+36s_2^3s_3^6s_4^4s_5-314s_1^3s_2s_3^5s_4^5s_5-250000s_1^3s_2^2s_3^3s_4s_5^5\]
\[ +191625s_1^2s_3^4s_4^4s_5^3-1536s_1^2s_3s_4^{10}+6384s_2^2s_3^4s_4^6s_5+2218750s_1s_2s_4^3s_5^6+54300s_2^5s_3s_4^3s_5^4\]
\[ -556s_1s_2^6s_3^3s_4^2s_5^3-135000s_1^4s_2s_3s_4^4s_5^4-83400s_1^3s_2^2s_3s_4^5s_5^3+82692s_1s_2s_3^4s_4^5s_5^2\]
\[ -32s_1^2s_2^5s_4^7s_5+225000s_2^3s_3^3s_5^6+12500s_1^4s_2^2s_4^3s_5^5+2048s_1s_4^{11}+14250s_1^2s_2^5s_4^2s_5^5\]
\[ +576s_1^3s_3^3s_4^7s_5+468750s_3s_4^3s_5^6-1012500s_1s_2s_3^4s_5^6+1404s_1s_2^2s_3^5s_4^5s_5-623s_1s_2^6s_3^4s_5^4\]
\[ -264s_1^4s_2^2s_4^8s_5+312500s_1^2s_2^2s_4s_5^7-1024s_1s_2^2s_4^{10}+56875s_2^4s_3^3s_4^2s_5^4+8750s_1^5s_2s_3s_4^5s_5^3\]
\[ +850000s_1s_2^2s_3s_4^3s_5^5+16s_1^2s_2^4s_3s_4^8+16000s_2s_4^7s_5^3-249750s_1^2s_3^5s_4^2s_5^4+544s_2^3s_3^3s_4^5s_5^2\]
\[ -10000s_1^5s_2s_3^2s_4^3s_5^4+1920s_1s_2s_3^2s_4^9+468750s_1s_2s_3s_4s_5^7-2322s_2s_3^6s_4^5s_5-5859375s_1^2s_4^2s_5^7\]
\[ -1050s_1^5s_2s_4^7s_5^2-3552s_2^3s_3^2s_4^7s_5-2176s_1^2s_2s_4^9s_5+18s_1s_2s_3^6s_4^6-18000s_1^3s_3^6s_4s_5^4\]
\[ +54s_1^2s_3^6s_4^5s_5+10800s_1^2s_2s_3^7s_5^4-137500s_2^3s_3^2s_4^2s_5^5-5270s_1^3s_3^4s_4^5s_5^2+4920s_1s_2^6s_4^3s_5^4\]\[ +247500s_1s_3^4s_4^3s_5^4-1195s_1^3s_2^2s_3^4s_4^4s_5^2+528s_1^3s_2^3s_3s_4^7s_5+1040625s_3^3s_4^4s_5^4\]
\[ +1562500s_1^3s_4^3s_5^6-117500s_1s_2^5s_4s_5^6-10750s_1^4s_2^2s_3^2s_4^4s_5^3-5920s_2^6s_3^3s_4s_5^4\]
\[ -6144s_4^{10}s_5-93750s_1^3s_2^2s_4^2s_5^6+14100s_1^2s_2s_3^2s_4^5s_5^3+128s_2^6s_4^7s_5-17500s_1^3s_2^4s_3^3s_5^5\]
\[ +176250s_1^2s_2^2s_3s_4^4s_5^4+216s_2^{10}s_5^5-4212s_1s_2^2s_3^6s_4^3s_5^2+1500000s_1^2s_3^3s_4s_5^6-1375000s_1^2s_2s_3s_4^2s_5^6\]\[ -486s_1s_2s_3^9s_5^3+32500s_1^3s_2^3s_4^4s_5^4-262500s_2s_3s_4^5s_5^4-146250s_2^5s_3^2s_4s_5^5+9250s_1^4s_2s_4^6s_5^3\]
\[ +3234375s_1s_3^2s_4^2s_5^6+32500s_1^2s_2^5s_3s_5^6+212500s_1^3s_2s_3^4s_4^2s_5^4+732s_1^3s_2s_3^6s_4^3s_5^2-4s_2^3s_3^5s_4^6\]
\[ +10825s_1^2s_2^6s_3^2s_5^5-18225s_1s_3^8s_5^4+8880s_1^3s_2^2s_4^7s_5^2+153750s_1^2s_2^4s_3^2s_4s_5^5-408s_1s_3^4s_4^8\]
\[ +32s_2^9s_4^3s_5^3-1024s_2s_3s_4^{10}-250000s_1^3s_3^4s_5^6+135000s_1^2s_3^6s_5^5-2343750s_2s_4^2s_5^7-25000s_2^4s_4^3s_5^5\]
\[ -2232s_3^4s_4^7s_5-26392s_1^2s_2^3s_3s_4^6s_5^2+1408s_1s_2^5s_4^6s_5^2-625s_1^6s_4^6s_5^3-48s_1^3s_3^2s_4^9\]
\[ +312s_1^2s_2s_3^3s_4^8-1518750s_3^4s_4^2s_5^5+1050s_1^3s_2^5s_4^3s_5^4+625s_1^3s_2^6s_5^6-64s_2^5s_3s_4^8+59s_2^8s_3^3s_5^4\]
\[ +156250s_2^3s_4s_5^7+32s_2^4s_3^3s_4^7-6s_1^2s_3^5s_4^7+252s_2s_3^5s_4^7+21200s_1^3s_4^8s_5^2+1875000s_2^2s_3^2s_5^7\]
\[ -2812500s_1s_3^3s_5^7+76800s_3s_4^8s_5^2-32s_1^3s_2^3s_4^9-134000s_1^2s_4^7s_5^3+4s_2^6s_3^6s_5^3-33552s_1s_2^2s_3^2s_4^6s_5^2\]
\[ -152s_1s_2^2s_3^4s_4^7-22016s_1s_3s_4^9s_5+s_1^2s_2^2s_3^5s_4^6-732s_1^2s_2^3s_3^6s_4s_5^3+15408s_2^4s_3s_4^6s_5^2\]
\[ +712500s_1^3s_2s_3s_4^3s_5^5+40000s_2^3s_3^4s_4^3s_5^3+1120s_2^7s_4^4s_5^3+1152s_1^3s_2s_4^{10}+56250s_1^3s_3^2s_4^4s_5^4\]
\[ -2015s_1^4s_3^5s_4^4s_5^2-10880s_1s_2s_4^8s_5^2+64s_2^8s_3^2s_4^2s_5^3-478125s_1s_2^4s_3^2s_5^6+150000s_1^4s_2s_3^2s_4^2s_5^5\]
\[ +37500s_1^3s_2s_4^5s_5^4+44640s_1s_2^3s_3s_4^5s_5^3+4302s_1s_3^5s_4^6s_5+77940s_1^3s_2s_3^2s_4^6s_5^2-625s_1^5s_2^2s_4^4s_5^4\]
\[ +576s_2^5s_3^2s_4^6s_5-187800s_1s_3^3s_4^5s_5^3-14000s_2^6s_4^2s_5^5-1408s_2^4s_4^8s_5-48880s_1s_2^4s_3^3s_4^3s_5^3\]
\[ +128s_1s_2^4s_4^9-252s_2^9s_3s_4s_5^4+264s_1s_2^8s_4^2s_5^4-390625s_1s_2^2s_5^8+7500s_1^3s_2^3s_3^4s_4s_5^4+8s_2^7s_3^3s_4^3s_5^2\]
\[ -111500s_1^3s_3s_4^6s_5^3-175500s_1s_2^2s_3^5s_5^5+245000s_1s_4^6s_5^4-7290s_2s_3^8s_4s_5^3+2610s_1^5s_3s_4^8s_5\]
\[ +303750s_1s_2^4s_3s_4^2s_5^5-162s_1s_2s_3^7s_4^4s_5-6250s_1^3s_2^4s_4s_5^6+17500s_1^5s_3^3s_4^4s_5^3-50000s_2^4s_3s_4s_5^6\]
\[ -50000s_1^3s_2^3s_3^2s_5^6-15720s_2s_3^3s_4^6s_5^2-160000s_1^4s_3^3s_4^3s_5^4-2656s_1s_2^3s_4^7s_5^2+32s_1s_2^7s_4^5s_5^2\]
\[ -1824s_2^6s_3s_4^5s_5^2+36s_1^3s_3^7s_4^4s_5-14650s_1^4s_3s_4^7s_5^2+5120s_2^2s_4^9s_5+60000s_2^4s_3^4s_5^5+93750s_2^6s_3s_5^6\]
\[ +12150s_2^2s_3^7s_5^4-7328s_2^5s_4^5s_5^3-7500s_1^4s_2s_3^4s_4^3s_5^3-1787500s_1^2s_3^2s_4^3s_5^5-200000s_1^2s_2^3s_4^3s_5^5\]
\[ -9765625s_5^9+332500s_1s_2^3s_3^3s_4s_5^5+3228s_1s_2^3s_3^5s_4^2s_5^3-16736s_1s_2^5s_3s_4^4s_5^3-15000s_2^3s_3s_4^4s_5^4\]
\[ -729s_3^9s_4^2s_5^2+140000s_1^2s_2s_3^3s_4^3s_5^4+1050s_1^2s_2^7s_4s_5^5-9720s_1s_2^3s_3^6s_5^4-34s_2^7s_3^4s_4s_5^3\]
\[ +315000s_1^3s_2^2s_3^2s_4^3s_5^4-36s_1s_2^4s_3^7s_5^3+18750s_1^5s_4^5s_5^4+1250000s_1^2s_2s_3^2s_5^7+850000s_1^2s_2^2s_3^3s_5^6\]
\[ -54000s_1s_2^2s_4^5s_5^4+83900s_1^2s_2^2s_3^4s_4^3s_5^3-8750s_1^3s_2^5s_3s_4s_5^5+3600s_2^8s_4s_5^5-9s_1^2s_2^2s_3^6s_4^4s_5\]
\[ -216s_1^5s_4^{10}-73800s_2^3s_3^5s_4s_5^4+927500s_1^2s_3s_4^5s_5^4+10000s_1^4s_2^3s_3^2s_4s_5^5+5600s_2^3s_4^6s_5^3\]
\[ +125000s_2^2s_4^4s_5^5+810s_1^2s_3^8s_4s_5^3+512s_2^3s_3s_4^9+252s_1^4s_2s_3s_4^9+243s_3^8s_4^4s_5.\]

\subsection{Battaglini formulas;  the Kummer surface}
Let $Q_a= \textrm{diag}(a_1, a_2, a_3, a_4)$ and $Q_b= \textrm{diag}(b_1, b_2, b_3, b_4)$. \\
We assume that these matrices 
Define:
\[ Q = xQ_a + y Q_b = \textrm{diag}(p, q, r, s), \]
\[ p = xa_1 + yb_1, \hspace{3pt}  q = xa_2 + yb_2, \hspace{3pt}r  = xa_3 + yb_3, \hspace{3pt} s = xa_4 + yb_4.\] 
Then the Battaglini family of quadrics are:
\[ Q =   (p(X^1)^2 + q(X^2)^2  + r(X^3)^2 + s(X^4)^2)( p(Y^1)^2 + q(Y^2)^2  + r(Y^3)^2 + s(Y^4)^2) \]
\[ - ( pX^1Y^1 + qX^2Y^2  + rX^3Y^3 + sX^4Y^4)^2\]
\[  =   pq(X^{12})^2 + pr(X^{13})^2 + ps(X^{14})^2 + rs(X^{34})^2+ qs(X^{42})^2 + qr(X^{23})^2, \]
\[ = pqP^2 + pr Q^2 + ps R^2 + rs X^2 + qs Y^2 + qr Z^2, \]
\[ X^{ij} = X^i Y^j - Y^i X^j, \]
\[ P = X^{12}, Q = X^{13}, R = X^{14}, X = X^{34}, Y = X^{42}, Z = X^{23}. \]
Here the $X^{ij}$ obey the Klein quadric relation:
\[ 0 = X^{12}X^{34} + X^{13}X^{42} + X^{14}X^{23} =  PX + QY + RZ.\]
This gives:
\[ Q = x^2 Q_{aa} + xy Q_{ab} + y^2 Q_{bb}, \]
\[ Q_{aa} =  a_1a_2P^2 + a_1a_3Q^2 + a_1a_4R^2 + a_3a_4X^2+ a_4a_2Y^2 + a_2a_3Z^2, \]
\[ Q_{ab}  = Q_{ba} = c_{12}P^2+ c_{13}Q^2 + c_{14}R^2 + c_{34}X^2+ c_{42}Y^2 + c_{34}Z^2,\]
\[ Q_{bb} =  b_1b_2P^2 + b_1b_3Q^2 + b_1b_4R^2 + b_3b_4X^2+ b_4b_2Y^2 + b_2b_3Z^2, \]
\[ c_{ij} = a_i b_j + b_i a_j.\]
The endomorphism corresponding to $Q_{ab}$ is:
\[  Q'_{ab} = c_{12}Px+ c_{13}Qy + c_{14} Rz + c_{34}Xp+ c_{42}Yq + c_{23}Zr.\]
So its matrix is:
\[ Q'_{ab} = \begin{array}{|cccccc|}0&0&0&c_{12}&0&0\\0&0&0&0&c_{13}&0\\0&0&0&0&0&c_{14}\\c_{34}&0&0&0&0&0\\0&c_{42}&0&0&0&0\\0&0&c_{23}&0&0&0\end{array}.\]
Its characteristic polynomial is:
\[ f(s) = (s^2 - c_{12}c_{34})(s^2 - c_{13} c_{42})(s^2 - c_{14} c_{23}).\]
This has at least a pair of repeated roots, if and only if $a_i b_j = \pm b_i a_j$, for some $i \ne j$.
\eject\noindent
We now calculate the Kummer surface in twistor space, determined by the Battaglini quadric.
Given a non-zero twistor $Z = (Z^1, Z^2, Z^3, Z^4)$, the  corresponding quadric cone in twistor space, with vertex $Z$ has the equation:
\[ 0 = c_{12}(Z^1 Y^2 - Z^2 Y^1)^2+ c_{13}(Z^1 Y^3 - Z^3 Y^1)^2 + c_{14}(Z^1 Y^4 - Z^4 Y^1)^2\]
\[  + c_{34}(Z^3 Y^4 - Z^4Y^3)^2+ c_{42}(Z^4 Y^2 - Z^2 Y^4)^2 + c_{34}(Z^3 Y^4 - Z^4 Y^3)^2  =  \sum_{i = 1}^4\sum_{j = 1}^4 h_{ij} Y^iY^j, \]
\[ h_{11}  = c_{12}u + c_{13}v +  c_{14}w, \]
\[ h_{22}  = c_{12}t + c_{23}v +  c_{42}w, \]
\[ h_{33} = c_{13}t + c_{23}u + c_{34}w, \]
\[ h_{44} = c_{14}t + c_{42}u + c_{34}v. \]
\[ h_{ij} = - c_{ij} Z^i Z^j,  \hspace{10pt} \textrm{when $i \ne j$}.\]
Here we have put $t = (Z^1)^2$, $u = (Z^2)^2$, $v = (Z^3)^2$ and $w = (Z^4)^2$.
Geometrically this cone describes the intersection of the $\alpha$-plane of $Z$, a projective plane inside the Klein quadric, with the Battaglini quadric.
The symmetric matrix $h$ is necessarily degenerate and has adjoint of rank at most one.  Explicitly this adjoint is:
\[ H^{ij}  =  K(t, u, v, w) Z^i Z^j, \]
\[ \hspace{-95pt} 2K(t, u, v, w) = \hspace{7pt}  [t, u, v, w]  \hspace{7pt} \begin{array}{|cccc|} 2c_{14}c_{12}c_{13}&c_{12}(c_{42}c_{13}+c_{14}c_{23}) &c_{13}(c_{12}c_{34}+c_{14}c_{23})&c_{14}(c_{42}c_{13}+c_{12}c_{34})\\c_{12}(c_{42}c_{13}+ c_{14}c_{23})&2c_{12}c_{42}c_{23}&c_{23}(c_{12}c_{34}+c_{13}c_{42})& c_{42}(c_{23}c_{14}+c_{34}c_{12})\\c_{13}(c_{12}c_{34}+c_{14}c_{23})& c_{23}(c_{12}c_{34}+c_{13}c_{42})&2c_{13}c_{23}c_{34}&c_{34}(c_{42}c_{13} + c_{23}c_{14})\\c_{14}(c_{42}c_{13}+c_{12}c_{34})&c_{42}(c_{23}c_{14}+c_{34}c_{12})&c_{34}(c_{42}c_{13} + c_{23}c_{14})&2c_{14}c_{42}c_{34}\end{array}\hspace{7pt} \begin{array}{|c|} t\\u\\v\\w\end{array}\hspace{3pt}.\]
This is a non-degenerate quadric in the $(t, u, v, w)$ projective space,  provided that $a_i b_j - b_i a_j$ is non-zero, when $i \ne j$.   Then the quadric cone of $Z$ degenerates to a pair of planes if and only if $K(t, u, v, w) = 0$.   We would like to know when this pair of planes is a repeated pair.    This entails that the matrix $h_{ij}$ be of rank one.  In particular we need the relation:
\[ 0 = h_{12} h_{34} - h_{14}h_{32} = (c_{12} c_{34} - c_{14} c_{32})Z^1Z^2Z^3Z^4 .\]
So unless all of the quantities $c_{ij} c_{kl} - c_{il} c_{kj}$ vanish for all permutations $(ijkl)$ of the numbers $(1234)$, we must have at least one $Z$ zero. 
\eject\noindent   Henceforth, we assume that the parameters $c_{ij}$ are generic.   When $Z^1$ vanishes,   the only possibly non-zero entry of the first row and column of the matrix $h$ is the first one.  If that entry does not vanish, then it is easy to see that $Z $ must vanish identically, contrary to our hypothesis that $Z$ be non-zero.  So when $Z^1 = 0$, we need the condition $h_{11} = 0$.   Then the matrix $h$ has rank one or less provided the adjoint of its $(1, 1)$-minor vanishes.   Computing this adjoint, we find it is a matrix of rank one with entries $k_1(t, u,v, w) Z^i Z^j$, for $2 \le i, j \le 4$, where we have:
\[ k_1 = c_{23}c_{42}u  + c_{23}c_{34}v+c_{42}c_{34}w.\]
Also $K(t, u, v, w) = k_1 h_{11}$ when $t$ vanishes.  
\begin{itemize} \item So the cone is a repeated pair of planes when:
\[0 =  t = c_{12}u + c_{13}v +  c_{14}w = c_{23}c_{42}u  + c_{23}c_{34}v+c_{42}c_{34}w.\]
This gives the projective point:
\[ (t, u, v, w) = (0, c_{34}(c_{23}c_{14} - c_{42} c_{13}),  c_{42}(c_{34}c_{12} - c_{23}c_{14}), c_{23}(c_{13}c_{42}  - c_{12} c_{34})).\]
\item Similarly, the cone is a repeated pair of planes when:
\[ 0 = u = c_{12}t + c_{23}v +  c_{42}w = c_{14}c_{13}t + c_{13}c_{34}v +c_{14}c_{34}w.\]
This gives the projective point:
\[ (t, u, v, w) = (c_{34}( c_{23}c_{14}  - c_{42}c_{13}), 0, c_{14}(c_{42}c_{13} - c_{12}c_{34}), c_{13}(c_{12}c_{34} -  c_{23}c_{13})).\]
\item Similarly, the cone is a repeated pair of planes when:
\[ 0 = v = c_{13}t + c_{23}u + c_{34}w = c_{14}c_{12}t +c_{12}c_{42}u + c_{14}c_{42}w.\]
This gives the projective point:
\[ (t, u, v, w) = (c_{42}(c_{34}c_{12} - c_{23}c_{14}),  c_{14}(c_{42}c_{13} - c_{12}c_{34}), 0, c_{12}(c_{14}c_{23} - c_{42}c_{13})).\]
\item Similarly, the cone is a repeated pair of planes when:
\[ 0 = w = c_{14}t + c_{42}u + c_{34}v = c_{12}c_{13}t  + c_{12} c_{23} u + c_{13}c_{23}v.\]
This gives the projective point:
\[ (t, u, v, w) = (c_{23}(c_{13}c_{42} - c_{12}c_{34}), c_{13}(c_{12} c_{34} - c_{14}c_{23}), c_{12}(c_{14}c_{23} - c_{42}c_{13}), 0).\]
\end{itemize}
These give after taking square roots, sixteen singular points for the twistor surface.   These sixteen points lie in quartets on the singular cones of the pencil of quadrics given by $Q_a$ and $Q_b$.   In the present language the quadrics $Q_a$ and $Q_b$ have the respective equations:
\[   a_1 t + a_2 u + a_3 v + a_4 w = 0, \]
\[   b_1 t + b_2 u + b_3 v + b_4 w = 0. \]
Then for example one of these cones is:
\[ 0 = (a_1b_2 - a_2 b_1)u + (a_1b_3 - a_3 b_1)v + (a_1 b_4 - a_4 b_1)w =0.\]
This has vertex $(1, 0, 0, 0)$ and polar plane with respect to the pencil $t = 0$.  Then the first projective point given above, $ (0, c_{34}(c_{23}c_{14} - c_{42} c_{13}),  c_{42}(c_{34}c_{12} - c_{23}c_{14}), c_{23}(c_{13}c_{42}  - c_{12} c_{34}))$ lies on the intersection of this cone with the plane $t = 0$,  as is easily checked.\\\\
Summarizing:  the condition that the cone defined by the intersection of a twistor $\alpha$-plane with the Battaglini quadric should degenerate to a product of planes is the condition $K((Z^1), (Z^2)^2,  (Z^3)^2, (Z^4)^2) = 0$; this is a quartic surface in the projective three-space, a Kummer surface.  It then has sixteen singular points, where the two planes of the product coincide.     
\subsection{The cross-ratio}
Let an ordered quartuple $(\alpha, \beta, \gamma, \delta)$ of points be given in a complex vector space $\mathbb{X}$ of dimension two.   Then their cross-ratio  $C(\alpha, \beta, \gamma, \delta)$  is the triple:
\[ (x, y, z) = C(\alpha, \beta, \gamma, \delta) = ((\alpha\wedge\beta)(\gamma\wedge \delta), (\alpha\wedge\gamma)(\delta\wedge \beta),  (\alpha\wedge\delta)(\beta\wedge \gamma)).\]
This is a point in the vector space $\mathbb{L}^2 \oplus   \mathbb{L}^2 \oplus \mathbb{L}^2 $, where $\mathbb{L}$ is the one-dimensional complex vector space $\mathbb{L} = \Omega^2(\mathbb{X})$.  The points $(x, y, z)$ of the image of the cross-ratio lie in the plane $x + y + z = 0$.      Note that at least one pair of the four points is linearly dependent if and only if the product $xyz$ vanishes.   Under the permutation $(\alpha, \beta, \gamma, \delta) \rightarrow (\beta, \alpha, \gamma, \delta)$, we get $(x, y, z) \rightarrow (-x, -z, -y)$.  Under the permutation $(\alpha, \beta, \gamma, \delta) \rightarrow (\gamma, \beta, \alpha, \delta)$, we get $(x, y, z) \rightarrow (-z, -y, -x)$.  Under the permutation $(\alpha, \beta, \gamma, \delta) \rightarrow (\delta, \beta, \gamma, \alpha)$, we get $(x, y, z) \rightarrow (- y, -x, -z)$.    So even permutations permute the triple, whereas odd permutations permute and simultaneously change all the signs of the triple.
\\\\
Define:
\[ (J, K)  = \left(2^{-1}(x^2 + y^2 + z^2), 3^{-1}(x^3 + y^3 + z^3)\right) \in   \mathbb{L}^4 \oplus \mathbb{L}^6.\]
Then, since $x + y + z = 0$, we have the relations:
\[  J + xy + yz + zx = K + xyz =  0.\]
So $x$, $y$ and $z$ are the roots of the cubic:
\[ s^3 - Js - K = 0.\]
Here $s \in \mathbb{L}^2$.
The lambda invariant of the points $(\alpha, \beta, \gamma, \delta)$ is then the ratio of the pair $ (J^3, K^2) \in \mathbb{L}^{12} \oplus \mathbb{L}^{12}$.   Notice that this ratio is a pure (extended) complex number and is well-defined provided $J$ and $K$ do not both vanish.    The exceptional case when $\lambda$ is undefined, when $J$ and $K$ both vanish is the case that $x, y$ and $z$ are all zero, so is the case that  at least one of the four points $\alpha, \beta, \gamma$ or $\delta$ vanishes, or at least three of the four points are pairwise linearly dependent.   The cubic has a repeated pair of roots,  but not all three roots coincident, if and only if $\lambda$ is well-defined and takes the value $\displaystyle{\lambda = \frac{27}{4}}$.   In this case the four points are said to be harmonic.  There is another special value of $\lambda$,  $\lambda = 0$ (i.e $J = 0$), corresponding to $x$, $y$ and $z$ being not all zero and proportional to the cube roots of unity.  In this case the four points are said to be equianharmonic. 

If now  the two points $\alpha$ and $\beta$ of $\mathbb{X}$ are linearly independent, we may write $\gamma = p\alpha + q\beta$ and $\delta = r \alpha + s\beta$, for complex numbers $(p, q, r, s)$.   Then  we have:
\[ (x,\hspace{5pt}  y, \hspace{5pt} z) = (\alpha\wedge \beta)^2(ps - qr,\hspace{5pt}  qr, \hspace{5pt}   -ps) = (- t - u, \hspace{5pt} u, \hspace{5pt} t),  \hspace{10pt} t = - ps (\alpha \wedge \beta)^2,    \hspace{10pt} u = qr (\alpha\wedge \beta)^2. \]
This gives the formulas:
\[ J =  t^2 + u^2 +  tu,  \hspace{10pt} K =  tu(t + u).\]
The equianharmonic case is the case that $t = \omega u$, where $\omega^2 + \omega + 1 = 0$.  The harmonic case is the case that $t = u$, or $t = - 2u$, or $u = -2t$.
\eject\noindent
When the four points $(\alpha, \beta, \gamma, \delta)$ are each non-zero elements of $\mathbb{X}$, they give rise to an homogeneous quartic $f(\pi)$ in a variable $\pi \in \mathbb{X}$ by the formula:  
\[ f(\pi) = (\pi \wedge \alpha)(\pi\wedge \beta)(\pi \wedge \gamma)(\pi\wedge \delta).\]
Here $f$ takes values in $\mathbb{L}^4$ and has its roots proportional to the quartet of points.    Note that the $\lambda$ invariant is invariant under permutations of the points, and under scalings of the points, so depends only of the quartic $f$ (up to scale), not on its roots. \\\\ The three-dimensional vector space $\mathbb{X} \odot \mathbb{X}$ has a natural (invertible) symmetric bilinear form $G(A, B)$, defined for any $A$ and $B$ in  $\mathbb{X} \odot \mathbb{X}$ and taking values in $\mathbb{L}^2$, such that $G(\pi \otimes \pi, \rho\otimes \rho) = (\pi\wedge \rho)^2$ for any $\pi$ and $\rho$ in $\mathbb{X}$.   Then there is a unique symmetric bilinear form $F$, taking values in $\mathbb{L}^4$, on the space $\mathbb{X} \odot \mathbb{X}$, such that $F$ is trace-free with respect to the metric $G$ and such that $F(\pi\otimes \pi, \pi\otimes\pi) = f(\pi)$,  for any $\pi \in \mathbb{X}$.  The characteristic polynomial of $F$ is the cubic $s^3 - 2^{-1}sA - 3^{-1}B$, whose roots are $\frac{1}{6} (y - x),  \frac{1}{6}( x - z)$ and $\frac{1}{6}(z - y)$ (here $s$ takes values in $\mathbb{L}^2$).    Here we have:
\[  A = \textrm{tr}((FG^{-1})^2) \in \mathbb{L}^4, \hspace{10pt} B =  \textrm{tr}((FG^{-1})^3)\in \mathbb{L}^6.\]
Also we have the relations:
\[ (6A)^2 = J^2, \hspace{10pt} (72 B)^2  =   4J^3 - 27K^2.\]
The harmonic case is the case that $B = 0$.  \\The equianharmonic case is the case that $A = 0$.

\end{document}